\documentclass[conference]{IEEEtran}
\usepackage{etex}
\usepackage{pictexwd,dcpic}
\usepackage[usenames,dvipsnames]{color}
\usepackage{amsmath}
\usepackage{amsfonts}
\usepackage{amssymb}
\usepackage{centernot}
\usepackage{calc}
\usepackage{graphicx}
 \usepackage{enumitem}
\usepackage{multirow}
\usepackage{framed}
\usepackage{hyperref}
\usepackage{xspace}
\usepackage{tabularx}
\usepackage{tabulary}
\usepackage{tabu}
\usepackage{footnote}
\usepackage[shortcuts]{extdash}
\usepackage{bill}

\makesavenoteenv{table}

\newcommand{\rbac}{\ensuremath{\mathsf{RBAC}}\xspace}
\newcommand{\abac}{\ensuremath{\mathsf{ABAC}}\xspace}
\newcommand{\file}{\ensuremath{\mathsf{F}}\xspace}
\newcommand{\filekey}{\ensuremath{\mathsf{FK}}\xspace}
\newcommand{\rolekey}{\ensuremath{\mathsf{RK}}\xspace}
\newcommand{\readp}{\ensuremath{\mathsf{Read}}\xspace}
\newcommand{\writep}{\ensuremath{\mathsf{Write}}\xspace}
\newcommand{\rwp}{\ensuremath{\mathsf{RW}}\xspace}

\newcommand{\G}{\mathbb{G}}
\newcommand{\Z}{\mathbb{Z}}

\newcommand{\true}{\textsc{true}}
\newcommand{\false}{\textsc{false}}

\newcommand{\mm}[1]{\ensuremath{\mathbf{#1}}}
\newcommand{\SymGen}{\ensuremath{\mm{Gen}^\mm{Sym}}\xspace}
\newcommand{\SymEnc}{\ensuremath{\mm{Enc}^\mm{Sym}}\xspace}
\newcommand{\SymDec}{\ensuremath{\mm{Dec}^\mm{Sym}}\xspace}

\newcommand{\PKGen}{\ensuremath{\mm{Gen}^\mm{Pub}}\xspace}
\newcommand{\PKEnc}{\ensuremath{\mm{Enc}^\mm{Pub}}\xspace}
\newcommand{\PKDec}{\ensuremath{\mm{Dec}^\mm{Pub}}\xspace}
\newcommand{\SigGen}{\ensuremath{\mm{Gen}^\mm{Sig}}\xspace}
\newcommand{\SigSign}{\ensuremath{\mm{Sign}^\mm{Sig}}\xspace}
\newcommand{\SigVerify}{\ensuremath{\mm{Ver}^\mm{Sig}}\xspace}
\newcommand{\IBEMSKGen}{\ensuremath{\mm{MSKGen}^\mm{IBE}}\xspace}
\newcommand{\IBEKeyGen}{\ensuremath{\mm{KeyGen}^\mm{IBE}}\xspace}
\newcommand{\IBEEnc}{\ensuremath{\mm{Enc}^\mm{IBE}}\xspace}
\newcommand{\IBEDec}{\ensuremath{\mm{Dec}^\mm{IBE}}\xspace}
\newcommand{\IBSMSKGen}{\ensuremath{\mm{MSKGen}^\mm{IBS}}\xspace}
\newcommand{\IBSKeyGen}{\ensuremath{\mm{KeyGen}^\mm{IBS}}\xspace}
\newcommand{\IBSSign}{\ensuremath{\mm{Sign}^\mm{IBS}}\xspace}
\newcommand{\IBSVerify}{\ensuremath{\mm{Ver}^\mm{IBS}}\xspace}
\newcommand{\enckey}{\ensuremath{\mm{k}^\mm{enc}}\xspace}
\newcommand{\deckey}{\ensuremath{\mm{k}^\mm{dec}}\xspace}
\newcommand{\verkey}{\ensuremath{\mm{k}^\mm{ver}}\xspace}
\newcommand{\sigkey}{\ensuremath{\mm{k}^\mm{sig}}\xspace}

\newcommand{\ext}[4]{#1 & #2 & #3 & #4}
\newcommand{\key}[4]{#1 & #2 & #3 & #4}
\newcommand{\enc}[4]{#1 & #2 & #3 & #4}
\newcommand{\ciph}[4]{#1 & #2 & #3 & #4}
\newcommand{\dec}[4]{#1 & #2 & #3 & #4}

\newcommand\ds[1]{\textsf{#1}}

\newtheorem{theorem}{Theorem}
\newtheorem{proposition}{Proposition}

\begin{document}

\title{On the Practicality of Cryptographically Enforcing Dynamic Access
Control Policies in the Cloud \\ \smaller[1] (Extended Version)}

\author{
\IEEEauthorblockN{William C.\ Garrison III}
\IEEEauthorblockA{University of Pittsburgh}
\and
\IEEEauthorblockN{Adam Shull}
\IEEEauthorblockA{Indiana University}
\and
\IEEEauthorblockN{Steven Myers}
\IEEEauthorblockA{Indiana University}
\and
\IEEEauthorblockN{Adam J.\ Lee}
\IEEEauthorblockA{University of Pittsburgh}
}


\maketitle

\begin{abstract}
The ability to enforce robust and dynamic access controls on cloud-hosted data
while simultaneously ensuring confidentiality with respect to the cloud itself
is a clear goal for many users and organizations.  To this end, there has been
much cryptographic research proposing the use of (hierarchical) identity-based
encryption, attribute-based encryption, predicate encryption, functional
encryption, and related technologies to perform robust and private access
control on untrusted cloud providers. However, the vast majority of this work
studies static models in which the access control policies being enforced do not
change over time.  This is contrary to the needs of most practical applications,
which leverage dynamic data and/or policies.


In this paper, we show that the cryptographic enforcement of dynamic access controls on untrusted platforms
incurs computational costs that are likely prohibitive in practice.
Specifically, we develop
lightweight constructions for enforcing role-based access controls (i.e.,
$\rbac_0$) over cloud-hosted files using identity-based and traditional
public-key cryptography. This is done under a threat model as close as possible
to the one assumed in the cryptographic literature.  We prove the correctness of
these constructions, and leverage real-world $\rbac$ datasets and recent
techniques developed by the access control community to experimentally analyze,
via simulation, their associated computational costs. This analysis shows that
supporting revocation, file updates, and other state change functionality is
likely to incur prohibitive overheads in even minimally-dynamic, realistic scenarios. We identify a
number of bottlenecks in such systems, and fruitful areas for future work that
will lead to more natural and efficient constructions for the cryptographic
enforcement of dynamic access controls. Our findings naturally extend to the use
of more expressive cryptographic primitives (e.g., HIBE or ABE) and richer
access control models (e.g., $\rbac_1$ or \abac).
\end{abstract}


\section{Introduction}

In recent years, numerous cryptographic schemes have been developed to support
access control on the (untrusted) cloud. One of the most expressive of these is
attribute-based encryption (ABE)~\cite{Goyal2006}, which is a natural fit for
enforcing attribute-based access control (\abac) policies~\cite{abac}. However,
the practical implications of using these types of cryptographic schemes to
tackle realistic access control problems are largely unexplored. In particular,
much of the literature concerns static scenarios in which data and/or access
control policies are rarely, if ever, modified
(e.g.,~\cite{SW05,Goyal2006,BSW07,Moller12,OSW07,GJPS08,KSW08}).  Such scenarios
are not representative of real-world systems, and oversimplify issues associated
with key management and revocation that can carry substantial practical
overheads. In this paper, we explore \emph{exactly} these types of issues in an
attempt to understand the computational overheads of using advanced
cryptographic techniques to enforce dynamic access controls over objects stored
on untrusted platforms.  Our primary result is negative: we demonstrate that
prohibitive computational burdens are likely to be incurred when supporting
practical, dynamic workloads.

The push to develop and use cryptography to support adaptive access control on
the cloud is natural. Major cloud providers such as Google, Microsoft, Apple,
and Amazon are providing both large-scale, industrial services and
smaller-scale, consumer services. Similarly, there are a number of user-focused
cloud-based file sharing services, such as Dropbox, Box, and Flickr. However,
the near-constant media coverage of data breaches has raised both consumer and
enterprise concerns regarding the privacy and integrity of cloud-stored data.
Among the widely-publicized stories of external hacking and data disclosure are
releases of private photos \cite{CelebGate14}. Some are even state-sponsored
attacks against cloud organizations themselves, such as Operation Aurora, in
which Chinese hackers infiltrated providers like Google, Yahoo, and
Rackspace~\cite{aurora,googleblog}. Despite the economic benefits and
ease-of-use provided by outsourcing data management to the cloud, this practice
raises new questions  regarding the maintenance and enforcement of the access
controls that users have come to expect from file sharing systems.

Although advanced cryptographic primitives seem well-suited for protecting
\emph{point states} in many access control paradigms, supporting the
\emph{transitions} between protection states that are triggered by
administrative actions in a dynamic system requires addressing very subtle
issues involving key management, coordination, and key/policy consistency.
While there has been some work seeking to provide a level of dynamism for these
types of advanced cryptographic primitives, this work is not without issues.
For instance, techniques have been developed to support key
revocation~\cite{Boldyreva2008} and delegated
re-encryption~\cite{Green:2007:IPR:1419765.1419791,sahai2012dynamic}.
Unfortunately, these techniques are not compatible with hybrid
encryption---which is necessary from an efficiency perspective---under
reasonable threat models.

In this paper, we attempt to tease out these types of critical details by
exploring the cryptographic enforcement of a widely-deployed access control
model: role-based access control (specifically, $\rbac_0$~\cite{rbac}).  In
particular, we develop two constructions for cryptographically enforcing dynamic
$\rbac_0$ policies in untrusted cloud environments: one based on standard
public-key cryptographic techniques, and another based on identity-based
encryption/signature (IBE/IBS)
techniques~\cite{Boneh2003,SW05,Cha:2003:ISG:648120.746918}.  By studying
$\rbac_0$ in the context of these relatively efficient cryptographic schemes, we
can effectively \emph{lower-bound the costs} that would be associated with
supporting richer access controls (e.g., \abac) by using more advanced---and
more expensive---cryptographic techniques exhibiting similar administrative and
key delegation structures (e.g., ABE).

We use tools from the access control
literature~\cite{Hinrichs:2013:AAC:2510170.2510419} to prove the correctness of
our $\rbac_0$ constructions.  To quantify the costs of using these constructions
in realistic access control scenarios, we leverage a stochastic modeling and
simulation-based approach developed to support access control suitability
analysis~\cite{garrison2014sacmat}. Our simulations are driven by real-world
$\rbac$ datasets that allow us to explore---in a variety of environments where
the $\rbac_0$ policy and files in the system are subject to dynamic change---the
costs associated with using these constructions. In doing so, we uncover several
design considerations that must be addressed, make explicit the complexities of
managing the transitions that occur as policies or data are modified at runtime,
and demonstrate the often excessive overheads of relying solely on advanced
cryptographic techniques for enforcing dynamic access controls. This provides us
with a number of insights toward the development of more effective cryptographic
access controls. Through our analysis, we make the following contributions:

\begin{itemize}[itemsep=0pt,leftmargin=*]

\item We demonstrate that the cryptographic enforcement of role-based access
controls on the cloud incurs overheads that are likely prohibitive in realistic
dynamic workloads.  For instance, we show that removing a single user from a
role in a moderately-sized organization can require hundreds or thousands of IBE
encryptions!  Since our constructions are designed to lower-bound deployment
costs (given current cryptographic techniques), this indicates that
cryptographic access controls are likely to carry prohibitive costs for even
mildly dynamic scenarios.

\item Prior work often dismisses the need for an access control reference
monitor when using cryptographically-enforced access controls
(e.g.,~\cite{Goyal2006,BSW07,GJPS08,OSW07}). We discuss the necessity of some
minimal reference monitor on the cloud when supporting dynamic,
cryptographically-enforced access controls, and we outline other design
considerations that must be addressed in dynamic environments.

\item We develop constructions that use either the IBE/IBS or public-key
cryptographic paradigms to enable dynamic outsourced $\rbac_0$ access controls.
In an effort to lower-bound deployment costs, our constructions exhibit design
choices that emphasize efficiency over the strongest possible security (e.g.,
using lazy rather than online re-encryption,
cf.\ \cref{sec:implementation-details}), but are easily extended to
support stronger security guarantees (albeit at additional costs).  These
constructions further highlight practical considerations that are often
overlooked in the literature, or that prevent the application of techniques
designed to enhance the dynamism of advanced cryptographic techniques.

\item Having established the infeasibility of enforcing even the relatively
simple $\rbac_0$ in dynamic scenarios, we discuss the increase in costs that
would be associated with more expressive cryptographically-enforced access
control such as hierarchical $\rbac$ ($\rbac_1$) using
HIBE~\cite{Gentry02,BonehBoyen2005}, or attribute-based access control ($\abac$)
using ABE.

\end{itemize}

The remainder of this paper is organized as follows. In \cref{sec:rw}, we
discuss relevant related work. \Cref{sec:bg} documents our system model and
assumptions, and provides background on $\rbac_0$ and the cryptographic
techniques used in this paper. In \cref{sec:constructions}, we describe our
IBE/IBS construction in detail, and overview the key differences between it and
our PKI-based construction. \Cref{sec:analysis} presents theorems stating the
correctness of our constructions, as well as experimental results showing the
overheads incurred by our constructions when applied to real-world $\rbac$
datasets. In \cref{sec:discussion}, we identify interesting directions for
future work informed by our findings. \Cref{sec:conclusions} details our
conclusions.


\section{Related Work} \label{sec:rw}

\subsection{Access Control}

Access control is one of the most fundamental aspects of computer security, with
instances occurring pervasively throughout most computer systems: relational
databases often provide built-in access control commands; network administrators
implement access controls, e.g., firewall rules and router ACLs; operating
systems provide access control primitives that enable users to protect their
private files; and web applications and other frameworks typically implement
purpose-specific access controls to control access to the information that they
manage. The literature describes a diversity of access control systems
supporting policies including basic access control lists~\cite{saltzer},
cryptographically-enforced capabilities~\cite{kerberos},
group\=/~\cite{KrishnanNSW11}, role\=/~\cite{rbac}, and attribute-based~\cite{abac}
controls.  Despite
this diversity, a central theme in most access control work is the reliance on a
fully-trusted reference monitor to check compliance with the policy to be
enforced prior to brokering access to protected resources. This dependency on a
trusted reference monitor is problematic, however, when resources are stored on
(potentially) untrusted infrastructure.

Distributed or decentralized approaches to access control have also
been well studied in the literature and in practice.  Work in the
trust management space (e.g.,~\cite{BlazeFL96, ellison1999rfc,
  li2002oakland, becker2010jcs}) allows the specification of
\emph{declarative access control policies} for protecting resources,
which are satisfied using \emph{digital credentials} of various forms.
For instance, a research portal may allow free access to publications,
provided that the requester is a graduate student at an accredited
university.  This allows the portal to delegate trust: provided that a
requestor can produce a proof-of-ownership for a ``graduate student''
attribute certificate issued by an accredited university, she
will be permitted access.  We note that these approaches need not rely
on heavyweight certificate infrastructures; recent work has provided
similar functionality using lightweight cryptographic bearer
credentials~\cite{birgisson2014ndss}.  Further, widely-deployed
identity management solutions (e.g., OAuth~\cite{oauth}) can also be
viewed as simplified trust management approaches that offload identity
verification to a third party, receiving only a ``token'' attesting to
a requestor's identity.
In all cases, however, a trusted reference monitor is still required
to validate that the presented credentials actually satisfy the policy
protecting a resource.


In this paper, by contrast, we investigate the implications of using
cryptography to enforce access controls on cloud-based storage infrastructure,
where the provider is not trusted to view file contents.

\subsection{Cryptography}

We assume the reader is familiar with basic concepts from symmetric-key and
public-key cryptography, and many references exist
(e.g.,~\cite{LindellKatz2014}) discussing these topics. Starting with the
development of practical identity-based encryption (IBE)
schemes~\cite{Boneh2003}, there has been considerable work on the
development of cryptographic systems that directly support a number of access
control functionalities, with examples including hierarchical
IBE~\cite{Horwitz2002,Gentry02}, attribute-based encryption~\cite{SW05}, and
functional encryption~\cite{Sahai2010}. At a high level, these encryption
schemes encrypt data to a policy, so that only those who have
secret keys satisfying the policy can decrypt. What varies between these types of schemes is the expressiveness
of the policies that are supported. With IBE and traditional public-key
encryption, one can encrypt to a given target individual, and only
that individual can decrypt. With attribute-based encryption, a ciphertext can
be encrypted to a certain policy, and can be decrypted only by individuals whose
secret keys satisfy that policy. With functional encryption, a certain function
is embedded in the ciphertext, and when one ``decrypts,'' one does not retrieve
the underlying value, but rather a function of the encrypted value and the
decryptor's secret key. One underlying motivation in all of the above work is
the ability to enforce access controls on encrypted data.

Each cryptographic scheme has its own associated costs, but they can
be broadly categorized as follows.  Symmetric cryptography is orders
of magnitude faster than traditional public-key encryption, and
traditional public-key encryption is an order of magnitude faster than
pairing-based cryptography, in which the pairing operation itself
typically carries the largest cost.\footnote{We will exclude
  lattice-based systems, due to the difficulty in determining
  appropriate security parameters.  This, amongst other factors, makes
  such generic comparisons difficult.} The vast majority of IBE, IBS,
HIBE and ABE schemes are pairing-based cryptographic schemes. IBE
schemes use a small constant number of pairings in either
encryption or decryption. In contrast, ABE schemes use a number of
pairings that is a function of the policy being encoded, and thus,
assuming minimally expressive access policies, have computational
costs substantially greater than IBE.

Much of the work on these advanced cryptographic systems allows for data to be
stored on the cloud, but it does not address the issue of revocation or dynamic
modification of the access control structure being used to store data on the
cloud. This can, of course, be done by downloading the data, decrypting it, and
then re-encrypting under a new policy, but this is communication intensive, and
potentially computationally intensive too. Further, for large files, clients
making the changes in the access structure may not be able to support the entire
file locally (e.g., smartphones). Therefore, there has been some work done in
considering delegated encryption and revocation in these models
(e.g.,~\cite{Green:2007:IPR:1419765.1419791,Boldyreva2008,LibertV09,Green:2011:ODA:2028067.2028101,sahai2012dynamic,SeoE13,ParkLL13}).

\subsection{Cryptographic Access Controls}

There has been significant work on using cryptography as an access control
mechanism, starting with seminal works such as that by Gudes~\cite{Gudes80}.
This work describes how access controls can be enforced using cryptography, but
does not address many practical issues such as key distribution and management,
policy updates, and costs. Furthermore, as the work's motivation is a local file
system, the access control system must be trusted with the keys (and trusted to
delete them from memory as soon as possible). Work by Akl and
Taylor~\cite{AklTay83} addresses some of the key management issues by proposing
a \emph{key assignment scheme}: a system for deriving keys in a hierarchical
access control policy, rather than requiring users higher in the hierarchy to
store many more keys than those lower in the hierarchy. Again, this work does
not consider key distribution or policy updates. Later work in key hierarchies
by Atallah et al.~\cite{AtallahBFF09} proposes a method that allows policy
updates, but in the case of revocation, all descendants of the affected node in
the access hierarchy must be updated, and the cost of such an operation is not
discussed. Continued work in key assignment schemes has improved upon the
efficiency of policy updates; see \cite{CramptonMW06} for a survey of such
schemes that discusses tradeoffs such as how much private vs.\ public
information must be stored and how much information must be changed for policy
updates. Much of this work focuses on the use of symmetric-key cryptography, and
so its use for the cloud is potentially limited.

De Capitani di Vimercati et al.~\cite{diVimercati07,VimercatiFJPS10} describe a
method for cryptographic access controls on outsourced data using double
encryption (one layer by the administrator and one by the service). An extension
to this work also enforces write privileges~\cite{DeCapitani2013}. However, this
solution requires a high degree of participation by the cloud provider or third
party, and the work does not address the high cost of such operations as
deleting users (which can incur cascading updates). Ibraimi's
thesis~\cite{so78287} proposes methods for outsourcing data storage using
asymmetric encryption. However, the proposed method for supporting revocation
requires a trusted mediator and keyshare escrow to verify all reads against a
revocation list (and does not address revoked users reusing cached keyshares).
Furthermore, policy updates require an active entity to re-encrypt all affected
files under the new policy. Similarly, work by Nali et al.~\cite{NaliAM04}
enforces \rbac using public-key cryptography, but requires a
series of active security mediators.

Crampton has shown that cryptography is sufficient to enforce RBAC
policies~\cite{Crampton10} and general interval-based access control
policies~\cite{Crampton11}, but revocation and policy updates are not considered
(i.e., the constructions are shown only for static policies). Ferrara et
al.~\cite{FerraraFW13} formally define a cryptographic game for proving the
security of cryptographically-enforced \rbac systems and prove that such
properties can be satisfied using an ABE-based construction. This construction
has since been extended to provide policy privacy and support writes with less
trust on the provider~\cite{FerraraCSF15}. The latter is accomplished by
eliminating the reference monitor that checks if a write is allowed and instead
accepting each write as a new version; versions must then be verified when
downloaded for reading to determine the most recent permitted version (the
provider is trusted to provide an accurate version ordering). However, these
works do not consider the costs and other practical considerations for using
such a system in practice (e.g., lazy vs.\ active re-encryption, hybrid
encryption). In this paper, we consider exactly these types of issues.

Pirretti et al. \cite{Pirretti06} have shown that distributed file systems and social networks can
use ABE-based constructions to perform practical access control, but they leave dynamic revocation as future work.


\section{Background and Assumptions} \label{sec:bg}

Our goal is to understand the practical costs of leveraging public-key
cryptographic primitives to implement outsourced dynamic access controls in the
cloud. In this section, we (i) define the system and threat models in which we
consider this problem, (ii) specify the access control model that we propose to
enforce, and (iii) define the classes of cryptographic primitives that will be
used in our constructions.

\subsection{System and Threat Models}

\begin{figure}[t]
\centering
\includegraphics[width=0.74\columnwidth]{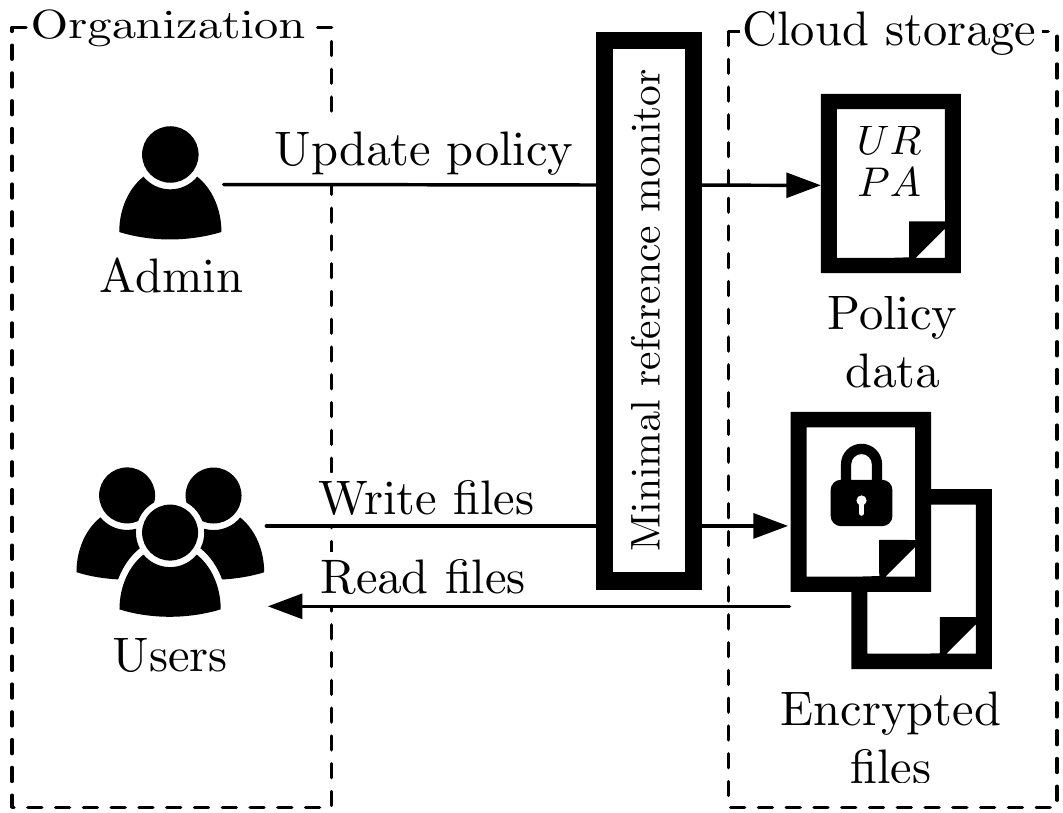}
\caption{Diagram of a cloud storage system}
\label{Fig:Admin-Cloud-Client}
\end{figure}

The environment that we consider---which is based on the untrusted
cloud provider typically assumed in the cryptographic literature---is
depicted in \cref{Fig:Admin-Cloud-Client}.  The system consists of
three main (classes of) entities: \emph{access control
  administrators}, \emph{users/clients}, and \emph{cloud storage
  providers}.  In particular, we consider a model in which a single
storage provider is contracted by an organization.  This is analogous
to companies contracting with providers like Microsoft (via OneDrive
for Business) or Dropbox (via Dropbox Business) to outsource
enterprise storage, or individuals making use of cloud platforms like
Apple iCloud or Google Drive for hosting and sharing personal media.
Further, this simplifies the overall system design by eliminating the
need for a secondary mechanism that synchronizes cryptographic
material and other metadata.

\textbf{Assumptions.} The \emph{cloud storage provider} is contracted
to manage the storage needs of a (perhaps virtual) organization.  This
includes storing the files hosted in the cloud, as well as any
metadata associated with the access control policies protecting these
files.  We assume that the cloud is \emph{not} trusted to view the
contents of the files that it stores. However, it is trusted to ensure
the availability of these files, and to ensure that only authorized
individuals update these files. File access is assumed to occur
directly though the cloud provider's API, with read access permissions
being enforced cryptographically on the client side, and write access
permissions being enforced by a minimal reference monitor on the cloud
provider that validates client signatures that prove write privileges
prior to file updates.\footnote{Note that this eliminates the
  possibility of a purely symmetric-key approach: the ability to
  validate, e.g., symmetric-key MACs would also allow the cloud
  provider to \emph{modify} these MACs.}  In short, the storage provider
ensures file system consistency by preventing unauthorized updates,
yet cannot read or make legitimate modifications to files or metadata.

\emph{Access control administrators} are tasked with managing the
protection state of the storage system. That is, they control the
assignment of access permissions, which entails the creation,
revocation, and distribution of cryptographic keys used to protect
files in a role-based manner. Metadata to facilitate key distribution
is stored in a cryptographically-protected manner on the cloud
provider.  \emph{Users} may download any file stored on the storage
provider, but may decrypt, read, and (possibly) modify only the files
for which they have been issued the appropriate (role-based) keys. All
files are encrypted and signed prior to being uploaded to the cloud storage
provider.  Finally, we assume that all parties can communicate via
pairwise-authenticated and private channels (e.g., SSL/TLS tunnels).


\textbf{Implications.}  To simplify presentation and analysis, the
above threat model does leave some degree of trust in the cloud
provider (albeit far less than is routinely placed in these providers
today).  In particular, the cloud provider is trusted to verify
digital signatures prior to authorizing write operations.  This could
be avoided by using a versioning file system, allowing all writes, and
relying on clients to find the most recent version of a file that has
a valid signature prior to accessing that file.  Similarly, it is
possible---although prohibited by our threat model---for a malicious
provider to ``roll back'' the filesystem to a prior state by replacing
current files and metadata with previous versions. We note that it is
possible to detect (e.g., via comparison with off-cloud metadata) or
prevent (e.g., by splitting metadata and file storage across multiple
providers) this issue, and thus this prohibition could be dropped.
Further, we do not consider the denial-of-service threat of a user
overwhelming the storage provider with spurious file downloads; in
practice, this is easily addressed by using unguessable (perhaps
cryptographically-produced) file names, or lightweight authorization tokens.
However, all of these
types of relaxations come with additional complexity.  As we will
demonstrate, the costs associated with cryptographic enforcement of
dynamic access controls are likely prohibitive, \emph{even under the above
  threat model.}  This, effectively, lower-bounds the costs entailed
by weaker threat models (which require more complex mechanisms).  For
the bulk of this paper, we will therefore focus on the above threat
model, leaving discussion of further relaxations to
\cref{sec:discussion}.

\subsection{Access Control Model}

In this paper, we focus on cryptographic enforcement of a role-based access
control ($\rbac$) system, given the prevalence of this type of access control
system in both the research literature and commercial systems. $\rbac$ systems
simplify permission management through the use of abstraction: roles describe
the access permissions associated with a particular (class of) job function,
users are assigned to the set of roles entailed by their job responsibilities,
and a user is granted access to an object if they are assigned to a role that is
permitted to access that object. In this paper, we will investigate
cryptographic implementations of the simplest $\rbac$ formulation:
$\rbac_0$~\cite{rbac}. More formally, the state of an $\rbac_0$ system can be
described as follows:
\begin{itemize}[itemsep=0pt]

\item $U$ is a set of users,

\item $R$ is a set of roles,

\item $P$ is a set of permissions (e.g., $\langle file, op \rangle$),

\item $PA \subseteq R \times P$ is the permission assignment relation, and

\item $UR \subseteq U \times R$ is the user assignment relation.

\end{itemize}
The authorization predicate $auth: U \times P \rightarrow \mathbb{B}$ determines
whether user $u$ can use permission $p$ and is defined as follows:
\[auth(u,p) = \exists r : [(u,r) \in UR] \wedge [(r,p) \in PA]\]

Many variants of $\rbac$ exist, but we focus on the use of $\rbac_0$ as it is
conceptually the simplest of these variants yet still provides adequate
expressive power to be interesting for realistic applications.  Generalizing this model to richer $\rbac$ variants (e.g., $\rbac_1$) and attribute-based access control (\abac) is discussed in \cref{sec:discussion-generalizing}.

\subsection{Cryptographic Primitives}
\label{sec:crypto}

Both of our constructions make use of symmetric-key authenticated encryption
($\SymGen$, $\SymEnc$, $\SymDec$). Our PKI scheme uses public-key encryption and
digital signatures ($\PKGen$, $\PKEnc$, $\PKDec$, $\SigGen$, $\SigSign$,
$\SigVerify$). While many attribute-based encryption (ABE) schemes are being
developed to support policy constructions of varying expressivity, $\rbac_0$
does not require this level of sophistication. To this end, we instead use
identity-based encryption (IBE):
\begin{itemize}[leftmargin=*,itemsep=0pt]

\item $\IBEMSKGen(1^n)$: Takes security parameter $n$; generates public
parameters (which are implicit parameters to every other IBE algorithm) and
master secret key $msk$.

\item $\IBEKeyGen(ID,msk)$: Generates a decryption key $k_{ID}$ for identity $ID$.

\item $\IBEEnc_{ID}(M)$: Encrypts message $M$ under identity $ID$.

\item $\IBEDec_{k_{ID}}(C)$: Decrypts ciphertext $C$ using key $k_{ID}$;
correctness requires that $\forall\;ID$ if $k_{ID} = \IBEKeyGen(ID)$ then
$\forall\;M, \IBEDec_{k_{ID}}(\IBEEnc_{ID}(M))=M$.

\end{itemize}

We also use identity-based signature (IBS) schemes:
\begin{itemize}[leftmargin=*,itemsep=0pt]

\item $\IBSMSKGen(1^n)$: Takes security parameter $n$; generates public
parameters (which are implicit parameters to every other IBS algorithm) and
master secret key $msk$.

\item $\IBSKeyGen(ID,msk)$: Generates a signing key $s_{ID}$ for identity $ID$.

\item $\IBSSign_{ID, s_{ID}}(M)$: Generates a signature $sig$ on message $M$ if
$s_{ID}$ is a valid signing key for $ID$.

\item $\IBSVerify_{ID}(M,sig)$: Verifies whether $sig$ is a valid signature on
message $M$ for identity $ID$; requires that $\forall\;ID$ \\
if $s_{ID} = \IBSKeyGen(ID)$ then \\
$\forall\;M, \IBSVerify_{ID}(M, \IBSSign_{ID, s_{ID}}(M))=1$.

\end{itemize}

IBE (resp.\ IBS) schemes build upon traditional public-key schemes by allowing
any desired string to act as one's encryption (resp.\ verification) key. This
requires the introduction of a third party who can generate the decryption and
signing keys corresponding to these identity strings. This third party, who
holds the master keys, is able to produce decryption or signing keys for anyone,
and thus the system has inbuilt escrow. In our use of these systems, the $\rbac$
administrator(s) will act as this third party. Since administrators
traditionally have the power to access/assign arbitrary permissions, this escrow
is not a weakness. In practice, if this is still a concern, threshold/secret
splitting schemes can be used to distribute trust amongst several individuals. However, such schemes would
increase the cryptographic costs of operations associated with the master key.


\section{Construction} \label{sec:constructions}

While cryptographic access control enforcement has been studied in the past, the
focus has been almost entirely on techniques that are best suited for
\emph{mostly static} scenarios lacking a trusted reference monitor
(e.g.,~\cite{Goyal2006,Moller12}), in which the policies to be enforced and
files to be protected change very little over time. As such, the particulars
associated with \emph{securely} managing policy change and the associated
overheads have been largely under-explored. In this
\lcnamecref{sec:constructions}, we begin with a strawman construction for
cryptographic access control enforcement, and use it to highlight a variety of
limitations and design considerations that must be addressed. We conclude with a
detailed description of our IBE/IBS and PKI constructions for $\rbac_0$, which address these
issues.

\subsection{A Strawman Construction}

At first blush, it seems conceptually simple to provision a
cryptographically-enforced $\rbac_0$ system. We now overview such a system,
which will allow us to highlight a variety of issues that arise as a result.
This strawman construction will make use of IBE/IBS; the use of a more
traditional PKI is a straightforward translation. We assume that the
administrator holds the master secret keys for the IBE/IBS systems.

\begin{itemize}[itemsep=0pt,leftmargin=*]

\item \textbf{Registration.} Each user, $u$, of the system must carry out an
initial registration process with the administrator. The result of this process
is that the user will obtain identity-based encryption and signing keys $k_u
\leftarrow \IBEKeyGen(u)$ and $s_u \leftarrow \IBSKeyGen(u)$ from the
administrator.

\item \textbf{Role Administration.} For each role, $r$, the administrator will
generate identity-based encryption and signing keys $k_r \leftarrow
\IBEKeyGen(r)$ and $s_r \leftarrow \IBSKeyGen(r)$. For each user $u$ that is a
member of $r$ (i.e., for each $(u, r) \in UR$ in the $\rbac_0$ state), the
administrator will create and upload a tuple of the form:
\[ \langle \rolekey, u, r, \IBEEnc_u(k_r, s_r), \IBSSign_{SU} \rangle. \]
This tuple provides $u$ with cryptographically-protected access to the
encryption and signing keys for $r$, and is signed by the administrator. Here,
$\IBSSign_{SU}$ at the end of the tuple represents an IBS signature by identity
$SU$ (the administrator), and \rolekey is a sentinel value indicating that this
is a role key tuple.

\item \textbf{File Administration.} For each file $f$ to be shared with a role
$r$ (i.e., for each $(r, \langle f, op \rangle) \in PA$ in the $\rbac_0$ state),
the administrator will create and upload a tuple:
\[ \langle \file, r, \langle fn, op \rangle, \IBEEnc_r(f), SU, \IBSSign_{SU} \rangle.
\]
This tuple contains a copy of $f$ that is encrypted to members of $r$. Here,
$fn$ represents the name of the file $f$, while $op$ is the permitted
operation---either \readp or \writep. As before, $\IBSSign_{SU}$ is a signature
by the administrator, and \file is a sentinel value indicating that this is a
file tuple.

\item \textbf{File Access.} If a user $u$ who is authorized to read a file $f$
(i.e., $\exists r : (u,r) \in UR \wedge (r, \langle f, \readp \rangle) \in PA$)
wishes to do so, she must (i) download an \rolekey tuple for the role $r$ and an
\file tuple for $f$; (ii) validate the signatures on both tuples; (iii) decrypt
the role key $k_r$ from the \rolekey tuple using their personal IBE key $k_u$;
and (iv) decrypt the file $f$ from the \file tuple using the role key $k_r$.

Writes to a file are handled similarly. If $u$ is authorized to write a file $f$
via membership in role $r$ (i.e., $\exists r : (u,r) \in UR \wedge (r, \langle
f, \writep \rangle) \in PA$), she can upload a new \file tuple $\langle \file,
r, \langle fn, \writep \rangle, \IBEEnc_r(f^\prime), \IBSSign_{r} \rangle$. If
the signature authorizing the write ($\IBSSign_{r}$) can be verified by the
cloud provider, the existing \file tuple for $f$ will be replaced.

\end{itemize}

This construction describes a cryptographic analog to $\rbac_0$. The $UR$
relation is encoded in the collection of \rolekey tuples, while the $PA$
relation is encoded in the collection of \file tuples. The authorization
relation of $\rbac_0$ is upheld cryptographically: to read a file $f$, a user
$u$ must be able to decrypt a tuple granting her the permissions associated with
a role $r$, which can be used to decrypt a tuple containing a copy of $f$
encrypted to role $r$.

\subsection{Design Considerations} \label{sec:designconsider}

While conceptually straightforward, the strawman construction is by no means a
complete solution. We now use this construction as a guide to discuss a number
of design tradeoffs that must be addressed to support cryptographic enforcement
of dynamic $\rbac_0$ states.


\textbf{PKI vs.\ IBE.} Basing an $\rbac_0$ system on IBE and IBS allows
for a simple mapping from encryption keys to roles in $\rbac_0$: The
name of the role is the public-key used to encrypt under that
role. This is conceptually simpler than what is achieved by
traditional public key or symmetric encryption,
which may help limit certain key management issues in software.  IBE-based constructions also generalize to richer access control models (e.g., enforced using HIBE or ABE), which we explore in \cref{sec:discussion}.  That
said, rich infrastructure has been developed to support public key
cryptography, which may make the systems support issues inherent in
these constructions easier to manage.  To this end, we present
constructions based on both IBE and public key cryptography.


\textbf{Inefficiency Concerns.} The strawman construction exhibits two
key issues with respect to efficiency. First, IBE (like public-key
cryptography) is not particularly well-suited for the bulk encryption
of large amounts of data.  As such, the performance of this
construction would suffer when large files are shared within the
system.  Second, this construction requires a duplication of effort
when a file, say $f$, is to be shared with multiple roles, say $r_1$
and $r_2$. That is, $f$ must actually be encrypted twice: once with
$r_1$ and once with $r_2$. We note that this also leads to consistency
issues between roles when $f$ is updated. Fortunately, both of these
concerns can be mitigated via the use of hybrid cryptography. Rather
than storing \file tuples of the form:
\[ \langle \file, r, \langle fn, op \rangle, \IBEEnc_r(f), SU, \IBSSign_{SU} \rangle
\]
We can instead store the following tuples, where $k~\leftarrow ~\SymGen$ is a
symmetric key:
\begin{gather*}
\langle \filekey, r, \langle fn, op \rangle, \IBEEnc_r(k), SU, \IBSSign_{SU} \rangle
\\
\langle \file, fn, \SymEnc_k(f), r, \IBSSign_r \rangle
\end{gather*}
The \filekey tuples are similar to the file encryption tuples in the
strawman construction, except that the ciphertext portion of the tuple now
includes an IBE-encrypted symmetric key rather than an IBE-encrypted file.
\file tuples contain a symmetric-key-encrypted (using an authenticated mode)
version of the file $f$, and are IBS-signed using the role key of the last
authorized updater. This adjustment to the metadata improves the efficiency of
bulk encryption by using symmetric-key cryptography, and greatly reduces
the duplication of effort when sharing a file with multiple roles: a single
\file tuple can be created for the file along with multiple \filekey tuples
(i.e., one per role).

\textbf{Handling Revocation.} The strawman construction can neither revoke a
permission from a role, nor remove a user from a role. The former case can be
handled by versioning the \file and \filekey tuples stored within the system,
and the latter case handled by adding role versioning to the role key tuples and
\filekey tuples in the system:
\begin{gather*}
\langle \rolekey, u, (r, v_r), \IBEEnc_u(k_{(r,v_r)}, s_{(r,v_r)}),
\IBSSign_{SU} \rangle \\
\langle \filekey, r, \langle fn, op \rangle, v, \IBEEnc_{(r,v_r)}(k), SU,
\IBSSign_{SU} \rangle \\
\langle \file, fn, v, \SymEnc_k(f), (r,v_r), \IBSSign_{(r, v_r)} \rangle
\end{gather*}
Here, $v$ represents a version number for the symmetric key used to encrypt a
file. Role names have been replaced with tuples that include the role name
(e.g., $r$), \emph{as well as a version number ($v_r$)}. Removing a permission
from a role entails re-keying and re-encrypting the file (i.e., creating a new
\file tuple), and creating new \filekey tuples for each role whose access to the
file has \emph{not} been revoked. The roles increment their previous role
number. Similarly, removing a user $u$ from a role $r$ entails deleting $u$'s
\rolekey tuple for $r$, generating new role keys for $r$ (with an incremented
version number) and encoding these into new \rolekey tuples for each user
remaining in $r$, and re-versioning all files to which the role $r$ holds some
permission. We note that both of these processes must be carried out by an
administrator, as only administrators can modify the $\rbac_0$ state. There is
much nuance to these processes, and we defer a full discussion to
\cref{sec:implementation-details}.

\textbf{Online, Lazy, and Proxy Re-Encryption.} Supporting revocation leads to
an interesting design choice: should files be re-encrypted immediately upon
re-key, or lazily re-encrypted upon their next write? From a confidentiality
standpoint, forcing an administrator---or some daemon process running on her
behalf---to re-encrypt files immediately upon re-key is preferential, as it
ensures that users who have lost the ability to access a file cannot later read
its contents. On the other hand, this comes with a potentially severe efficiency
penalty in the event that many files are re-keyed due to changes to some role,
as access to these files must be locked while they are downloaded, re-encrypted,
and uploaded. In this paper, we opt for a \emph{lazy re-encryption} strategy, in
which files are re-encrypted by the next user to write to the file
(cf.,~\cref{sec:implementation-details}). We note that such a scheme is not
appropriate for all scenarios, but substantially reduces the computational
burden on the cloud when allowing for dynamic updates to the $\rbac_0$ state
(cf.,~\cref{sec:results-experimental}).  Similarly, if a client is powerful enough to download a source file and decrypt it to view the material, it presumably is powerful enough to perform the roughly computationally equivalent operation of re-encrypting it. Note that a single client is unlikely to need to re-encrypt large numbers of files, unlike the cloud if a lazy re-encryption strategy were not used.
Adapting our construction to instead use online re-encryption is a straightforward extension.

While appealing on the surface, IBE schemes that support proxy re-encryption, or
revocation (e.g.,~\cite{Green:2007:IPR:1419765.1419791,Boldyreva2008}) are not
suitable for use in our scenario. These types of schemes would seemingly allow
us to remove our reliance on lazy re-encryption, and have the cloud locally
update encryptions when a permission is revoked from a role, or a role from a
user. This would be done by creating an updated role name, using proxy
re-encryption to move the file from the old role name to the updated one, and
then revoking all keys for the old file. The \emph{significant} issue, here, is
that such schemes do not address how one would use them with hybrid encryption.
We do not believe that a reasonable threat model can assume that even a limited
adversary would be unable to cache all the symmetric keys for files she has
access to. \emph{Thus, using proxy re-encryption on the \rolekey and \filekey tuples
and not the \file tuples would allow users to continue to access files to which
their access has been revoked, and so our construction would still require
online or lazy re-encryption of the files themselves.}

As a final note, we acknowledge that key-homomorphic PRFs~\cite{BonehLMR13} could
be combined with revocation and proxy re-encryption schemes, solving the
revocation problem completely on the cloud in the hybrid model. However, current
technology does not solve the computational effort, as costs of current
key-homomorphic PRFs are comparable or greater than the IBE and PKI technologies
in consideration.

\textbf{Multiple Levels of Encryption.} We note that our construction has levels
of indirection between \rolekey, \filekey, and \file tuples that mirror the
indirection between users, roles, and permissions in $\rbac_0$. This indirection
could be flattened to decrease the number of cryptographic operations on the
critical path to file access; this would be akin to using an access matrix to
encode $\rbac_0$ states. While this is possible, it has been shown to cause
computational inefficiencies when roles' memberships or permissions are
altered~\cite{garrison2014codaspy}; in our case this inefficiency would be
amplified due to the cryptographic costs associated with these updates.

\textbf{Other Issues and Considerations.} Our constructions are measured without
concern for concurrency-related issues that would need to be addressed in
practice. We note, however, that features to handle concurrency would be largely
independent of the proposed cryptography used to enforce the $\rbac_0$ policies.
As such, we opt for the analysis of the conceptually-simpler schemes presented
in this paper. Finally, our analysis is agnostic to the underlying achieved
security guarantees and hardness assumptions of the public-key and IBE schemes.
Production implementations would need to consider these issues.

\subsection{Detailed IBE/IBS Construction} \label{sec:implementation-details}

We now flesh out the strawman and previously-discussed enhancements. This
produces a full construction for enforcing $\rbac_0$ protections over an
evolving collection managed by a minimally-trusted cloud storage provider.

\subsubsection{Overview and Preliminaries}

We reiterate that the administrators act as the Master Secret Key Generator of
the IBE/IBS schemes. Users add files to the system by IBE-encrypting these files
to the administrators, using hybrid cryptography and \file tuples.
Administrators assign permissions (i.e., $\langle file, op \rangle$ pairs) to
roles by distributing symmetric keys using \filekey tuples. Role keys are
distributed to users using \rolekey tuples. Recall the format of these tuples is
as follows:
\begin{gather*}
\langle \rolekey, u, (r, v_r), \IBEEnc_u(k_{(r,v_r)}, s_{(r,v_r)}),
\IBSSign_{SU} \rangle \\
\langle \filekey, r, \langle fn, op \rangle, v, \IBEEnc_{(r,v_r)}(k), SU,
\IBSSign_{SU} \rangle \\
\langle \file, fn, v, \SymEnc_k(f), (r,v_r), \IBSSign_{(r, v_r)} \rangle
\end{gather*}
Note that symmetric keys and role keys are associated with version information
to handle the cases where a user is removed from a role or a permission is
revoked from a role.

We assume that files have both read and write permissions associated with them.
However, we cannot have write without read, since writing requires decrypting
the file's symmetric key, which then can be used to decrypt and read the stored
file. Thus we only assign either \readp{} or \rwp{}, and only revoke \writep{}
(\readp{} is retained) or \rwp{} (nothing is retained). When a user wishes to
access a file, she determines which of her roles has access to the permission in
question. She then decrypts the role's secret key using her identity, and then
decrypts the symmetric key for the file using the role's secret key, and finally
uses the symmetric key to decrypt the symmetrically-encrypted ciphertext in
question.

\subsubsection{Full Construction}
\label{sec:const}


\begin{figure*}
\renewcommand{\labelitemi}{}
\footnotesize
\setlist{nosep,leftmargin=*}
\setlist[1]{itemsep=1em}
\begin{framed}
\begin{minipage}[t]{.48\textwidth}
\begin{itemize}

	\item $addU(u)$
	\begin{itemize}
		\item Add $u$ to USERS	
		\item Generate IBE private key $k_u \leftarrow \IBEKeyGen(u)$ and IBS private key $s_u \leftarrow \IBSKeyGen(u)$ for the new user $u$
		\item Give $k_u$ and $s_u$ to $u$ over private and authenticated channel
  \end{itemize}

	\item $delU(u)$
	\begin{itemize}
		\item For every role $r$ that $u$ is a member of:
		\begin{itemize}
			\item $revokeU(u,r)$
		\end{itemize}
	\end{itemize}

	\item $addP_u(fn,f)$
	\begin{itemize}
		\item Generate symmetric key $k \leftarrow \SymGen$
		\item Send $\langle \file, fn, 1, \SymEnc_k(f), u, \IBSSign_{u} \rangle$ and $\langle \filekey$, $SU$, $\langle fn, \rwp\rangle$, 1, $\IBEEnc_{SU}(k)$, $u$, $\IBSSign_{u}\rangle$ to R.M.
		\item The R.M. receives $\langle \file, fn, 1, c, u, sig\rangle$ and $\langle \filekey$, $SU$, $\langle fn, \rwp{}\rangle$, 1, $c'$, $u$, $sig'\rangle$ and verifies that the tuples are well-formed and the signatures are valid, i.e.,
		$\IBSVerify_{u}(\langle \file, fn, 1, c, u\rangle, sig)=1$ and \\
		$\IBSVerify_{u}(\langle \filekey$, $SU$, $\langle fn, \rwp\rangle$, 1, $c'$, $u\rangle$, $sig') = 1$.

		\item If verification is successful, the R.M. adds $(fn,1)$ to FILES and stores $\langle \file, fn, 1, c, u, sig\rangle$ and $\langle \filekey$, $SU$, $\langle fn, \rwp{}\rangle$, 1, $c'$, $u$, $sig'\rangle$
	\end{itemize}

	\item $delP(fn)$
	\begin{itemize}
		\item Remove $(fn,v_{fn})$ from FILES
		\item Delete $\langle \file, fn, -, -, -, -\rangle$ and all $\langle \filekey$, $-$, $\langle fn, -\rangle$, $-$, $-$, $-$, $-\rangle$
	\end{itemize}

	\item $addR(r)$
	\begin{itemize}
		\item Add $(r,1)$ to ROLES
		\item Generate IBE private key $k_{(r,1)} \leftarrow \IBEKeyGen((r,1))$ and IBS private key $s_{(r,1)} \leftarrow \IBSKeyGen((r,1))$ for role $(r,1)$
		\item Send $\langle \rolekey$, $SU$, $(r,1)$, $\IBEEnc_{SU}\left(k_{(r,1)}, s_{(r,1)}\right)$, $\IBSSign_{SU}\rangle$ to R.M.
	\end{itemize}

	\item $delR(r)$
	\begin{itemize}
	  \item Remove $(r,v_r)$ from ROLES
		\item Delete all $\langle \rolekey, -, (r,v_r), -, -\rangle$ 
		\item For all permissions $p = \langle fn,op\rangle$ that $r$ has access to:
		\begin{itemize}
			\item $revokeP(r, \langle fn,\rwp\rangle)$
		\end{itemize}
	\end{itemize}

	\item $assignU(u,r)$
	\begin{itemize}
		\item Find $\langle \rolekey$, $SU$, $(r,v_r)$, $c$, $sig\rangle$ with $\IBSVerify_{SU}(\langle \rolekey$, $SU$, $(r,v_r)$, $c\rangle$, $sig) = 1$
		\item Decrypt keys $(k_{(r,v_r)},s_{(r,v_r)}) = \IBEDec_{k_{SU}}(c)$
		\item Send $\langle \rolekey$, $u$, $(r,v_r)$, $\IBEEnc_u\left(k_{(r,v_r)}, s_{(r,v_r)}\right)$, $\IBSSign_{SU}\rangle$ to R.M.
	\end{itemize}

	\item $revokeU(u,r)$
	\begin{itemize}
		\item Generate new role keys $k_{(r,v_r+1)} \leftarrow \IBEKeyGen((r,v_r+1))$, $s_{(r,v_r+1)} \leftarrow \IBSKeyGen((r,v_r+1))$
		\item For all $\langle \rolekey$, $u'$, $(r,v_r)$, $c$, $sig\rangle$ with $u' \neq u$ and $\IBSVerify_{SU}(\langle \rolekey$, $u'$, $(r,v_r)$, $c\rangle$, $sig) = 1$:
		\begin{itemize}
			\item Send $\langle \rolekey$, $u'$, $(r,v_r+1)$, $\IBEEnc_{u'}\left(k_{(r,v_r+1)}, s_{(r,v_r+1)}\right)$, $\IBSSign_{SU}\rangle$ to R.M.
		\end{itemize}
		\item For every $fn$ such that there exists $\langle \filekey$, $(r,v_r)$, $\langle fn,op\rangle$, $v_{fn}$, $c$, $SU$, $sig\rangle$ with $\IBSVerify_{SU}(\langle \filekey$, $(r,v_r)$, $p$, $v_{fn}$, $c$, $SU\rangle$, $sig) = 1$:
		\begin{itemize}
			\item For every $\langle \filekey$, $(r,v_r)$, $\langle fn,op'\rangle$, $v$, $c'$, $SU$, $sig\rangle$ with $\IBSVerify_{SU}(\langle \filekey$, $(r,v_r)$, $\langle fn,op'\rangle$, $v$, $c'$, $SU\rangle$, $sig) = 1$:
			\begin{itemize}
				\item Decrypt key $k = \IBEDec_{k_{(r,v_r)}}(c')$
				\item Send $\langle \filekey$, $(r,v_r+1)$, $\langle fn,op'\rangle$, $v$, $\IBEEnc_{(r,v_r+1)}(k)$, $SU$, $\IBSSign_{SU}\rangle$ to R.M.
			\end{itemize}
			\item Generate new symmetric key $k' \leftarrow \SymGen$ for $p$
			\item For all $\langle \filekey$, $id$, $\langle fn,op'\rangle$, $v_{fn}$, $c''$, $SU$, $sig\rangle$ with $\IBSVerify_{SU}(\langle \filekey$, $id$, $\langle fn,op'\rangle$, $v_{fn}$, $c''$, $SU\rangle$, $sig) = 1$:
			\begin{itemize}
				\item Send $\langle \filekey$, $id$, $\langle fn,op'\rangle$, $v_{fn}+1$, $\IBEEnc_{id}(k')$, $SU$, $\IBSSign_{SU}\rangle$ to R.M.
			\end{itemize}
			\item Increment $v_{fn}$ in FILES, i.e., set $v_{fn} := v_{fn}+1$
		\end{itemize}
		\item Increment $v_r$ in ROLES, i.e., set $v_r := v_r+1$
		\item Delete all $\langle \rolekey$, $-$, $(r,v_r)$, $-$, $-\rangle$
		\item Delete all $\langle \filekey$, $(r,v_r)$, $-$, $-$, $-$, $-$, $-\rangle$
	\end{itemize}
\end{itemize}
\end{minipage}
\hfill
\begin{minipage}[t]{.48\textwidth}
\begin{itemize}
	\item $assignP(r,\langle fn, op\rangle)$
	\begin{itemize}
		\item For all $\langle \filekey$, $SU$, $\langle fn, \rwp\rangle$, $v$, $c$, $id$, $sig\rangle$ with $\IBSVerify_{id}(\langle \filekey$, $SU$, $\langle fn, \rwp{}\rangle$, $v$, $c$, $id\rangle$, $sig) = 1$:
		\begin{itemize}
			\item If this adds \writep permission to existing \readp permission, i.e., $op = \rwp$ and there exists $\langle \filekey$, $(r,v_r)$, $\langle fn, \readp\rangle$, $v$, $c'$, $SU$, $sig\rangle$ with $\IBSVerify_{SU}(\langle \filekey$, $(r,v_r)$, $\langle fn, op'\rangle$, $v$, $c'$, $SU\rangle$, $sig) = 1$:
			\begin{itemize}
				\item Send $\langle \filekey$, $(r,v_r)$, $\langle fn, \rwp\rangle$, $v$, $c'$, $SU$, $\IBSSign_{SU}\rangle$ to R.M.
				\item Delete $\langle \filekey$, $(r,v_r)$, $\langle fn, \readp\rangle$, $v$, $c'$, $SU$, $sig\rangle$
			\end{itemize}
			\item If the role has no existing permission for the file, i.e., there does not exist $\langle \filekey$, $(r,v_r)$, $\langle fn, op'\rangle$, $v$, $c'$, $SU$, $sig\rangle$ with $\IBSVerify_{SU}(\langle \filekey$, $(r,v_r)$, $\langle fn, op'\rangle$, $v$, $c$, $SU\rangle$, $sig) = 1$:
			\begin{itemize}
				\item Decrypt key $k = \IBEDec_{k_{SU}}(c)$
				\item Send $\langle \filekey$, $(r,v_r)$, $\langle fn, op\rangle$, $v$, $\IBEEnc_{(r,v_r)}(k)$, $SU$, $\IBSSign_{SU}\rangle$ to R.M.
			\end{itemize}
		\end{itemize}
	\end{itemize}

	\item $revokeP(r,\langle fn, op\rangle)$
	\begin{itemize}
		\item If $op = \writep{}$:
		\begin{itemize}
			\item For all $\langle \filekey$, $(r,v_r)$, $\langle fn, \rwp\rangle$, $v$, $c$, $SU$, $sig\rangle$ with $\IBSVerify_{SU}(\langle \filekey$, $(r,v_r)$, $\langle fn, \rwp\rangle$, $v$, $c$, $SU\rangle$, $sig) = 1$:
			\begin{itemize}
				\item Send $\langle \filekey$, $(r,v_r)$, $\langle fn, \readp\rangle$, $v$, $c$, $SU$, $\IBSSign_{SU}\rangle$ to R.M.
				\item Delete $\langle \filekey$, $(r,v_r)$, $\langle fn, \rwp\rangle$, $v$, $c$, $SU$, $sig\rangle$
			\end{itemize}
		\end{itemize}
		\item If $op = \rwp{}$:
		\begin{itemize}
			\item Delete all $\langle \filekey$, $(r,v_r)$, $\langle fn, -\rangle$, $-$, $-$, $-\rangle$
			\item Generate new symmetric key $k' \leftarrow \SymGen$
			\item For all $\langle \filekey$, $r'$, $\langle fn, op'\rangle$, $v_{fn}$, $c$, $SU$, $sig\rangle$ with $\IBSVerify_{SU}(\langle \filekey$, $r'$, $\langle fn, op'\rangle$, $v$, $c$, $SU\rangle$, $sig) = 1$:
			\begin{itemize}
				\item Send $\langle \filekey$, $r'$, $\langle fn, op'\rangle$, $v_{fn}+1$, $\IBEEnc_{r'}(k')$, $SU$, $\IBSSign_{SU}\rangle$ to R.M.
			\end{itemize}
			\item Increment $v_{fn}$ in FILES, i.e., set $v_{fn} := v_{fn}+1$
		\end{itemize}
	\end{itemize}

	\item $read_u(fn)$
	\begin{itemize}
		\item Find $\langle \file, fn, v, c, id, sig\rangle$ with valid ciphertext $c$ and valid signature $sig$, i.e., $\IBSVerify_{id}(\langle \file, fn, 1, c, id\rangle, sig)=1$
		\item Find a role $r$ such that the following hold:
		\begin{itemize}
			\item $u$ is in role $r$, i.e., there exists $\langle \rolekey$, $u$, $(r,v_r)$, $c'$, $sig\rangle$ with $\IBSVerify_{SU}(\langle \rolekey$, $u$, $(r,v_r)$, $c'\rangle$, $sig) = 1$
			\item $r$ has read access to version $v$ of $fn$, i.e., there exists $\langle \filekey$, $(r,v_r)$, $\langle fn, op\rangle$, $v$, $c''$, $SU$, $sig'\rangle$ with $\IBSVerify_{SU}(\langle \filekey$, $(r,v_r)$, $\langle fn, op\rangle$, $v$, $c''$, $SU\rangle$, $sig') = 1$
		\end{itemize}
		\item Decrypt role key $k_{(r,v_r)} = \IBEDec_{k_u}(c')$
		\item Decrypt file key $k = \IBEDec_{k_{(r,v_r)}}(c'')$
		\item Decrypt file $f = \SymDec_k(c)$
	\end{itemize}

	\item $write_u(fn, f)$
	\begin{itemize}
		\item Find a role $r$ such that the following hold:
		\begin{itemize}
			\item $u$ is in role $r$, i.e., there exists $\langle \rolekey$, $u$, $(r,v_r)$, $c$, $sig\rangle$ with $\IBSVerify_{SU}(\langle \rolekey$, $u$, $(r,v_r)$, $c\rangle$, $sig) = 1$
			\item $r$ has write access to the newest version of $fn$, i.e., there exists $\langle \filekey$, $(r,v_r)$, $\langle fn, \rwp\rangle$, $v_{fn}$, $c'$, $SU$, $sig'\rangle$ and $\IBSVerify_{SU}(\langle \filekey$, $(r,v_r)$, $\langle fn, \rwp\rangle$, $v$, $c'$, $SU\rangle$, $sig') = 1$
		\end{itemize}
		\item Decrypt role key $k_{(r,v_r)} = \IBEDec_{k_u}(c)$
		\item Decrypt file key $k = \IBEDec_{k_{(r,v_r)}}(c')$
		\item Send $\langle \file, fn, v_{fn}, \SymEnc_k(f), (r,v_r), \IBSSign_{(r,v_r)}\rangle$ to R.M.
		\item The R.M. receives $r$ and $\langle \file, fn, v, c'', (r,v_r), sig''\rangle$ and verifies the following:
		\begin{itemize}
			\item The tuple is well-formed with $v = v_{fn}$
			\item The signature is valid, i.e., $\IBSVerify_{(r,v_r)}(\langle \file$, $fn$, $v$, $c'', (r,v_r)\rangle$, $sig'') = 1$
			\item $r$ has write access to the newest version of $fn$, i.e., there exists $\langle \filekey$, $(r,v_r)$, $\langle fn, \rwp\rangle$, $v_{fn}$, $c'$, $SU$, $sig'\rangle$ and $\IBSVerify_{SU}(\langle \filekey$, $(r,v_r)$, $\langle fn, \rwp\rangle$, $v_{fn}$, $c'$, $SU\rangle$, $sig') = 1$
		\end{itemize}
		\item If verification is successful, the R.M. replaces $\langle \file, fn, -, -, -, -\rangle$ with $\langle \file, fn, v_{fn}, c'', (r,v_r), sig''\rangle$
	\end{itemize}
\end{itemize}
\end{minipage}
\end{framed}
\caption{Implementation of $\rbac_0$ using IBE and IBS}
\label{fig:rbacibe}
\end{figure*}

\Cref{fig:rbacibe} lists every $\rbac_0$ operation and shows how each can be
implemented using IBE, IBS, and the metadata structures described previously.
This figure uses the following notation: $u$ is a user, $r$ is a role,
$p$ is a permission, $fn$ is a file name, $f$ is a file, $c$ is a ciphertext
(either IBE or symmetric), $sig$ is an IBS signature, and $v$ is a version
number. Users are listed in a file called USERS. The identity corresponding to a role $r$ is $(r,v)$, where $v$ is a
positive integer representing the version number. We use $v_r$ to denote the
latest version number for role $r$. Roles and versions are stored as $(r,v_r)$
pairs in a file called ROLES, which is publicly viewable and can only be changed
by the administrator. Similarly, we use $v_{fn}$ to denote the latest version number for
the file with name $fn$. Filenames and versions are stored as $(fn,v_{fn})$
pairs in a file called FILES, which is publicly viewable and can only be changed
by the admin or reference monitor (R.M.). $SU$ is the superuser identity
possessed by the administrators. We use ``$-$'' to represent a wildcard.
$\IBSSign_{id}$ at the end of a tuple represents an IBS signature by identity
$id$ over the rest of the tuple. The subscript after an operation name
identifies who performs the operation if it is not performed by an
administrator.

Many operations described in \cref{fig:rbacibe} are straightforward given the
discussion earlier in this section. To demonstrate some of the more complicated
aspects of this construction, we now describe the procedure to revoke a role
from a user, which demonstrates several types of re-keys as well as our notion
of lazy re-encryption. The procedure for removing a user $u$ from a role $r$
consists of three steps: (i) re-keying $r$, (ii) re-encrypting existing file keys
stored in \filekey tuples to the new role key, and (iii) re-keying all files
accessible by $r$.

To re-key a role $r$, we must transition from \((r, v_r)\) to \((r, v_r + 1)\),
generating new IBE keys for this new role version. The old \rolekey tuples for
$r$ are deleted, and each remaining member \(u^\prime\) of role $r$ is given the
new \rolekey tuples of the form of \(\langle \rolekey, u^\prime, (r, v_r + 1),
c, \IBSSign_{SU} \rangle\), where \(c\) contains the new IBE/IBS keys encrypted
to \(u^\prime\)'s identity key. Next, all (symmetric) file keys encrypted to
\((r, v_r)\) in \filekey tuples are replaced with file keys encrypted to \((r,
v_r + 1)\). This allows the remaining members of \(r\) to retain access to
existing files, while preventing the revoked user $u$ from accessing any file
keys that he has not already decrypted and cached.

Finally, each file to which \(r\) has access must be re-keyed to prevent $u$
from accessing future updates to this file \emph{using cached symmetric keys}.
For each file \(f\), a new symmetric key is generated via \SymGen. This key is
then encrypted for each role \(r^\prime\) that has access to \(f\) (including
\(r\)), and new \filekey tuples \(\langle \filekey, r^\prime, \langle f, op
\rangle, v+1, c^\prime, \IBSSign_{SU} \rangle\) are uploaded \emph{alongside}
existing \(\langle \filekey, r^\prime, \langle f, op \rangle, v, c,
\IBSSign_{SU} \rangle\) tuples. Here, \(v+1\) is the new file key version, \(c\)
is the existing encrypted file key, and \(c^\prime\) is the new file key
IBE-encrypted to identity \(r^\prime\). The next time \(f\) is read, the key
contained in \(c\) will be used for decryption; the next time \(f\) is written,
the key contained in \(c^\prime\) will be used for encryption. This process
obviates the need for a daemon to re-encrypt all files at revocation time, but
prevents the revoked user $u$ from accessing any future modifications to these
files using cached symmetric file keys.

\subsection{PKI Construction Overview} \label{sec:implementation-pki}


\begin{figure*}
\renewcommand{\labelitemi}{}
\footnotesize
\setlist{nosep,leftmargin=*}
\setlist[1]{itemsep=1ex}
\begin{framed}
\begin{minipage}[t]{.49\textwidth}
\begin{itemize}

	\item $addU(u)$
	\begin{itemize}
		\item User $u$ generates encryption key pair $(\enckey_u, \deckey_u) \leftarrow \PKGen$ and signature key pair $(\verkey_u, \sigkey_u) \leftarrow \SigGen$
		\item User $u$ sends $\enckey_u,\verkey_u$ to admin
		\item Admin adds $(u,\enckey_u,\verkey_u)$ to USERS
  \end{itemize}

	\item $delU(u)$
	\begin{itemize}
		\item For every role $r$ that $u$ is a member of:
		\begin{itemize}
			\item $revokeU(u,r)$
		\end{itemize}
	\end{itemize}

	\item $addP_u(fn,f)$
	\begin{itemize}
		\item Generate symmetric key $k \leftarrow \SymGen$
		\item Send $\langle \file, fn, 1, \SymEnc_k(f), u, \SigSign_u\rangle$ and $\langle \filekey$, $SU$, $\langle fn, \rwp{}\rangle$, 1, $\PKEnc_{\enckey_{SU}}(k)$, $u$, $\SigSign_u\rangle$ to R.M.
		\item The R.M. receives $\langle \file, fn, 1, c, u, sig\rangle$ and $\langle \filekey$, $SU$, $\langle fn, \rwp\rangle$, 1, $c'$, $u$, $sig'\rangle$ and verifies that the tuples are well-formed and the signatures are valid, i.e.,
		$\SigVerify_{\verkey_u}(\langle \file, fn, 1, c, u\rangle, sig)=1$ and \\
		$\SigVerify_{\verkey_u}(\langle \filekey$, $SU$, $\langle fn, \rwp\rangle$, 1, $c'$, $u\rangle$, $sig') = 1$.
		\item If verification is successful, the R.M. adds $(fn,1)$ to FILES and stores $\langle \file, fn, 1, c, u, sig\rangle$ and $\langle \filekey$, $SU$, $\langle fn, \rwp{}\rangle$, 1, $c'$, $u$, $sig'\rangle$
	\end{itemize}

	\item $delP(fn)$
	\begin{itemize}
		\item Remove $(fn,v_{fn})$ from FILES
		\item Delete $\langle \file, fn, -, -, -, -\rangle$ and all $\langle \filekey$, $-$, $\langle fn, -\rangle$, $-$, $-$, $-$, $-\rangle$
	\end{itemize}

	\item $addR(r)$
	\begin{itemize}
		\item Generate encryption key pair $(\enckey_{(r,1)}, \deckey_{(r,1)}) \leftarrow \PKGen$ and signature key pair $(\verkey_{(r,1)}, \sigkey_{(r,1)}) \leftarrow \SigGen$
		\item Add $(r,1,\enckey_{(r,1)},\verkey_{(r,1)})$ to ROLES
		\item Send $\langle \rolekey$, $SU$, $(r,1)$, $\PKEnc_{\enckey_{SU}}\left(\deckey_{(r,1)}, \sigkey_{(r,1)}\right)$, $\SigSign_{SU}\rangle$ to R.M.
	\end{itemize}

	\item $delR(r)$
	\begin{itemize}
		\item Remove $(r,v_r,-,-)$ from ROLES
		\item Delete all $\langle \rolekey, -, (r,v_r), -, -\rangle$
		\item For all permissions $p = \langle fn,op\rangle$ that $r$ has access to:
		\begin{itemize}
			\item $revokeP(r, \langle fn,\rwp\rangle)$
		\end{itemize}
	\end{itemize}

	\item $assignU(u,r)$
	\begin{itemize}
		\item Find $\langle \rolekey$, $SU$, $(r,v_r)$, $c$, $sig\rangle$ with $\SigVerify_{\verkey_{SU}}(\langle \rolekey$, $SU$, $(r,v_r)$, $c\rangle$, $sig) = 1$
		\item Decrypt keys $(\deckey_{(r,v_r)},\sigkey_{(r,v_r)}) = \PKDec_{\deckey_{SU}}(c)$
		\item Send $\langle \rolekey$, $u$, $(r,v_r)$, $\PKEnc_{\enckey_u}\left(\deckey_{(r,v_r)}, \sigkey_{(r,v_r)}\right)$, $\SigSign_{SU}\rangle$ to R.M.
	\end{itemize}

	\item $revokeU(u,r)$
	\begin{itemize}
		\item Generate new role keys $(\enckey_{(r,v_r+1)}, \deckey_{(r,v_r+1)}) \leftarrow \PKGen$, $(\verkey_{(r,v_r+1)}, \sigkey_{(r,v_r+1)}) \leftarrow \SigGen$
		\item For all $\langle \rolekey$, $u'$, $(r,v_r)$, $c$, $sig\rangle$ with $u' \neq u$ and $\SigVerify_{\verkey_{SU}}(\langle \rolekey$, $u'$, $(r,v_r)$, $c\rangle$, $sig) = 1$:
		\begin{itemize}
			\item Send $\langle \rolekey$, $u'$, $(r,v_r+1)$, $\PKEnc_{\enckey_{u'}}\left(\deckey_{(r,v_r+1)}, \sigkey_{(r,v_r+1)}\right)$, $\SigSign_{SU}\rangle$ to R.M.
		\end{itemize}
		\item For every $fn$ such that there exists $\langle \filekey$, $(r,v_r)$, $\langle fn,op\rangle$, $v_{fn}$, $c$, $SU$, $sig\rangle$ with $\SigVerify_{\verkey_{SU}}(\langle \filekey$, $(r,v_r)$, $p$, $v_{fn}$, $c$, $SU\rangle$, $sig) = 1$:
		\begin{itemize}
			\item For every $\langle \filekey$, $(r,v_r)$, $\langle fn,op'\rangle$, $v$, $c'$, $SU$, $sig\rangle$ with $\SigVerify_{\verkey_{SU}}(\langle \filekey$, $(r,v_r)$, $\langle fn,op'\rangle$, $v$, $c'$, $SU\rangle$, $sig) = 1$:
			\begin{itemize}
				\item Decrypt key $k = \PKDec_{\deckey_{(r,v_r)}}(c')$
				\item Send $\langle \filekey$, $(r,v_r+1)$, $\langle fn,op'\rangle$, $v$, $\PKEnc_{\enckey_{(r,v_r+1)}}(k)$, $SU$, $\SigSign_{SU}\rangle$ to R.M.
			\end{itemize}
			\item Generate new symmetric key $k' \leftarrow \SymGen$ for $p$
			\item For all $\langle \filekey$, $id$, $\langle fn,op'\rangle$, $v_{fn}$, $c''$, $SU$, $sig\rangle$ with $\SigVerify_{\verkey_{SU}}(\langle \filekey$, $id$, $\langle fn,op'\rangle$, $v_{fn}$, $c''$, $SU\rangle$, $sig) = 1$:
			\begin{itemize}
				\item Send $\langle \filekey$, $id$, $\langle fn,op'\rangle$, $v_{fn}+1$, $\PKEnc_{\enckey_{id}}(k')$, $SU$, $\SigSign_{SU}\rangle$ to R.M.
			\end{itemize}
			\item Increment $v_{fn}$ in FILES, i.e., set $v_{fn} := v_{fn}+1$
		\end{itemize}
		\item Update $r$ in ROLES, i.e., replace $(r,v_r,\enckey_{(r,v_r)},\verkey_{(r,v_r)})$ with $(r,v_r+1,\enckey_{(r,v_r+1)},\verkey_{(r,v_r+1)})$
		\item Delete all $\langle \rolekey$, $-$, $(r,v_r)$, $-$, $-\rangle$
		\item Delete all $\langle \filekey$, $(r,v_r)$, $-$, $-$, $-$, $-$, $-\rangle$
	\end{itemize}
\end{itemize}
\end{minipage}
\hfill
\begin{minipage}[t]{.48\textwidth}
\begin{itemize}
	\item $assignP(r,\langle fn, op\rangle)$
	\begin{itemize}
		\item For all $\langle \filekey$, $SU$, $\langle fn, \rwp{}\rangle$, $v$, $c$, $id$, $sig\rangle$ with $\SigVerify_{\verkey_{id}}(\langle \filekey$, $SU$, $\langle fn, \rwp\rangle$, $v$, $c$, $id\rangle$, $sig) = 1$:
		\begin{itemize}
			\item If this adds \writep permission to existing \readp permission, i.e., $op = \rwp$ and there exists $\langle \filekey$, $(r,v_r)$, $\langle fn, \readp\rangle$, $v$, $c'$, $SU$, $sig\rangle$ with $\SigVerify_{\verkey_{SU}}(\langle \filekey$, $(r,v_r)$, $\langle fn, op'\rangle$, $v$, $c'$, $SU\rangle$, $sig) = 1$:
			\begin{itemize}
				\item Send $\langle \filekey$, $(r,v_r)$, $\langle fn, \rwp\rangle$, $v$, $c'$, $SU$, $\SigSign_{SU}\rangle$ to R.M.
				\item Delete $\langle \filekey$, $(r,v_r)$, $\langle fn, \readp\rangle$, $v$, $c'$, $SU$, $sig\rangle$
			\end{itemize}
			\item If the role has no existing permission for the file, i.e., there does not exist $\langle \filekey$, $(r,v_r)$, $\langle fn, op'\rangle$, $v$, $c'$, $SU$, $sig\rangle$ with $\SigVerify_{\verkey_{SU}}(\langle \filekey$, $(r,v_r)$, $\langle fn, op'\rangle$, $v$, $c$, $SU\rangle$, $sig) = 1$:
			\begin{itemize}
				\item Decrypt key $k = \PKDec_{\deckey_{SU}}(c)$
				\item Send $\langle \filekey$, $(r,v_r)$, $\langle fn, op\rangle$, $v$, $\PKEnc_{\enckey_{(r,v_r)}}(k)$, $SU$, $\SigSign_{SU}\rangle$ to R.M.
			\end{itemize}
		\end{itemize}
	\end{itemize}

	\item $revokeP(r,\langle fn, op\rangle)$
	\begin{itemize}
		\item If $op = \writep{}$:
		\begin{itemize}
			\item For all $\langle \filekey$, $(r,v_r)$, $\langle fn, \rwp\rangle$, $v$, $c$, $SU$, $sig\rangle$ with $\SigVerify_{\verkey_{SU}}(\langle \filekey$, $(r,v_r)$, $\langle fn, \rwp\rangle$, $v$, $c$, $SU\rangle$, $sig) = 1$:
			\begin{itemize}
				\item Send $\langle \filekey$, $(r,v_r)$, $\langle fn, \readp\rangle$, $v$, $c$, $SU$, $\SigSign_{SU}\rangle$ to R.M.
				\item Delete $\langle \filekey$, $(r,v_r)$, $\langle fn, \rwp\rangle$, $v$, $c$, $SU$, $sig\rangle$
			\end{itemize}
		\end{itemize}
		\item If $op = \rwp{}$:
		\begin{itemize}
			\item Delete all $\langle \filekey$, $(r,v_r)$, $\langle fn, -\rangle$, $-$, $-$, $-\rangle$
			\item Generate new symmetric key $k' \leftarrow \SymGen$
			\item For all $\langle \filekey$, $r'$, $\langle fn, op'\rangle$, $v_{fn}$, $c$, $SU$, $sig\rangle$ with $\SigVerify_{\verkey_{SU}}(\langle \filekey$, $r'$, $\langle fn, op'\rangle$, $v$, $c$, $SU\rangle$, $sig) = 1$:
			\begin{itemize}
				\item Send $\langle \filekey$, $r'$, $\langle fn, op'\rangle$, $v_{fn}+1$, $\PKEnc_{\enckey_{r'}}(k')$, $SU$, $\SigSign_{SU}\rangle$ to R.M.
			\end{itemize}
			\item Increment $v_{fn}$ in FILES, i.e., set $v_{fn} := v_{fn}+1$
		\end{itemize}
	\end{itemize}

	\item $read_u(fn)$
	\begin{itemize}
		\item Find $\langle \file, fn, v, c, id, sig\rangle$ with valid ciphertext $c$ and valid signature $sig$, i.e., $\SigVerify_{\verkey_{id}}(\langle \file, fn, 1, c, id\rangle, sig)=1$
		\item Find a role $r$ such that the following hold:
		\begin{itemize}
			\item $u$ is in role $r$, i.e., there exists $\langle \rolekey$, $u$, $(r,v_r)$, $c'$, $sig\rangle$ with $\SigVerify_{\verkey_{SU}}(\langle \rolekey$, $u$, $(r,v_r)$, $c'\rangle$, $sig) = 1$
			\item $r$ has read access to version $v$ of $fn$, i.e., there exists $\langle \filekey$, $(r,v_r)$, $\langle fn, op\rangle$, $v$, $c''$, $SU$, $sig'\rangle$ with $\SigVerify_{\verkey_{SU}}(\langle \filekey$, $(r,v_r)$, $\langle fn, op\rangle$, $v$, $c''$, $SU\rangle$, $sig') = 1$
		\end{itemize}
		\item Decrypt role key $\deckey_{(r,v_r)} = \PKDec_{\deckey_u}(c')$
		\item Decrypt file key $k = \PKDec_{\deckey_{(r,v_r)}}(c'')$
		\item Decrypt file $f = \SymDec_k(c)$
	\end{itemize}

	\item $write_u(fn, f)$
	\begin{itemize}
		\item Find a role $r$ such that the following hold:
		\begin{itemize}
			\item $u$ is in role $r$, i.e., there exists $\langle \rolekey$, $u$, $(r,v_r)$, $c$, $sig\rangle$ with $\SigVerify_{\verkey_{SU}}(\langle \rolekey$, $u$, $(r,v_r)$, $c\rangle$, $sig) = 1$
			\item $r$ has write access to the newest version of $fn$, i.e., there exists $\langle \filekey$, $(r,v_r)$, $\langle fn, \rwp\rangle$, $v_{fn}$, $c'$, $SU$, $sig'\rangle$ and $\SigVerify_{\verkey_{SU}}(\langle \filekey$, $(r,v_r)$, $\langle fn, \rwp\rangle$, $v$, $c'$, $SU\rangle$, $sig') = 1$
		\end{itemize}
		\item Decrypt role key $\deckey_{(r,v_r)} = \PKDec_{\deckey_u}(c)$
		\item Decrypt file key $k = \PKDec_{\deckey_{(r,v_r)}}(c')$
		\item Send $\langle \file, fn, v_{fn}, \SymEnc_k(f), (r,v_r), \SigSign_{(r,v_r)}\rangle$ to R.M.
		\item The R.M. receives $\langle \file, fn, v, c'', (r,v_r), sig''\rangle$ and verifies the following:
		\begin{itemize}
			\item The tuple is well-formed with $v = v_{fn}$
			\item The signature is valid, i.e., $\SigVerify_{\verkey_{(r,v_r)}}(\langle \file$, $fn$, $v$, $c'', (r,v_r)\rangle$, $sig'') = 1$
			\item $r$ has write access to the newest version of $fn$, i.e., there exists $\langle \filekey$, $(r,v_r)$, $\langle fn, op\rangle$, $v_{fn}$, $c'$, $SU$, $sig'\rangle$ and $\SigVerify_{\verkey_{SU}}(\langle \filekey$, $(r,v_r)$, $\langle fn, op\rangle$, $v_{fn}$, $c'$, $SU\rangle$, $sig') = 1$
		\end{itemize}
		\item If verification is successful, the R.M. replaces $\langle \file, fn, -, -, -, -\rangle$ with $\langle \file, fn, v_{fn}, c'', (r,v_r), sig''\rangle$
	\end{itemize}
\end{itemize}
\end{minipage}
\end{framed}
\caption{Implementation of $\rbac_0$ using PKI}
\label{fig:rbacpki}
\end{figure*}

\Cref{fig:rbacpki} shows how traditional public-key cryptography can be
used in place of IBE/IBS to implement $\rbac_0$. In our PKI
construction, public-key encryption and signatures take the place of IBE and
IBS. Each role is assigned a public/private key pair rather than IBE/IBS keys.
The primary difference between the IBE and PKI constructions is that IBE/IBS
clients are given \emph{escrowed} IBE/IBS identity private keys by the role
administrator, while PKI clients generate their own public/private key pairs and
upload their public keys. Note that in both systems, the administrators have
access to all of the roles' private keys. Public keys (encryption and verification keys) for users and roles are stored in USERS and ROLES, respectively.

This figure uses the following notation: $u$ is a user, $r$ is a role,
$p$ is a permission, $fn$ is a file name, $f$ is a file, $c$ is a ciphertext
(either public-key or symmetric), $sig$ is an digital signature, and $v$ is a version
number. Users are listed in a file called USERS, which consists of $(u,\enckey_u,\verkey_u)$ tuples containing usernames and their public keys.
Each role key is assigned to a pair $(r,v)$, where $v$ is a
positive integer representing the version number. We use $v_r$ to denote the
latest version number for role $r$. Roles, versions, and their public keys are stored as $(r,1,\enckey_{(r,1)},\verkey_{(r,1)})$
tuples in a file called ROLES, which is publicly viewable and can only be changed
by the administrator. Similarly, we use $v_{fn}$ to denote the latest version number for
the file with name $fn$. Filenames and versions are stored as $(fn,v_{fn})$
pairs in a file called FILES, which is publicly viewable and can only be changed
by the admin or reference monitor (R.M.). $SU$ is the superuser identity
possessed by the administrators. We use ``$-$'' to represent a wildcard.
$\SigSign_{id}$ at the end of a tuple represents an a digital signature using key
$\sigkey_{id}$ over the rest of the tuple. The subscript after an operation name
identifies who performs the operation if it is not performed by an
administrator.


\section{Analysis} \label{sec:analysis}

We now describe our evaluation of the
\emph{suitability} of IBE/IBS and PKI constructions for enforcing $\rbac_0$
access controls. We utilize a workflow similar to that proposed in
\cite{garrison2014sacmat}, in which we first evaluate the candidates'
\emph{expressive power} (i.e., ability to represent the desired policy as it
evolves), then evaluate the \emph{cost} of using each candidate using Monte
Carlo simulation based on initial states obtained from real-world datasets.

\subsection{Qualitative Analysis}

We analyze the correctness and security guarantees of our implementations using
the access control expressiveness framework known as \emph{parameterized
expressiveness}~\cite{Hinrichs:2013:AAC:2510170.2510419}. In particular, we
ensure that the implementation properties of correctness, AC-preservation, and
safety are preserved by these constructions. \emph{Correctness} ensures that the
$\rbac_0$ state's image in our constructions answers queries exactly as the
original $\rbac_0$ system would, and that the same end state is reached by
either executing an $\rbac_0$ action natively and mapping the result into our
construction or by mapping the initial $\rbac_0$ state and executing the
action's image in our construction. \emph{AC-preservation} says that the
$\rbac_0$ system's authorization requests must be asked directly in the
simulating system. For instance, the policy must be simulated in such a way that
the $\rbac_0$ request ``Can subject \(s\) read file \(f\)?'' is asked directly
in the simulated state rather than being translated to any other queries.
Finally, \emph{safety} ensures that our constructions do not grant or revoke
unnecessary permissions during the simulation of a single $\rbac_0$ command.
That is, the intermediate states through which our constructions travel while
implementing an $\rbac_0$ command do not add or remove any granted requests
except those that must be added or removed as determined by the start and end
states of the $\rbac_0$ command.

For formal definitions of these properties,
see~\cite{Hinrichs:2013:AAC:2510170.2510419}. Using parameterized
expressiveness, we get the following results:
\begin{theorem}
\label{thm:ibe}
The implementation of $\rbac_0$ using IBE and IBS detailed in \cref{fig:rbacibe}
is correct, AC-preserving, and safe.
\end{theorem}


\begin{theorem}
\label{thm:pki}
The implementation of $\rbac_0$ using public key cryptographic techniques
 is correct, AC-preserving, and safe.
\end{theorem}


We now give an overview of the structure of and ideas behind the
proof of \cref{thm:ibe}.  This proof begins by formalizing the
IBE/IBS construction presented in \cref{sec:constructions}
using the parameterized expressiveness framework. We then provide a
formal mapping from $\rbac_0$ to our IBE/IBS system. We show that this
mapping preserves user authorization, meaning that a user is
authorized for a permission in $\rbac_0$ if and only if the user is
also authorized by the IBE/IBS construction.

The tricky part of this proof involves showing that changes to the
$\rbac_0$ state map correctly as changes to the IBE/IBS state.  This
means that changing the $\rbac_0$ state and then mapping to IBE/IBS
has the same effect as mapping to IBE/IBS and then changing the state
there in an equivalent way. Our use of version numbers in IBE/IBS
means that a single $\rbac_0$ state may map to multiple IBE/IBS
states; i.e., if a user is granted permissions that are later revoked,
the resulting $\rbac_0$ state will be the same as if the permissions
were never granted, but the IBE/IBS state will have different version
numbers as a result of the revocation. Therefore, we consider IBE/IBS
states that only differ in version numbers to be \emph{congruent}. We
show that the IBE/IBS state resulting from a change to the $\rbac_0$
state, followed by mapping to IBE/IBS, is congruent to one crafted by
first mapping to IBE/IBS, and then changing the IBE/IBS state in a
corresponding way.

The full proof of \cref{thm:ibe} can be found in
\cref{sec:ibeproof}.  The proof of
\cref{thm:pki}, which is very similar in structure, can be found in \cref{sec:pkiproof}.

\subsection{Algebraic Costs}

\Cref{tab:algcosts} lists the costs for each \rbac operation based on the system state. All costs are incurred by the user or administrator running the operation unless otherwise noted. In order to simplify the formulas, we employ a slight abuse of notation: we use the operation itself to represent its cost (e.g., $\IBEEnc$ is used to represent the cost of one $\IBEEnc$ operation).
We use the following notation:
\begin{itemize}
	\item $roles(u)$ is the set of roles to which user $u$ is assigned
	\item $perms(r)$ is the set of permissions to which role $r$ is assigned
	\item $users(r)$ is the set of users to which role $r$ is assigned
	\item $roles(p)$ is the set of roles to which permission $p$ is assigned
	\item $versions(p)$ is the number of versions of permission $p$
\end{itemize}



\begin{table*}
\small
\renewcommand{\arraystretch}{1.5}
\renewcommand{\tabularxcolumn}[1]{p{#1}}
\begin{tabularx}{\textwidth}{ r @{\,:\quad} X }
	$addU(u)$ & $\IBEKeyGen + \IBSKeyGen$ \\
	$delU(u)$ & $\sum_{r\in roles(u)} revokeU(u,r)$\\
	$addP(p)$ & $\IBEEnc+2 \cdot \IBSSign$ and $2\cdot \IBSVerify$ by R.M. \\
	$delP(p)$ & None\\
	$addR(r)$ & $\IBEKeyGen+\IBEEnc+ \IBSKeyGen+\IBSSign$ \\
	$delR(r)$ & $\sum_{p\in perms(r)} revokeP(p,r)$ \\
	$assignU(u,r)$ & $\IBEEnc+\IBEDec+\IBSSign+\IBSVerify$ \\
	$revokeU(u,r)$ & $\IBEKeyGen+\IBSKeyGen+ \left( |users(r)| + \sum_{p \in perms(r)} (versions(p) + |roles(p)|)\right)\big( \IBEEnc+ \IBSSign+ \IBSVerify\big)+ \left(\IBEDec\cdot\sum_{p \in perms(r)} versions(p) \right) $\\
	$assignP(p,r)$ & $versions(p)\cdot \left(\IBSSign + \IBSVerify\right)$; if $r$ has no permissions for the file then also $versions(p)\cdot \left(\IBEEnc+\IBEDec\right)$ \\
	$revokeP(p,r)$ & Revokes all access: $|roles(p)|\cdot \left(\IBEEnc+ \IBSSign+\IBSVerify\right)$;\newline
	Revokes only write access: $|versions(p)|\cdot \left(\IBSSign+ \IBSVerify\right)$ \\
	$read(fn)$ & $2\cdot \left(\IBEDec+ \IBSVerify\right)$ \\
	$write(fn,f)$ & $\IBSSign + 2\cdot \left(\IBEDec+ \IBSVerify\right)$ and  $2\cdot \IBSVerify$ by R.M.
\end{tabularx}
\caption{Algebraic costs of $\rbac_0$ operations in our IBE/IBS implementation}
\label{tab:algcosts}
\end{table*}

\subsection{Experimental Setup}

To evaluate the costs of using our constructions to enforce
$\rbac_0$, we utilize the simulation framework proposed
in~\cite{garrison2014sacmat}. 
We encode $\rbac_0$ as a
workload, with implementations in IBE/IBS and PKI as described in
\cref{sec:implementation-details,sec:implementation-pki}.  
Simulations are initialized from start states extracted from
real-world $\rbac$ datasets. We then generate traces of access control actions
using actor-specific continuous-time Markov chains, or \emph{actor machines}.
While this is a fairly simple model of actors' behaviors, it allows us
to easily investigate trends in costs. In particular, we are able to investigate
changes in the relative frequencies of the various administrative actions, and
the costs resulting from these changes.

\begin{figure}
\centering
\includegraphics[scale=0.40]{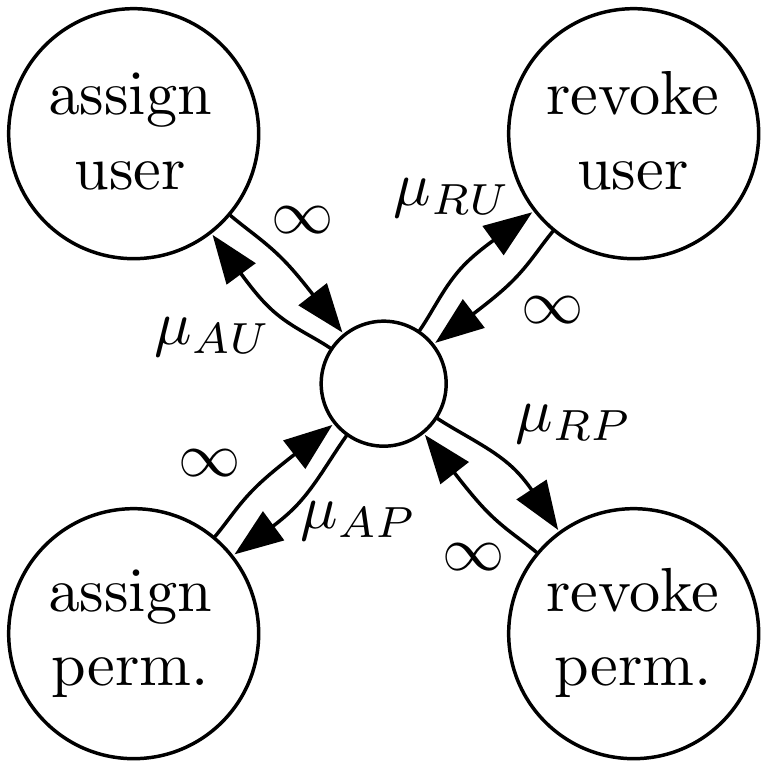}\\
\vspace{1em}
\resizebox{\columnwidth}{!}{
\begin{tabular}{c c c}
var & semantics & value \\\hline
\(R\) & administrative rate & \(0.1 \times \sqrt{|U|} / \text{day}\) \\
\(\mu_A\) & add bias & \(\left[0.7, 1.0\right]\) \\
\(\mu_U\) & UR bias & \(\left[0.3, 0.7\right]\) \\
\(\mu_{AU}\) & Rate of assignUser & \(\mu_A \times \mu_U \times R\) \\
\(\mu_{RU}\) & Rate of revokeUser & \((1 - \mu_A) \times \mu_U \times R\) \\
\(\mu_{AP}\) & Rate of assignPermission & \(\mu_A \times (1 - \mu_U) \times R\) \\
\(\mu_{RP}\) & Rate of revokePermission & \((1 - \mu_A) \times (1 - \mu_U) \times R\)
\end{tabular}
} 
\caption{Administrative actions in our experiments}
\label{fig:adminmachine}
\end{figure}

We simulate one-month periods in which the administrator of the system behaves
as described in the actor machine depicted in \cref{fig:adminmachine}. The
administrative workload increases with the number of users in the system, and we
randomly sample an \emph{add bias} parameter that describes the relative
proportion of assignment vs.\ revocation operations. We do not include
administrative actions that add or remove users or roles, due to the unlikely
occurrence of these actions on such short timescales (one-month simulations).

\begin{figure*}[t]
\centering\small
\begin{subfigure}{0.32\textwidth}\small
\includegraphics[width=\columnwidth]{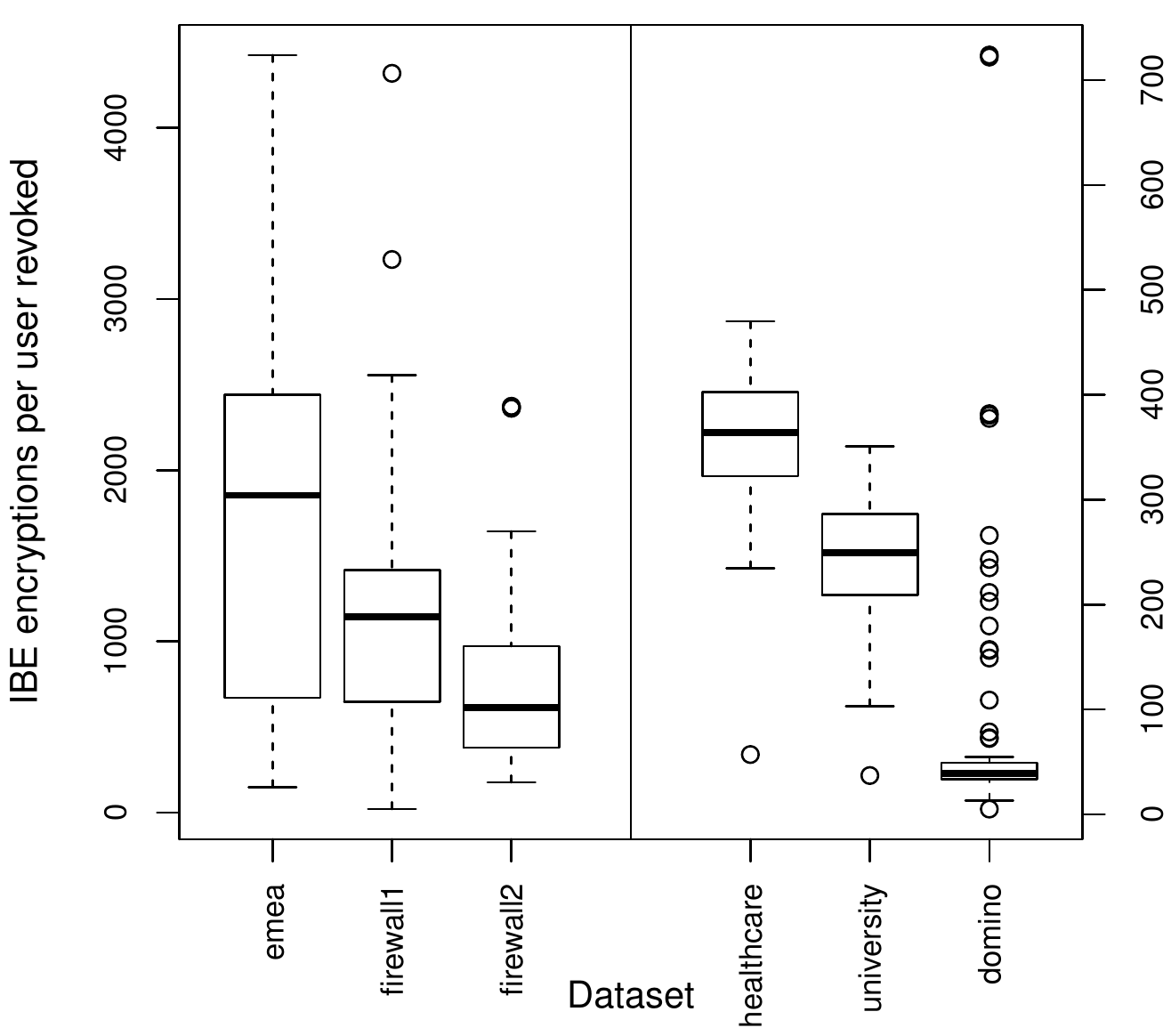}
\caption{IBE encs.\ per user revoked}
\label{chart:ibeuserbox}
\end{subfigure}
\hfill
\begin{subfigure}{0.32\textwidth}
\includegraphics[width=\columnwidth]{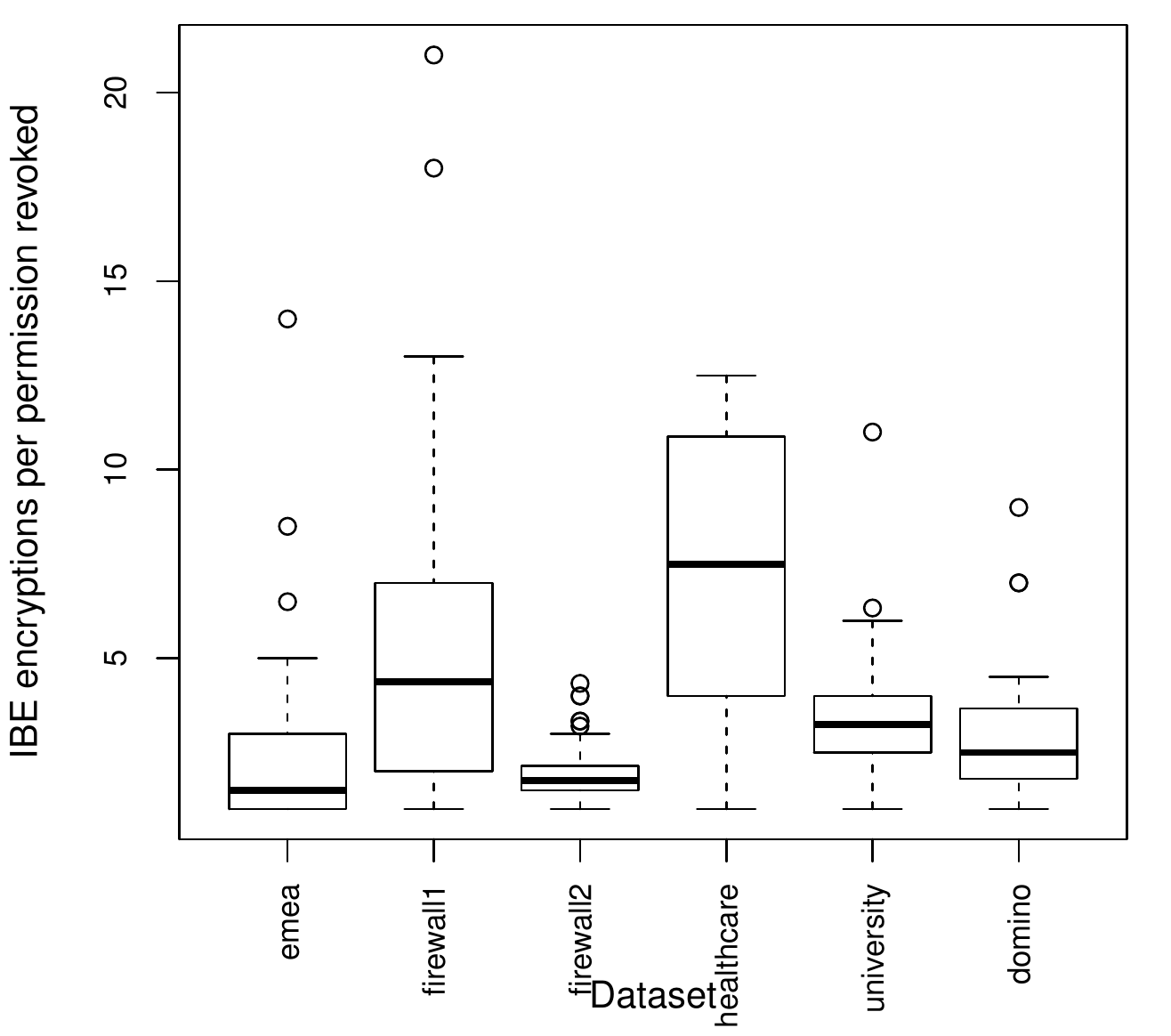}
\caption{IBE encs.\ per permission revoked}
\label{chart:ibepermbox}
\end{subfigure}
\hfill
\begin{subfigure}{0.32\textwidth}
\includegraphics[width=\columnwidth]{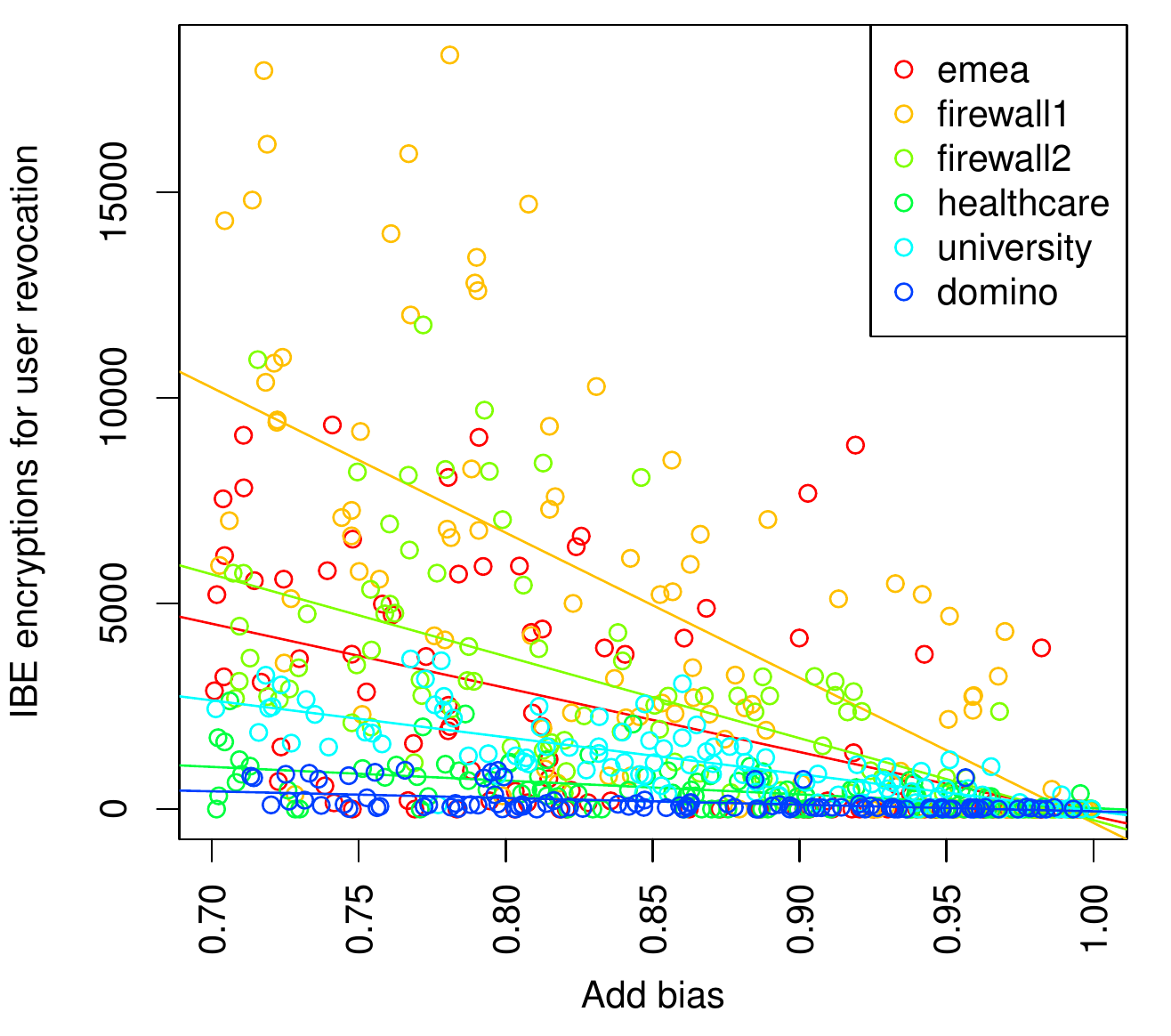}
\caption{IBE encs.\ for user revocation vs.\ add bias}
\label{chart:ibeuserscatter}
\end{subfigure}
\begin{subfigure}{0.32\textwidth}
\includegraphics[width=\columnwidth]{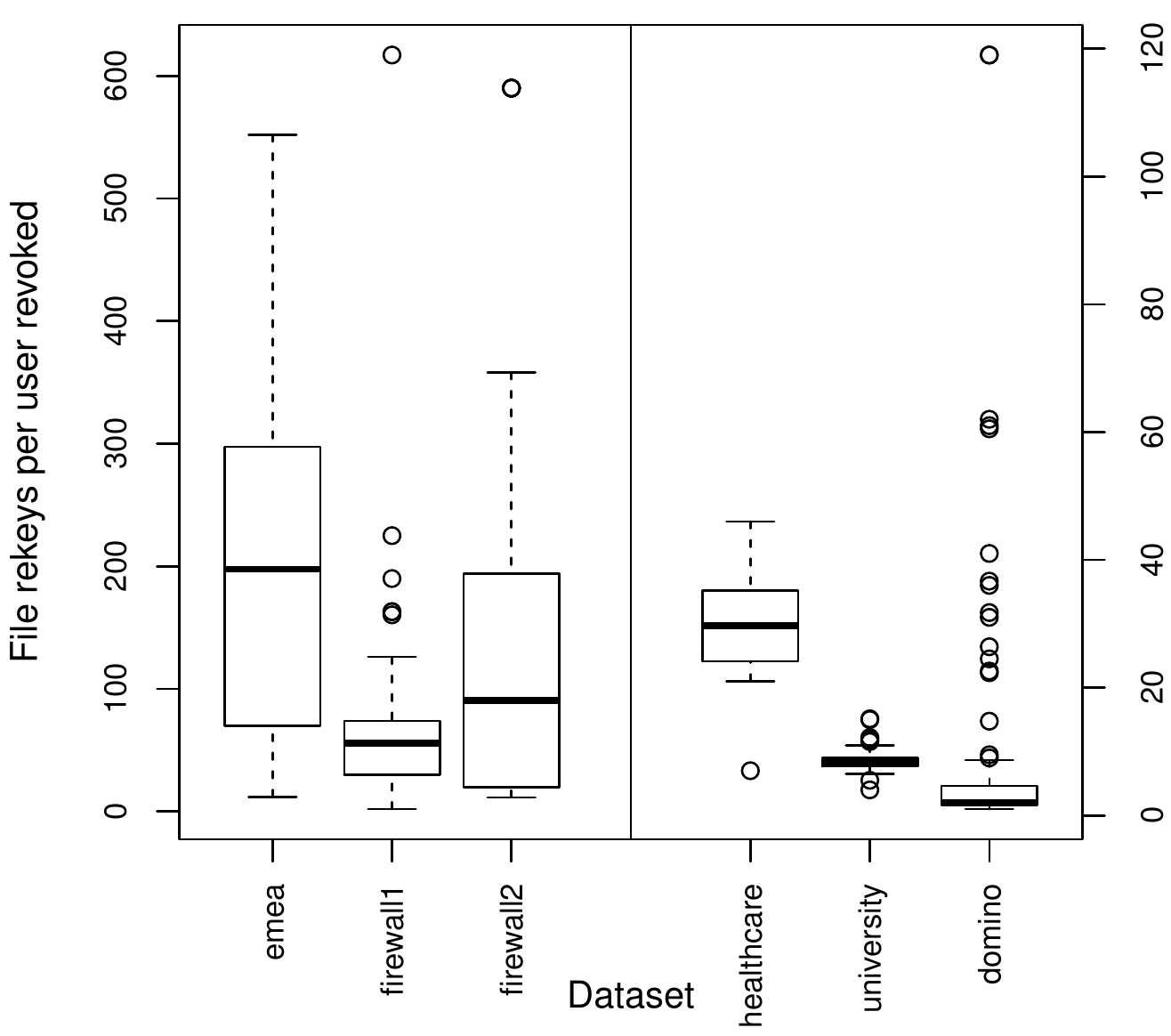}
\caption{File rekeys per user revoked}
\label{chart:rekeysuserbox}
\end{subfigure}
\hfill
\begin{subfigure}{0.32\textwidth}
\includegraphics[width=\columnwidth]{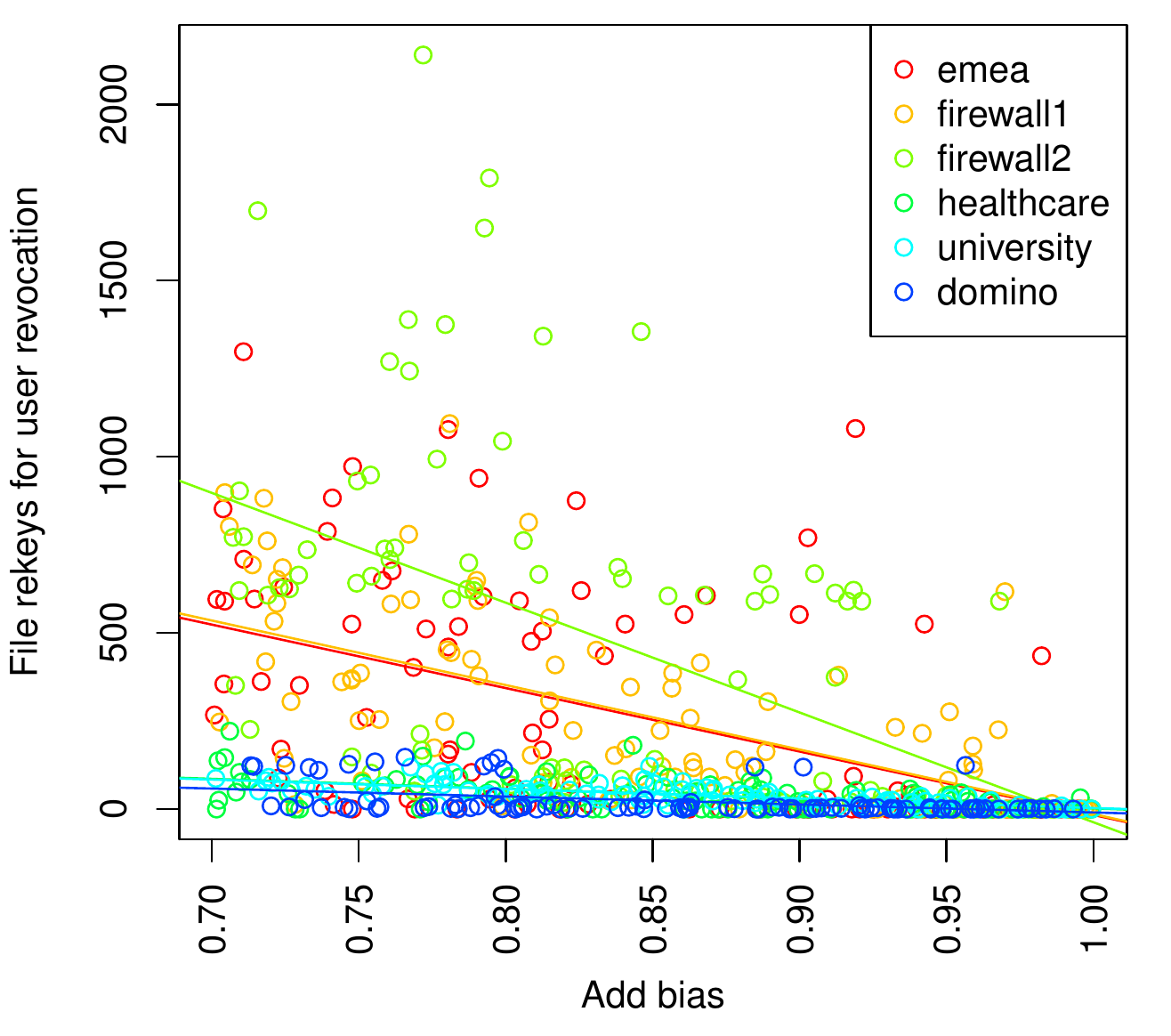}
\caption{File rekeys for user revoc.\ vs.\ add bias}
\label{chart:rekeysuserscatter}
\end{subfigure}
\hfill
\begin{subfigure}{0.32\textwidth}
\includegraphics[width=\columnwidth]{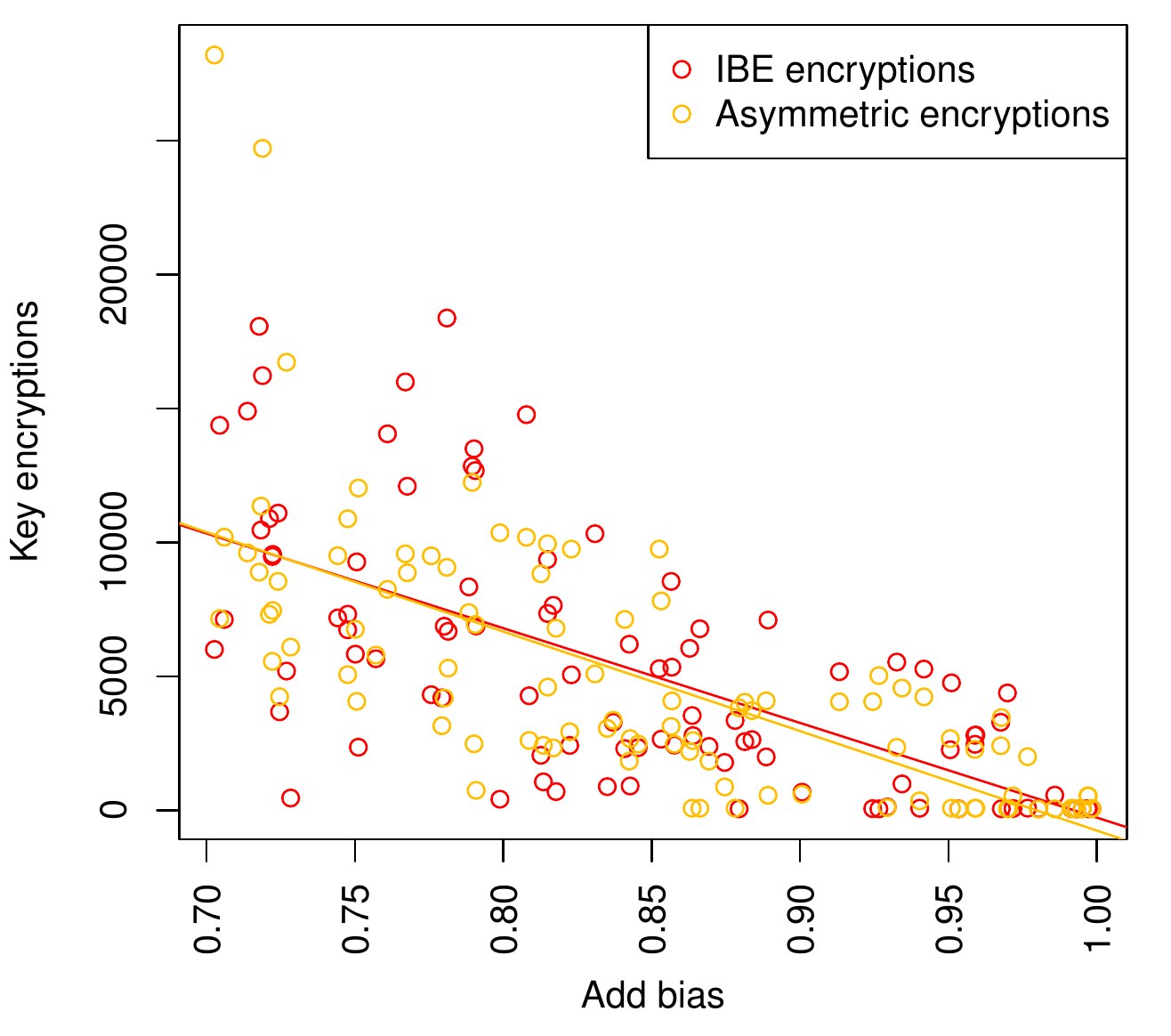}
\caption{Key encryptions vs.\ add bias (\ds{firewall1})}
\label{chart:encryptionsscatter}
\end{subfigure}
\caption{Results of running 100 one-month simulations on each dataset (each data point is a simulation)}
\label{chart:results}
\end{figure*}

{ \newcommand{\size}[1]{\textbar#1\textbar}

\begin{table*}
\centering\small
\begin{tabular}{c c c c c c c c c c c c c c}
&&&&&& \multicolumn{2}{c}{roles/user} & \multicolumn{2}{c}{users/role} & \multicolumn{2}{c}{perm./role} & \multicolumn{2}{c}{roles/perm.} \\
set & users & \size{P} & \size{R} & \size{UR} & \size{PA} & max & min & max & min & max & min & max & min \\\hline
\ds{domino} & 79 & 231 & 20 & 75 & 629 & 3 & 0 & 30 & 1 & 209 & 1 & 10 & 1 \\
\ds{emea} & 35 & 3046 & 34 & 35 & 7211 & 1 & 1 & 2 & 1 & 554 & 9 & 31 & 1 \\
\ds{firewall1} & 365 & 709 & 60 & 1130 & 3455 & 14 & 0 & 174 & 1 & 617 & 1 & 25 & 1 \\
\ds{firewall2} & 325 & 590 & 10 & 325 & 1136 & 1 & 1 & 222 & 1 & 590 & 6 & 8 & 1 \\
\ds{healthcare} & 46 & 46 & 13 & 55 & 359 & 5 & 1 & 17 & 1 & 45 & 7 & 12 & 1 \\
\ds{university} & 493 & 56 & 16 & 495 & 202 & 2 & 1 & 288 & 1 & 40 & 2 & 12 & 1 \\
\end{tabular}
\caption{Overview of the datasets used in our experiments}
\label{fig:datasets}
\end{table*}

} 

This administrative behavior model describes a range of realistic scenarios and
thus allows us to investigate the interactions in which we are interested. The
overall administrative rate is approximately \(\sqrt{\left|U\right|}\) (with
\(\left|U\right|\) the number of users), ranging from about 0.6 administrative
actions per day on our smallest dataset to 2.2 on the largest. We consider the
range of 0\% to 30\% of the administrative load consisting of revocations, since
in realistic scenarios permissions tend to be assigned at a greater rate than
they are revoked~\cite{SinclairSTJP07}.

To quantify the costs associated with our cryptographic constructions, we record
the number of instances of each cryptographic operation executed, including
counts or averages for traces of related operations (e.g., the average number of
IBE encryptions needed to revoke a role from a user).

As mentioned above, 
 simulation start states are extracted from real-world $\rbac$ datasets. 
These datasets are summarized in \cref{fig:datasets}.
All of
these datasets, aside from \ds{university}, were originally provided by
HP~\cite{EneHM+08}. The \ds{domino} dataset is from a Lotus Domino server,
\ds{emea} is from a set of Cisco firewalls, \ds{firewall1} and \ds{firewall2}
are generated from network reachability analysis, and \ds{healthcare} is a list
of healthcare permissions from the US Veteran's Administration. The
\ds{university} dataset describes a university's access control system, and was
developed by IBM~\cite{SYRG07,MCL+08}.

\subsection{Experimental Results} \label{sec:results-experimental}

\Cref{chart:results} presents a sampling of our results. First, we consider the
cost of performing revocations in our implementation of $\rbac_0$ using IBE/IBS.
\Cref{chart:ibeuserbox} shows the average number of IBE encryptions needed for a
single user revocation (i.e., removing a user from a role), and
\cref{chart:ibepermbox} shows the same for permission revocation (i.e., revoking
a permission from a role). This shows that revoking a permission can cost
several IBE encryptions, while user revocation incurs hundreds or thousands of
IBE encryptions, on average. We note that, by inspection of the code in
\cref{fig:rbacibe}, a user revocation also requires an equal number of IBS
signatures and verifications, a smaller number of IBE decryptions, and the
generation of new IBE and IBS keys for the role.

For our chosen distribution of administrative actions,
\cref{chart:ibeuserscatter} shows the total number of IBE encryptions performed
over a month for \emph{all} user revocations. As the add bias approaches 1, the
number of revocations (and thus the total number of IBE encryptions for user
revocation) approaches 0. However, even when only 5--10\% of administrative
actions are revocation, the number of monthly IBE encryptions under this
parameterization is often in the thousands.

In \cref{chart:rekeysuserbox}, we show the number of files that must be re-keyed
for a single user revocation. This highlights the benefit of utilizing lazy
re-encryption; if we had instead utilized \emph{active} re-encryption, each of
these files would need to be locked, downloaded, decrypted, re-encrypted, and
re-uploaded \emph{immediately} following revocation. In certain scenarios,
active re-encryption may be computationally feasible. For instance, in
\ds{university}, only $\approx$~10 files must be re-encrypted for the average
user revocation, adding less than 1\% to the total number of file encryptions
executed over the entire simulation, even at the highest rate of revocations
that we consider. However, in most other scenarios, a user revocation triggers
the re-key of tens or hundreds of files, such as in \ds{emea} or \ds{firewall2},
where active re-encryption increases the total number of file encryptions by
63\% and 12\%, respectively (at 20--30\% revocation rate). Thus, in most
scenarios, active re-encryption is likely to be infeasible, as discussed in
\cref{sec:designconsider}.

Given the administrative behavior model depicted in \cref{fig:adminmachine},
\cref{chart:rekeysuserscatter} shows the total number of file re-keys that take
place over a month for the purpose of user revocation. For scenarios with very
user- and permission-dense roles (e.g., \ds{firewall1} and \ds{firewall2}), we
see several times as many re-keys as \emph{total files}, indicating that, on
average, each file is re-keyed multiple times per month for the purposes of user
revocation. This further enforces that inefficiencies that active re-encryption
would bring, as each file (on average) would be locked and re-encrypted by the
administrator multiple times per month.

Finally, we note that the costs for our IBE/IBS- and PKI-based constructions for
$\rbac_0$ are not notably different. For instance,
\cref{chart:encryptionsscatter} compares, for scenario \ds{firewall1}, the
number of IBE encryptions with the number of asymmetric encryptions executed
over each simulated month and reveals the same distribution in both IBE/IBS- and
PKI-based constructions. Given the similarity in the cost of these classes of
operations, we can conclude that these constructions are similarly expensive
from a computational standpoint.

\subsection{Converting Experimental Results to Real Costs}
\label{subsec:realcosts}

We now demonstrate how the costs of generic IBE encryptions turn into actual
computational costs for given schemes. Since any implementation's running time
is contingent on a myriad of variables (e.g., processor speed, memory, etc.) we
focus on the number of (pairing friendly) elliptic curve cryptographic
operations that need to be performed. We assume schemes are implemented using an
asymmetric (Type~3) pairing: $e : \G \times \hat{\G} \rightarrow \G_T$, where
$\G, \hat{\G}, \G_T$ are groups of prime order; this is more efficient than a
symmetric (Type~1) pairing~\cite{Galbraith:2008:PC:1450345.1450543}. Additive
notation is used in $\G$ and $\hat{G}$, while multiplicative notation is used in
$\G_T$.

We use multiplication in $\G$ as our cost unit, expressing the relative costs of
other operations in terms of this operation. The relative costs should be
somewhat stable across hardware and reasonable implementations. These relative
costs are given in \cref{Table:PairingCosts} and are based on data provided by
Ayo Akinyele, an ABE/pairing implementation expert at Zeutro LLC (personal
communication). 
Costs of addition in $\G, \hat{\G}$, and multiplication
in $\G_T$ are so low that we ignore them. These relative costs are based on the
implementation of RELIC v0.4~\cite{relic-toolkit}, using a Barreto-Naehrig curve
with a 256-bit base field, GMP for big number operations, and standard
configuration options for prime field arithmetic. For a point of reference, a
reasonable modern workstation running RELIC v0.4 on such curves will take
approximately 0.2 ms on average to compute a multiplication in $\G$.

\begin{table}[htb]
\centering\small
\begin{tabular}{l||c|c|c}
Operation & $\hat{\G}$ Multiply & $\G_T$ Exp. & Pairing ($e$) \\
\hline
$\G$ Multiplies & 4.5 & 9 & 9\\
\end{tabular}
\caption{Relative cost of Type 3 pairing operations in terms of multiplication
in $\G$ in RELIC v0.4}
\label{Table:PairingCosts}
\end{table}

To determine concrete costs, we consider three representative combinations of
IBE and IBS algorithms:

\textbf{BF+CC:} The IBE scheme from~\cite[Sec.\ 4.1]{Boneh2003} and the IBS
scheme from~\cite[Sec.\ 2]{Cha:2003:ISG:648120.746918}. Both are efficient and
are proven secure in the random oracle model.

\textbf{$\text{BB}_1$+PS:} The IBE scheme from~\cite[Sec.\
4]{Boneh:2011:ESI:2110135.2110139} and the IBS scheme from~\cite[Sec.\
4]{Paterson:2006:EIS:2171400.2171424}. These schemes are less efficient than
BF+CC but are proven secure in the standard model.

\textbf{LW+PS:} The IBE scheme from~\cite[App.\
C]{Lewko:2010:NTD:2128056.2128093} and the IBS scheme from~\cite[Sec.\
4]{Paterson:2006:EIS:2171400.2171424}. The IBE scheme here is less efficient but
has stronger security properties.

A table
documenting the individual costs of each basic IBE/IBS operation for
these schemes as well as several others can be found in \cref{sec:costs}.

\Cref{tab:opcosts} lists the cost of each additive $\rbac_0$, read and
write operation in terms of total ``multiplication units'' in
$\G$. That is, we sum the cost of cryptographic operations in terms of
multiplication units using the conversion factor in
\cref{Table:PairingCosts}. \Cref{tab:opcosts} specifies the costs
incurred by the invoker of the operation (either the admin or the
user) as well as the reference monitor.

\begin{table}[htb]
\centering\small
\begin{tabular}{clccc}
Incurred by              & Operation & BF+CC & $\text{BB}_1$+PS & LW+PS \\ \hline
\multirow{7}{*}{Invoker} & addU      & 5.5   & 14.5  & 32.5 \\
                         & addP      & 15    & 25    & 29 \\
                         & addR      & 18.5  & 33    & 55 \\
                         & assignU   & 41    & 63.5  & 103.5 \\
                         & assignP\footnote{Assumes permission is for new file; cost is per version of the file} & 41 & 63.5 & 103.5 \\
                         & read      & 56    & 90    & 162 \\
                         & write     & 58    & 96.5  & 168.5 \\ \hline
\multirow{2}{*}{R.M.}    & addP      & 38    & 54    & 54 \\
                         & write     & 38    & 54    & 54 \\
\hline
\end{tabular}
\caption{Costs of operations in terms of $\G$ multiplications}
\label{tab:opcosts}
\end{table}

\begin{figure*}[t]
	\centering
	\begin{subfigure}{0.49\textwidth}
		\includegraphics[width=\columnwidth]{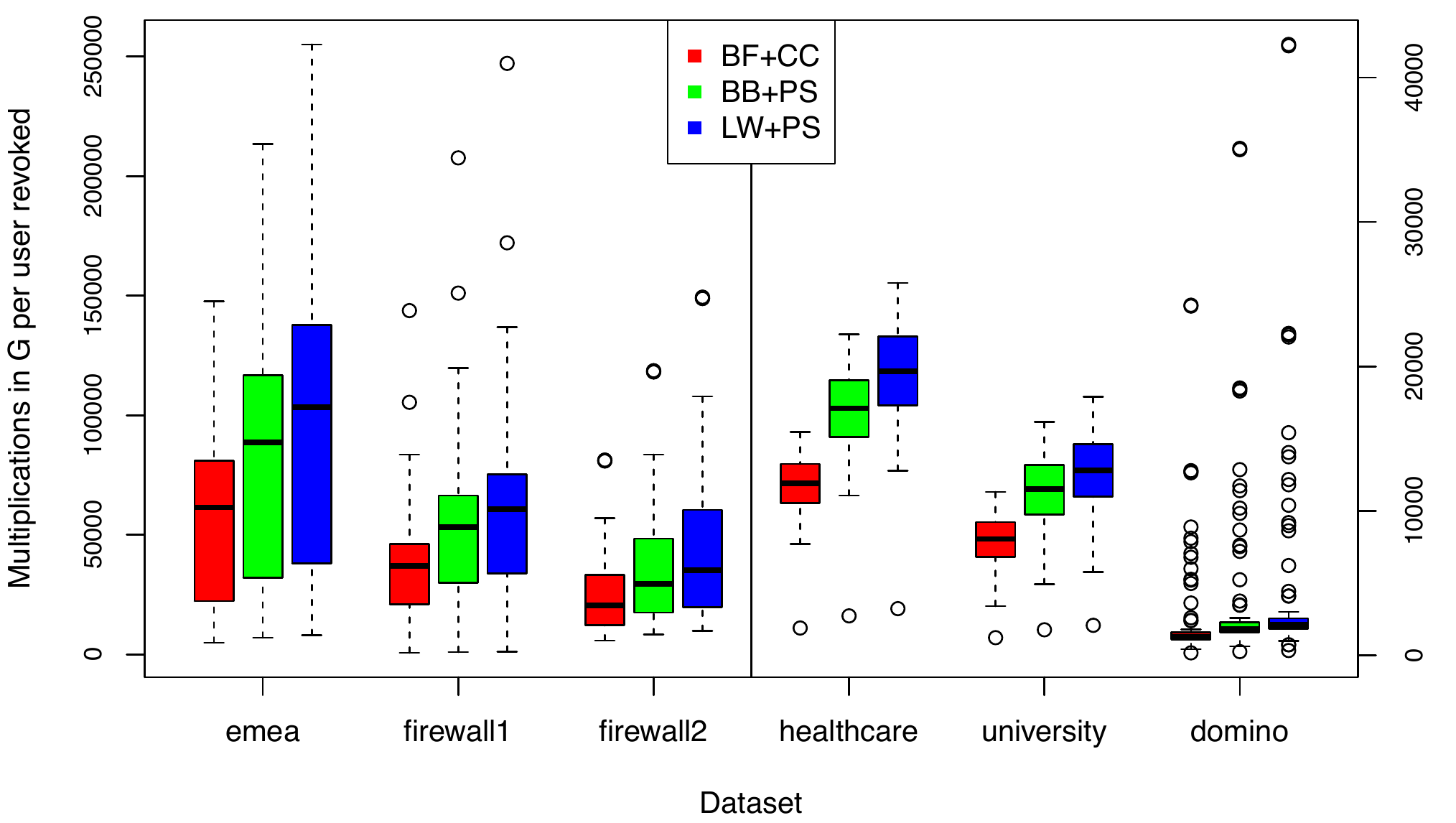}
		\caption{Multiplications in $\G$ per user revoked}
		\label{chart:multuserbox}
	\end{subfigure}
	\hfill
	\begin{subfigure}{0.49\textwidth}
		\includegraphics[width=\columnwidth]{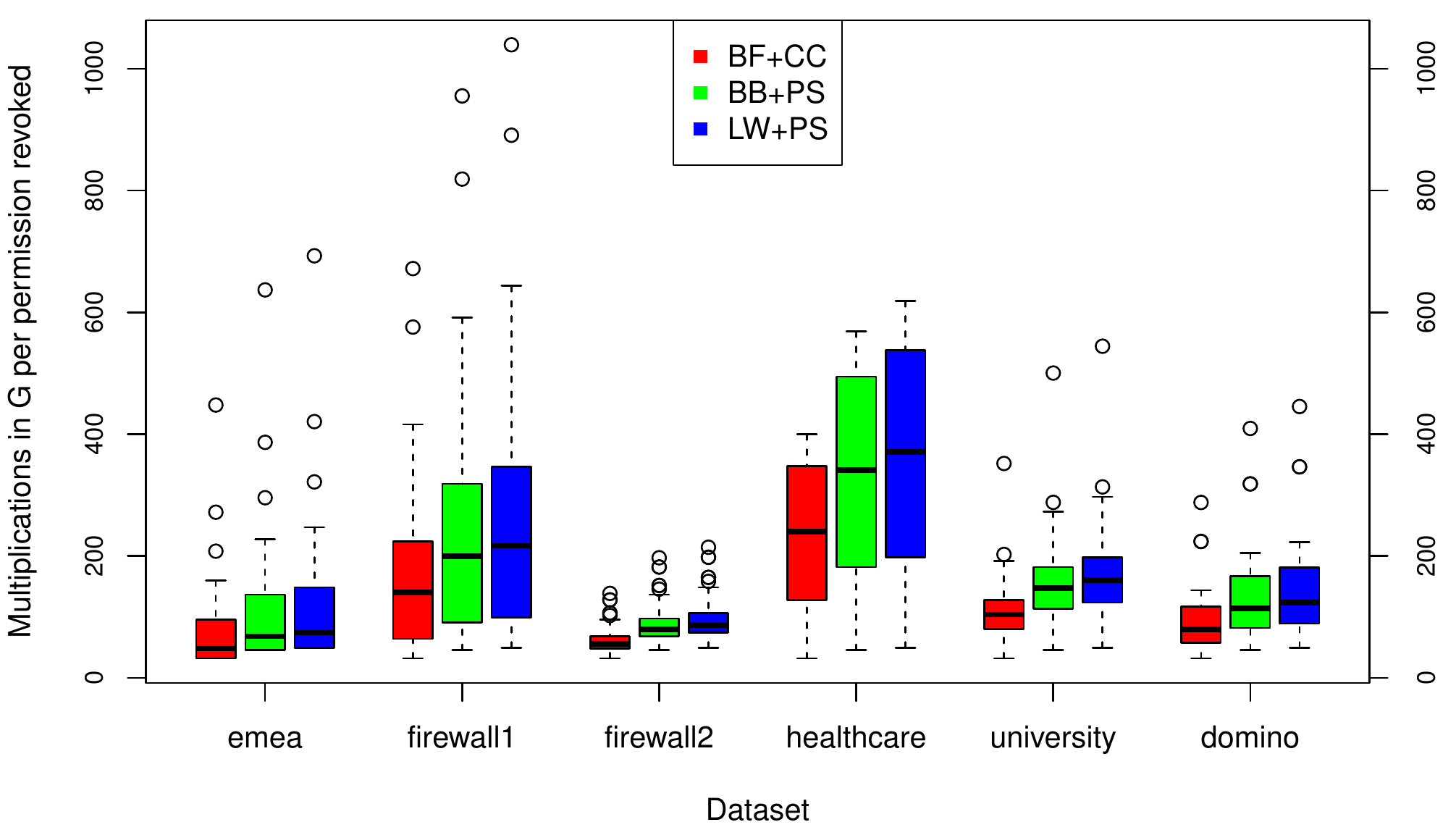}
		\caption{Multiplications in $\G$ per permission revoked}
		\label{chart:multpermbox}
	\end{subfigure}
	\caption{Costs of revocation in different IBE/IBS schemes, in terms of elliptic curve point multiplications in group $\G$}
	\label{chart:revokecosts}
\end{figure*}

The cost to delete a user/role or to revoke a user/permission depends on the
\rbac state at the time of revocation, so we cannot give definite costs for
these operations. Instead, we use the experimental results from
\cref{sec:results-experimental} to get an idea of how expensive revocation can
be. The results of this are in \cref{chart:revokecosts}, where we plot the costs
for each dataset using the three IBE/IBS combinations listed above.
\Cref{chart:multuserbox} shows the cost of revoking a user in terms of
multiplications in $\G$; \cref{chart:multpermbox} does the same for revoking a
permission. Note that for our datasets, a single user revocation usually costs
more than 10,000 multiplications in $\G$ ($\approx$~2~s.\ on a modern
workstation), and often costs more than 100,000 multiplications
($\approx$~20~s.) for some datasets. While not exceedingly huge, we remind the
reader that our costing does not account for many costs, such as concurrency,
communication, and storage costs. Further, our construction minimizes other
costs through the use of lazy re-encryption and hybrid encryption.

\section{Discussion} \label{sec:discussion}

There is no doubt that IBE and ABE can enable various forms of
cryptographic access control for data in the cloud. In fact, the
results presented in
\cref{chart:ibeuserscatter,chart:rekeysuserscatter,chart:encryptionsscatter}
show that in situations in which the system grows in a monotonic
manner (i.e., users and files are \emph{added} to the system and roles
are provisioned with \emph{new} permissions), there is no need for
revocation, re-keying, or complicated metadata management: IBE alone
can enforce $\rbac$ access controls on the cloud.
In fact, there are even implications or direct claims in the
literature that, in the static setting, the reference monitor can be
removed entirely (e.g.,~\cite{Goyal2006,GJPS08,Moller12}).  However,
this does not imply that IBE or ABE alone can entirely replace the use
of a reference monitor when implementing outsourced access controls:
\emph{it is not the case when dynamic controls are required}.

Specifically, this paper shows that IBE and PKI systems are
well-suited for implementing \emph{point states} of an $\rbac_0$
system. However, managing \emph{transitions} between these
states---specifically, supporting the removal of a user from a role,
the revocation of a permission from a role, and efficient updates to
files shared with multiple roles---requires non-trivial metadata
management and a small, minimally-trusted reference monitor that
verifies signatures prior to file deletion and replacement. In some of
the datasets that we analyzed, this could lead to thousands of IBE
encryptions (\cref{chart:ibeuserbox}) and over one hundred file
re-keys/re-encryptions (\cref{chart:rekeysuserbox}) when a single user
is removed from a role.

The above considerations lead to a tradeoff between confidentiality
and efficiency 
that must be weighed by both cryptographers and system designers.
There are two obvious ways that this can be accomplished: by altering
the threat model assumed, or developing cryptographic approaches that
are more amenable to the dynamic setting.  We now discuss both of these approaches, and comment on lessons learned during our analysis that can be applied to richer cryptographic access control, such as using HIBE to support $\rbac_1$, or ABE to support $\abac$.

\subsection{Alternate Threat Models}

Many of the overheads that we report on in the previous section result
from the threat model often implied by the
cryptographic literature (i.e., untrusted storage server, minimal
client-side infrastructure).  Altering this model can reduce the
cryptographic costs of enforcing dynamic access controls on the cloud.
Here we consider two such alternate models.


\textbf{Encryption/Decryption Proxy.} A large amount of overhead
  comes from relying the cloud storage provider to act as a
  (cryptographic) metadata broker, as well as a file store.  An
  alternative approach might make use of an encryption/decryption
  proxy server situated within an organization, using the cloud
  provider solely as a backing store for encrypted files.  This proxy
  would act as a traditional reference monitor, mediating all file
  access requests, downloading and decrypting files for authorized
  readers\footnote{Writes could be handled symmetrically.}, and
  returning plaintext to the user.  This would obviate the need for
  any cryptography beyond authenticated symmetric key encryption, and
  could make use of tried-and-true access control reference monitors.
  However, this approach carries an extra infrastructure overhead (the
  proxy server, itself) that could make it unappealing to
  \emph{individuals} hoping to enforce access controls over cloud
  hosted files.  Large \emph{organizations} may also have to deal with
  synchronizing access control policies and key material across
  multiple proxies in the event that file I/O demands outpace the
  abilities of a single server.

  \textbf{Trusted Hardware.} A more extreme approach to simplifying
  the cryptographic overheads of access control enforcement would be
  to use, e.g., an SGX enclave~\cite{sgx-programming, sgx} to carry out the work of the
  encryption/decryption proxy discussed above.  In this scenario,
  files could be stored encrypted on the cloud server, while file
  encryption keys and the access control policy to be enforced would
  be managed by a process running within an SGX enclave.  To access a
  file, a user would negotiate an authenticated channel (e.g., using
  public key cryptography) with this trusted process/reference
  monitor.  The reference monitor could then check the user's
  permission to access the file, and transmit the encrypted file and
  its associated key to the user using a session key that is unknown
  to any process outside of the SGX enclave.  This approach frees
  organizations from the overheads of running their own
  encryption/decryption proxies, but is not without its
  limitations. For instance, this approach will not work on
  commonly-used, storage-only services (e.g., Dropbox).  Further, this
  approach may be subject to architectural compromises or flaws (e.g.,
  memory integrity vulnerabilities) that cryptography-only solutions
  are not.


While these and other alterations to the threat model that we consider
can lead to decreased cryptographic overheads, each incurs other costs
or tradeoffs.  We now consider future research directions that may
decrease the costs associated with cryptography-only solutions to the
problem of outsourcing dynamic access controls.

\subsection{Future Directions}

Our experimentation and analysis has led to a number of interesting
directions for future work:

\begin{itemize}[itemsep=0pt,leftmargin=*]

\item \textbf{Revocation.} It is unclear how to use IBE to enforce even
$\rbac_0$ without incurring high costs associated with revocation-based state
changes. Given our use of hybrid cryptography for efficiency reasons, existing
schemes for revocation or proxy re-encryption
(e.g.,~\cite{Green:2007:IPR:1419765.1419791,Boldyreva2008}) cannot solve the
problem. Developing techniques to better facilitate these forms of revocation
and efficient use of hybrid encryption is an important area of future work.

\item \textbf{Trust Minimization.} Our construction makes use of a reference
monitor on the cloud to validate signatures prior to file replacement or
metadata update. Moving to file versioning (e.g., based on trusted timestamping
or block-chaining) rather than file replacement may result in a minimization of
the trust placed in this reference monitor, but at the cost of potential
confidentiality loss, since old key material may remain accessible to former
role members. It is important to better explore this tradeoff between reference
monitor trust and confidentiality guarantees.

\item \textbf{``Wrapper'' Minimization.} Our construction required the
management and use of three types of metadata structures to correctly implement
$\rbac_0$ using IBE or PKI technologies. It would be worth exploring whether
the core cryptography used to support outsourced access controls could be
enhanced to reduce the use of trusted management code needed to maintain these
sorts of structures.

\item \textbf{Deployability/Usability Costs.} We did not consider issues
related to the \emph{use} of the cryptographic tools underlying our
constructions. Further, our simulations do not separate our IBE- and PKI-based
constructions\footnote{Our PKI construction and its corresponding simulations
were omitted from this paper due to space limitations.} on the basis of
$\rbac_0$ implementation complexity. However, it may be the case that the
maturity of tools to support the use of PKIs or the conceptual simplicity of IBE
techniques tips the scales in one direction or the other. Developing reasonable
approaches for considering these types of tradeoffs would greatly inform future
analyses.

\end{itemize}

While this paper focused on the use of IBE/IBS and PKI schemes to
enforce $\rbac_0$ access controls, our findings translate in a
straightforward manner to the use of other cryptographic tools (e.g.,
HIBE or ABE/ABS) to implement more complex access control policies
(e.g., $\rbac_1$ or $\abac$).  We now discuss some lessons learned
when considering these richer access control models.



\subsection{Lessons Learned for More Expressive Systems}\label{sec:discussion-generalizing}
$\rbac_0$ and IBE were natural choices for our initial exploration of
the costs associated with using cryptography to implement dynamic
access control: $\rbac_0$ is a simple, but widely used, access control
system; roles in $\rbac_0$ have a natural correspondence to identities
in IBE; and the use of hybrid encryption allows us to easily share
resources between roles. Further, it seemed like an implementation of
$\rbac_0$ using IBE would be a jumping-off point for exploring the use
of hierarchical roles in $\rbac_1$ via an analogous use of
HIBE. However, many of the costs that we see with our IBE
implementation of $\rbac_0$ have analogues (or worse) in any
reasonable $\rbac_1$ or $\abac$ implementation that we foresee based
on respective cryptographic operations.

We first note that we assume that any reasonable cryptographic access
control system must make use of hybrid encryption. Without hybrid
encryption, we would need to continuously apply expensive asymmetric
operations to small ``blocks'' of a file that is to be encrypted.
Given the complexity of IBE/ABE encryption operations, the associated
overheads of this approach would be prohibitive, even for
moderately-sized files. Additionally, depending on the security
requirements of the application (e.g., Chosen Ciphertext Attack
security), even more complicated constructions than this simple
blocking will be required. The following observations may not apply to
an access control scheme where all files are small enough to do away
with the need of hybrid-encryption. However, the use cases for such
schemes seem limited.

A seemingly natural extension of our IBE-based $\rbac_0$ scheme to a
HIBE based $\rbac_1$ scheme exploits the fact that the HIBE can be
used to encode hierarchical relationships, such as those that exist
between roles in a $\rbac_1$ role hierarchy.  However, the costs of
this implementation proved to be considerable. A large initial problem
is that an $\rbac_1$ role hierarchy can be an arbitrary DAG structure,
while HIBE only supports trees. Yet, even limiting $\rbac_1$ to role
hierarchies that form a tree structure comes with serious costs. For
example, removing non-leaf roles in the hierarchy cascades
re-encryption down to all files at descendant leaves of the role, the
creation of new roles for each descendant node, and associated
rekeying. Similarly, practical operations like moving sub-trees in the
access structure can only be achieved by breaking the operation down
into addition and deletion of roles, which comes with the associated
costs of these primitive operations. We note that we have developed a
full $\rbac_1$ implementation using HIBE, which attempts to minimize
costs.  Unfortunately, a simple inspection of this implementation
shows that it would incur significantly more computational expense
than the $\rbac_0$ scheme discussed herein.

Similarly, one might hope that the expressiveness of the ABE
encryption schemes would allow us to naturally implement $\abac$
access control schemes. Further, there has been some initial work~\cite{sahai2012dynamic} supporting dynamic (restrictive) credentials and
revocations.  However, there is still significant work associated with
making a practical ABE implementation of $\abac$, and such schemes
will still have significant costs and meta-data to manage (as in our
IBE/$\rbac_0$ implementation). For example, revoking a secret-key in
an KP-ABE/$\abac$ setting requires the dynamic re-encryption of every
ciphertext whose attributes satisfy the policy in the revoked user's
key. Each attribute in each ciphertext that is re-encrypted must given
a new version, and then finally all users whose keys have policies
affected by the re-versioning of the attributes must be re-issued.
Further, there are $\abac$ design decisions that must be informed by
the ABE scheme being implemented. For example, suppose a single file
is to be accessed by multiple policies in a CP-ABE scheme. One can
support multiple policies $p_1,\dots,p_n$ as individual public-key
encryptions all encrypting the same hybrid key, or as a single
encryption supporting the disjunction of all previous policies, $p_1 \vee p_2 \vee \dots \vee p_n$. The cost trade-offs
are completely dependent on the ABE scheme used for the
implementation, as the cost of ABE encryption is highly dependent on the
policy encoded into the ciphertext.


\section{Conclusions} \label{sec:conclusions}

Advanced cryptographic techniques (e.g., IBE and ABE) are promising
approaches for cryptographically enforcing rich access controls in the
cloud. While prior work has focused on the \emph{types} of policies that
can be represented by these approaches, little attention has been
given to how policies may \emph{evolve} over time.  In this paper, we
move beyond cryptographically representing point states in an access
control system for cloud-hosted data, and study constructions that
cryptographically enforce dynamic (role-based) access controls.  We
provide evidence that, given the current state of the art, in
situations involving even a minimal amount of policy dynamism, the
cryptographic enforcement of access controls is likely to carry prohibitive
costs.  Further, these costs are seemingly amplified when enforcing
richer policies (e.g., $\rbac_1$ or $\abac$), requiring more stringent
security guarantees (e.g., online, rather than lazy, re-encryption),
or assuming more relaxed threat models.

To conduct our analysis, we developed IBE- and PKI-based constructions
that use hybrid cryptography to enforce dynamic $\rbac_0$ access
controls over files hosted on a third-party cloud storage provider.
In addition to proving the correctness of our constructions, we used
real-word $\rbac$ datasets to experimentally analyze their associated
cryptographic costs. Our findings indicate that IBE and ABE are a
natural fit to this problem in instances where users, roles, and
permissions increase monotonically, but incur very high
overheads---e.g., sometimes exceeding thousands of encryption
operations to support a single revocation---when updates and
revocation must be supported. In doing so, we have identified a number
of fruitful areas for future work that could lead to more natural
constructions for cryptographic enforcement of access control policies
in cloud environments.

\vspace{3mm}
\noindent \textbf{Acknowledgements.}  We would like to offer our thanks to our
paper shepherd, \'Ulfar Erlingsson, for his guidance in refining this paper.
This work was supported, in part, by the National Science Foundation under
awards CNS--1111149, CNS--1228697, and CNS--1253204.

\bibliographystyle{IEEEtranS}
\bibliography{refs-short}

\appendices

\section{Full IBE and IBS Costs}
\label[appendix]{sec:costs}

\begin{table*}[tb!]
		\centering\footnotesize
		\renewcommand{\arraystretch}{1.3}
		\begin{tabu}{cl|cccc|cccc|cccc|cccc|cccc}
			& & \multicolumn{4}{c|}{\mm{KeyGen}} & \multicolumn{4}{c|}{Key} & \multicolumn{4}{c|}{\mm{Enc}/\mm{Sign}} & \multicolumn{4}{c|}{Ciphertext/Sig.} & \multicolumn{4}{c}{\mm{Dec}/\mm{Ver}} \\ \hline
			Type & Scheme & $\G$ & $\hat{\G}$ & $\G_T$ & $e$ & $\G$ & $\hat{\G}$ & $\G_T$ & $\Z_p$ & $\G$ & $\hat{\G}$ & $\G_T$ & $e$& $\G$ & $\hat{\G}$ & $\G_T$ & $\Z_p$ & $\G$ & $\hat{\G}$ & $\G_T$ & $e$ \\ \hline
			\rowfont{\bfseries} \multirow{8}{*}{\textmd{IBE}} & BF \cite[Sec. 4.1]{Boneh2003} & \ext{0}{1}{0}{0} & \key{0}{1}{0}{0} & \enc{2}{0}{0}{1} & \ciph{1}{0}{1}{0} & \dec{0}{0}{0}{1} \\
			\rowfont{\bfseries} & $\text{BB}_1$ \cite[Sec. 4]{Boneh:2011:ESI:2110135.2110139} & \ext{0}{2}{0}{0} & \key{0}{2}{0}{0} & \enc{3}{0}{1}{0} & \ciph{2}{0}{1}{0} & \dec{0}{0}{0}{2} \\
			\rowfont{\bfseries} & LW \cite[App. C]{Lewko:2010:NTD:2128056.2128093} & \ext{0}{6}{0}{0} & \key{0}{6}{0}{0} & \enc{7}{0}{1}{0} & \ciph{6}{0}{1}{0} & \dec{0}{0}{0}{6} \\
			& $\text{BB}_2$ \cite[Sec. 5]{Boneh:2011:ESI:2110135.2110139} & \ext{0}{1}{0}{0} & \key{0}{1}{0}{1} & \enc{3}{0}{1}{0} & \ciph{2}{0}{1}{0} & \dec{1}{0}{0}{1} \\
			& W05 \cite[Sec. 4]{Waters:2005:EIE:2154598.2154608} & \ext{0}{2}{0}{0} & \key{0}{2}{0}{0} & \enc{3}{0}{1}{0} & \ciph{2}{0}{1}{0} & \dec{0}{0}{0}{2} \\
			& Gen \cite[Sec. 3]{Gentry:2006:PIE:2171160.2171198} & \ext{0}{1}{0}{0} & \key{0}{1}{0}{1} & \enc{2}{0}{2}{0} & \ciph{1}{0}{2}{0} & \dec{0}{0}{1}{1}\\
			& Boy \cite[Sec. 4]{Boyen:2006:AHI:2165316.2165333} & \ext{0}{5}{0}{0} & \key{0}{5}{0}{0} & \enc{6}{0}{1}{0} & \ciph{5}{0}{1}{0} & \dec{0}{0}{0}{5} \\
			& W09 \cite[Sec. 3]{Waters:2009:DSE:1615970.1616022} & \ext{0}{8}{0}{0} & \key{0}{8}{0}{1} & \enc{14}{0}{1}{0} & \ciph{9}{0}{1}{1} & \dec{0}{0}{1}{9} \\ \hline
			\rowfont{\bfseries} \multirow{5}{*}{\textmd{IBS}} & CC \cite[Sec. 2]{Cha:2003:ISG:648120.746918} & \ext{1}{0}{0}{0} & \key{1}{0}{0}{0} & \enc{2}{0}{0}{0} & \ciph{2}{0}{0}{0} & \dec{1}{0}{0}{2} \\
			\rowfont{\bfseries} & PS \cite[Sec. 4]{Paterson:2006:EIS:2171400.2171424} & \ext{1}{1}{0}{0} & \key{1}{1}{0}{0} & \enc{2}{1}{0}{0} & \ciph{1}{2}{0}{0} & \dec{0}{0}{0}{3} \\
			& Pat \cite[Sec. 3]{Paterson02id-basedsignatures} & \ext{1}{0}{0}{0} & \key{1}{0}{0}{0} & \enc{2}{1}{0}{0} & \ciph{1}{1}{0}{0} & \dec{1}{0}{1}{2} \\
			& Hes \cite[Sec. 2]{Hess:2002:EIB:646558.694902} & \ext{1}{0}{0}{1} & \key{1}{0}{1}{0} & \enc{1}{0}{1}{0} & \ciph{1}{0}{0}{1} & \dec{1}{0}{0}{2} \\
			& BLMQ \cite[Sec. 3]{Barreto:2005:EPI:2099921.2099958} & \ext{1}{0}{0}{0} & \key{1}{0}{0}{0} & \enc{1}{0}{1}{0} & \ciph{1}{0}{0}{1} & \dec{0}{1}{1}{1} \\
		\end{tabu}
	\caption{Operation costs and sizes in IBE and IBS schemes}
	\label{tab:ibecost}
\end{table*}

\Cref{tab:ibecost} lists the efficiency of several IBE and IBS schemes.
It lists the schemes discussed in \cref{subsec:realcosts} in bold, and also includes several other IBE/IBS schemes for comparison.

The table lists the cost of running \IBEKeyGen or \IBSKeyGen, the size of each private key, the cost of running \IBEEnc or \IBSSign, the size of each ciphertext/signature, and the cost of running \IBEDec or \IBSVerify.
For the algorithmic costs, columns $\G$ and $\hat{\G}$ represent the number of multiplications in those groups, while $\G_T$ represents the number of exponentiations.
Column $e$ represents the number of pairings that must be computed.
For the sizes, columns $\G$, $\hat{\G}$, $\G_T$, and $\Z_p$ represent the number of elements in each of those groups.


\section{Handling Differences in Versioning}
\label[appendix]{sec:versioning}
\allowdisplaybreaks

Because our IBE/IBS and PKI systems uses versioning to handle revocation, assigning and then revoking a user/permission will not result in the same state as if the user/permission were never assigned.
However, it will result in the same set of users having access to the latest versions of the same files, so the results of authorization requests will not be changed.
We consider such states, which are equal except for differences in versioning, to be \emph{congruent}, and represent this with the $\cong$ relation.
We also say that state mappings $\sigma$ and $\sigma'$ are congruent if $\sigma(x) \cong \sigma'(x)$ for all states $x$.

The definition of correctness from \cite{Hinrichs:2013:AAC:2510170.2510419} requires that $\alpha$ preserves $\sigma$, which means the following: For all $n \in \mathbb{N}$, states $x_0$, and labels $\ell_1,\dots,\ell_n$, let $y_0 = \sigma(x_0)$, $x_i = next(x_{i-1},\ell_i)$ for $i = 1,\dots,n$, and $y_i = terminal(y_{i-1},\alpha(y_{i-1},\ell_i))$ for $i = 1,\dots,n$. Then $\alpha$ preserves $\sigma$ means that $y_i = \sigma(x_i)$ for all $i = 1,\dots,n$.

We cannot achieve this in our system because of version numbers, e.g., if $\ell_1$ assigns a user to a role and then $\ell_2$ revokes that user from the role, $x_2$ will be equal to $x_0$ (and thus $\sigma(x_2)$ will be equal to $\sigma(x_0)$), but $y_2$ will have version numbers different from $y_0$.
Thus instead we will show that $y_i \cong \sigma(x_i)$ for all $i = 1,\dots,n$, which we define as $\alpha$ \emph{congruence-preserves} $\sigma$.

In \cite{garrison2014codaspy}, $\alpha$ preserves $\sigma$ is defined as
\begin{equation}
\label{eq}
\sigma\Big(next\big(x,\ell\big)\Big) = terminal\Big(\sigma(x),\alpha\big(\sigma(x),\ell\big)\Big)
\end{equation}
for every state $x$ and label $\ell$.
This implies the definition from \cite{Hinrichs:2013:AAC:2510170.2510419} by the following inductive argument:

\begin{proposition}
Let $x_0$ be a state, $\ell_1,\dots,\ell_n$ be labels, $y_0 = \sigma(x_0)$, $x_i = next(x_{i-1},\ell_i)$ for $i = 1,\dots,n$, and $y_i = terminal(y_{i-1},\alpha(y_{i-1},\ell_i))$ for $i = 1,\dots,n$. If $\sigma\Big(next\big(x,\ell\big)\Big) = terminal\Big(\sigma(x),\alpha\big(\sigma(x),\ell\big)\Big)$ for every state $x$ and label $\ell$, then $y_i = \sigma(x_i)$ for all $i = 1,\dots,n$.
\end{proposition}

\begin{IEEEproof}
By definition, $y_0 = \sigma(x_0)$.
Now assume that $y_i = \sigma(x_i)$.
Then by \cref{eq},
\begin{align*}
y_{i+1} &= terminal\Big(y_i,\alpha\big(y_i,\ell_{i+1}\big)\Big) \\
&= terminal\Big(\sigma(x_i),\alpha\big(\sigma(x_i),\ell_{i+1}\big)\Big) \\
&= \sigma\Big(next\big(x_i,\ell_{i+1}\big)\Big) = \sigma\big(x_{i+1}\big). & \IEEEQEDhere
\end{align*}
\end{IEEEproof}

However, an analogous proof with congruence instead of equality does not work because we cannot substitute $\sigma(x_i)$ for $y_i$ if they are not equal.
Thus
\[\sigma\Big(next\big(x,\ell\big)\Big) \cong terminal\Big(\sigma(x),\alpha\big(\sigma(x),\ell\big)\Big)\]
does not imply that $\alpha$ congruence-preserves $\sigma$.
This may occur, for instance, if one of the IBE/IBS labels does not work correctly when multiple versions of a file are present.

Instead we will show that
\begin{equation}
\label{cong}
\sigma'\Big(next\big(x,\ell\big)\Big) \cong terminal\Big(\sigma'(x),\alpha\big(\sigma'(x),\ell\big)\Big)
\end{equation}
for all states $x$, labels $\ell$, and state mappings $\sigma'$ congruent to $\sigma$.
This proves that $\alpha$ congruence-preserves $\sigma$ by the following inductive argument:

\begin{proposition}
Let $x_0$ be a state, $\ell_1,\dots,\ell_n$ be labels, $y_0 = \sigma(x_0)$, $x_i = next(x_{i-1},\ell_i)$ for $i = 1,\dots,n$, and $y_i = terminal(y_{i-1},\alpha(y_{i-1},\ell_i))$ for $i = 1,\dots,n$. If $\sigma\Big(next\big(x,\ell\big)\Big) \cong terminal\Big(\sigma(x),\alpha\big(\sigma(x),\ell\big)\Big)$ for every state $x$, label $\ell$, and state mapping $\sigma'$ congruent to $\sigma$, then $y_i \cong \sigma(x_i)$ for all $i = 1,\dots,n$.
\end{proposition}

\begin{IEEEproof}
By definition, $y_0 \cong \sigma(x_0)$.
Now assume that $y_i \cong \sigma(x_i)$.
Let $\sigma^*$ be the state mapping equivalent to $\sigma$ except that $\sigma^*(x_i) = y_i$.
Since $\sigma^*(x) = \sigma(x)$ for all $x \neq x_i$ and $\sigma^*(x_i) \cong \sigma(x_i)$, $\sigma^* \cong \sigma$.
Thus by \cref{cong},
\begin{align*}
y_{i+1} &= terminal\Big(y_i,\alpha\big(y_i,\ell_{i+1}\big)\Big) \\
&= terminal\Big(\sigma^*(x_i),\alpha\big(\sigma^*(x_i),\ell_{i+1}\big)\Big) \\
&\cong \sigma^*\Big(next\big(x_i,\ell_{i+1}\big)\Big) = \sigma^*\big(x_{i+1}\big) \cong \sigma\big(x_{i+1}\big). & \IEEEQEDhere
\end{align*}
\end{IEEEproof}

If we have an implementation $\langle \alpha,\sigma,\pi \rangle$ such that $\alpha$ congruence-preserves $\sigma$ and $\sigma$ preserves $\pi$, we say that the implementation is \emph{congruence-correct}.

\section{IBE/IBS Proof}
\label[appendix]{sec:ibeproof}
\allowdisplaybreaks

We first provide a formal definition of an access control system that uses IBE, IBS, and symmetric-key cryptography, and then show it implements $\rbac_0$, proving \cref{thm:ibe}.
\subsection{Our IBE/IBS System}

\subsubsection{Preliminaries}
\begin{itemize}
\item We use $m$ as the symmetric-key size, which is also the size of the IBE and IBS message spaces.
\item For signatures, we assume that hash-and-sign is used, where the message is hashed with a collision-resistant hash function and then signed using IBS.
\end{itemize}

\subsubsection{States}
\begin{itemize}
	\item USERS: a list of user names
	\item ROLES: a list of $(r,v_r)$ pairs containing role names and version numbers
	\item FILES: a list of $(fn,v_{fn})$ pairs containing file names and version numbers
	\item $FS$: the set of tuples (\rolekey, \filekey, or \file) stored on the filestore
\end{itemize}

\subsubsection{Request}
\begin{itemize}
	\item $u,p$ for whether user $u$ has permission $p$
\end{itemize}

\subsubsection{Queries}
\begin{itemize}
	\item $RK$ returns whether a user is in a role.
	Note that we do not verify the validity of the encrypted keys because the encryption is performed by the trusted admin, and the signature ensures integrity.
	\begin{multline*}
	RK(u,r) \triangleq \exists (c, sig).(\langle \rolekey, u, (r,v_r), c, sig\rangle \in FS \\
	\land sig = \IBSSign_{SU}(\langle \rolekey, u, (r,v_r), c\rangle))
	\end{multline*}
	Checking $RK$ requires one instance of \IBSVerify.
	\item $FK$ returns whether a role has a permission for the latest version of a file.
	As is the case $RK$, we do not need to verify the validity of the encrypted key.
	\begin{multline*}
	FK(r,\langle fn,op\rangle) \triangleq \; \exists (c, sig).( \\
	\langle \filekey, r, \langle fn,op\rangle, v_{fn}, c, SU, sig\rangle \in F \\
	\land sig = \IBSSign_{SU}(\langle \filekey, r, \langle fn,op\rangle, v_r, c, SU\rangle))
	\end{multline*}
	Checking $FK$ requires one instance of \IBSVerify.
	\item $Role(r) \triangleq \exists v.((r,v) \in ROLES)$
	\item $auth$ returns whether a user has a permission.
	\[auth(u,p) \triangleq \exists r.(RK(u,r) \land FK(r,p))\]
	Checking $auth$ requires two instances of \IBSVerify.
\end{itemize}

\subsubsection{Labels}
The labels used in this system are simply the operations in \cref{fig:rbacibe}.

\subsection{Implementing \texorpdfstring{$\rbac_0$}{RBAC0} using IBE/IBS}
\label{sec:ibeimp}

We use the definitions of congruence-preservation and congruence-correctness found in \cref{sec:versioning}.

\begin{theorem}
There exists an implementation $\langle \alpha,\sigma,\pi \rangle$ of $\rbac_0$ using IBE and IBS where:
\begin{itemize}
	\item $\alpha$ congruence-preserves $\sigma$ and preserves safety
	\item $\sigma$ preserves $\pi$
	\item $\pi$ is AC-preserving
\end{itemize}
Thus there exists a congruence-correct, AC-preserving, safe implementation of $\rbac_0$ using IBE and IBS.
\end{theorem}

\begin{IEEEproof}

The notation and conventions used here are listed in \cref{sec:const}.

\subsubsection{State mapping \texorpdfstring{$\sigma$}{sigma}}\mbox{}

{
\setlength\parindent{0pt}
For each $u \in U \cup \{SU\}$:
\begin{itemize}
	\item Add $u$ to USERS.
	\item Generate $k_u \leftarrow \IBEKeyGen(u)$ and $s_u \leftarrow \IBSKeyGen(u)$.
\end{itemize}

Let $FS = \{\}$. \\
Let ROLES and FILES be blank. \\
Run $\IBEMSKGen(m)$ to get IBE system parameters and master secret key $msk$. \\
Run $\IBSMSKGen(m)$ to get IBS system parameters and master secret key $msk'$.

For each $R(r) \in M$:
\begin{itemize}
	\item Add $(r,1)$ to ROLES.
	\item Let $FS = FS \cup \{\langle \rolekey$, $SU$, $(r,1)$, $\IBEEnc_{SU}\left(\IBEKeyGen_{msk}((r,1)),\IBSKeyGen_{msk'}((r,1))\right)$, $\IBSSign_{SU}\rangle\}$.
\end{itemize}

For each $P(fn) \in M$ where $fn$ is the name of file $f$:
\begin{itemize}
	\item Add $(fn,1)$ to FILES.
	\item Produce a symmetric key $k = \SymGen(m)$.
	\item Let $FS = FS \cup \{\langle \file, fn, 1, \SymEnc_k(f), SU, \IBSSign_{SU}\rangle\}$.
	\item Let $FS = FS \cup \{\langle \filekey$, $SU$, $\langle fn,\rwp\rangle$, $1$, $\IBEEnc_{SU}(k)$, $SU$, $\IBSSign_{SU}\rangle\}$.
\end{itemize}

For each $UR(u,r) \in M$:
\begin{itemize}
	\item Find $\langle \rolekey, SU, (r,1), c, sig\rangle \in FS$.
	\item Let $FS = FS \cup \{\langle \rolekey$, $SU$, $(r,1)$, $\IBEEnc_u\left(\IBEDec_{k_{SU}}(c)\right)$, $\IBSSign_{SU}\rangle\}$.
\end{itemize}

For each $PA(r,\langle fn,op\rangle)$:
\begin{itemize}
	\item Find $\langle \filekey, SU, \langle fn,\rwp\rangle, 1, c, SU, sig\rangle$.
	\item Let $FS = FS \cup \{\langle \filekey$, $(r,1)$, $\langle fn,op\rangle$, $1$, $\IBEEnc_{(r,1)}\left(\IBEDec_{k_{SU}}(c)\right)$, $SU$, $\IBSSign_{SU}\rangle\}$.
\end{itemize}

$output(FS, \text{ROLES}, \text{FILES})$
} 

\subsubsection{Query mapping \texorpdfstring{$\pi$}{pi}}

\begin{align*}
	\pi_{UR(u,r)}(T) &= RK(u,r) \in T \\
	\pi_{PA(r,p)}(T) &= FK(r,p) \in T \\
	\pi_{R(r)}(T) &= Role(r) \in T \\
	\pi_{auth(u,p)}(T) &= auth(u,p) \in T
\end{align*}

The query mapping $\pi$ is AC-preserving because it maps $auth(u,p)$ to \true{} for theory $T$ if and only if $T$ contains $auth(u,p)$.

\subsubsection{\texorpdfstring{$\sigma$}{sigma} preserves \texorpdfstring{$\pi$}{pi}}

This means that for every $RBAC_0$ state $x$, $Th(x) = \pi(Th(\sigma(x)))$.
To prove this, we show that for each $RBAC_0$ state $x$ and query $q$, $x \vdash q$ if and only if $\pi_q(Th(\sigma(x))) = \true$.

We consider each type of query separately.

\begin{itemize}

	\item \textbf{UR:}
	If $x \vdash UR(u,r)$ then $UR(u,r) \in Th(x)$, meaning that in $x$, $\langle u,r\rangle \in UR$.
	Thus in $\sigma(x)$, $v_r = 1$ and $\exists (c,sig).(\langle \rolekey, u, (r,1), c, sig\rangle \in FS \land sig = \IBSSign_{SU}(\langle \rolekey, u, (r,v_r), c\rangle))$.
	Hence $RK(u,r) \in Th(\sigma(x))$, so $\pi_{UR(u,r)}(Th(\sigma(x))) = \true.$

	If $x \nvdash UR(u,r)$ then $UR(u,r) \notin Th(x)$, meaning that in $x$, $\langle u,r\rangle \notin UR$.
	Thus in $\sigma(x)$, $v_r = 1$  and $\nexists (c,sig).(\langle \rolekey, u, (r,1), c, sig\rangle \in FS)$.
	Hence $RK(u,r) \notin Th(\sigma(x))$, so $\pi_{UR(u,r)}(Th(\sigma(x))) = \false.$

	\item \textbf{PA:}
	If $x \vdash PA(r,p)$ with $p = \langle fn,op\rangle$, then $PA(r,p) \in Th(x)$, meaning that in $x$, $\langle r,p\rangle \in PA$.
	Thus in $\sigma(x)$, $v_{fn} = 1$ and $\exists (c,sig).(\langle \filekey, r, \langle fn,op\rangle, v_{fn}, c, SU, sig\rangle \in FS \land sig = \IBSSign_{SU}(\langle \filekey, r, \langle fn,op\rangle, v_{fn}, c, SU\rangle))$.
	Hence $FK(r,p) \in Th(\sigma(x))$, so $\pi_{PA(r,p)}(Th(\sigma(x))) = \true.$

	If $x \nvdash PA(r,p)$ with $p = \langle fn,op\rangle$, then $PA(r,p) \notin Th(x)$, meaning that in $x$, $\langle r,p\rangle \notin PA$.
	Thus in $\sigma(x)$, $v_{fn} = 1$ and $\nexists (c,sig).(\langle \filekey, r, \langle fn,op\rangle, v_{fn}, c, SU, sig\rangle \in FS)$.
	Hence $FK(r,p) \notin Th(\sigma(x))$, so $\pi_{PA(r,p)}(Th(\sigma(x))) = \false.$

	\item \textbf{R:}
	If $x \vdash R(r)$ then $R(r) \in Th(x)$, meaning that in $x$, $r \in R$.
	Thus in $\sigma(x)$, $(r,1) \in \text{ROLES}$.
	Hence $Role(r) \in Th(\sigma(x))$, so $\pi_{R(r)}(Th(\sigma(x))) = \true.$

	If $x \nvdash R(r)$, then $R(r) \notin Th(x)$, meaning that in $x$, $r \notin R$.
	Thus in $\sigma(x)$, $\nexists v.((r,v) \in \text{ROLES})$.
	Hence $Role(r) \notin Th(\sigma(x))$, so $\pi_{R(r)}(Th(\sigma(x))) = \false.$

	\item \textbf{auth:}
	If $x \vdash auth(u,p)$ then $auth(u,p) \in Th(x)$, so there exists $r$ such that $UR(u,r) \in Th(x) \land PA(r,p) \in Th(x)$.
	Since $\sigma$ preserves $\pi$ for $UR$ and $PA$ queries, $RK(u,r) \in Th(\sigma(x)) \land FK(r,p) \in Th(\sigma(x))$.
	Hence $auth(u,p) \in Th(\sigma(x))$, so $\pi_{auth(u,p)}(Th(\sigma(x))) = \true.$

	If $x \nvdash auth(u,p)$ then $auth(u,p) \notin Th(x)$, so $\nexists r.(UR(u,r) \in Th(x) \land PA(r,p) \in Th(x))$.
	Since $\sigma$ preserves $\pi$ for $UR$ and $PA$ queries, $\nexists r.(RK(u,r) \in Th(\sigma(x)) \land FK(r,p) \in Th(\sigma(x)))$.
	Hence $auth(u,p) \notin Th(\sigma(x))$, so $\pi_{auth(u,p)}(Th(\sigma(x))) = \false.$

\end{itemize}

\subsubsection{Label mapping \texorpdfstring{$\alpha$}{alpha}} The label mapping $\alpha$ simply maps any $\rbac_0$ label, regardless of the state, to the IBE/IBS label of the same name found in \cref{fig:rbacibe}. The only difference is that in IBE/IBS, $addP$ takes as input a filename and file instead of a permission and $delP$ takes as input a filename instead of a permission.

\subsubsection{\texorpdfstring{$\alpha$}{alpha} congruence-preserves \texorpdfstring{$\sigma$}{sigma}}

We consider each type of $RBAC_0$ label separately.
We let $\sigma'$ be a state mapping congruent to $\sigma$ and let $x' = next(x,\ell)$ be the result of executing label $\ell$ in state $x$.
While key generation and encryption algorithms are normally randomized, for determining equality of states we assume that they are deterministic.

\begin{itemize}

	\item \textbf{addU:}
	If $\ell$ is an instance of $addU(u)$, then $x' = x \cup U(u)$.
	Thus
	\begin{align*}
	\sigma'(x') &= \sigma'\big(x \cup U(u)\big) = \sigma'(x) \cup \text{USERS}(u) \\
	&= next\big(\sigma'(x), addU(u)\big) \\
	&= terminal\big(\sigma'(x),\alpha(\sigma'(x),\ell)\big).
	\end{align*}

	\item \textbf{delU:}
	If $\ell$ is an instance of $delU(u)$, then $x' = x \setminus (U(u) \cup \{UR(u,r) \mid UR(u,r) \in x\})$.
	Let $T = \{(r,c,sig) \mid \langle \rolekey, u, (r,v_r), c, sig\rangle \in FS\}$ and $T' = \{r \mid \exists(c,sig).((r,c,sig) \in T)\}$.
	Let $\{r_1, r_2, \ldots, r_n\}$ be the elements of $T'$ in arbitrary order.
	Then
	\begin{align*}
		\sigma'(x') &= \sigma'\big(x \setminus (U(u) \cup \{UR(u,r) \mid UR(u,r) \in x\})\big) \\
		&= \sigma'(x) \setminus \text{USERS}(u) \\
		&\quad \setminus \big\{FS\big(\langle \rolekey, u, (r,v_r), c, sig\rangle\big) \mid (r,c,sig) \in T\big\} \\
		&\cong terminal\big(\sigma'(x) \setminus \text{USERS}(u), \\
		&\quad revokeU(u,r_1) \circ revokeU(u,r_2) \\
		&\quad \circ \cdots \circ revokeU(u,r_n)\big) \\
		&= next\big(\sigma'(x), delU(u)\big) \\
		&= terminal\big(\sigma'(x),\alpha(\sigma'(x),\ell)\big).
	\end{align*}

	\item \textbf{addR:}
	If $\ell$ is an instance of $addR(r)$, then $x' = x \cup R(r)$.
	Thus
	\begin{align*}
		\sigma'(x') &= \sigma'\big(x \cup R(r)\big) \\
		&= \sigma'(x) \cup \text{ROLES}(r,1) \cup FS\Big(\Big\langle \rolekey, SU, (r,1), \\
		&\quad \IBEEnc_{SU}\Big(\IBEKeyGen_{msk}((r,1)), \\
		&\quad \IBSKeyGen_{msk'}((r,1))\Big), \IBSSign_{SU}\Big\rangle\Big) \\
		&= next\big(\sigma'(x), addR(r)\big) \\
		&= terminal\big(\sigma'(x),\alpha(\sigma'(x),\ell)\big).
	\end{align*}

	\item \textbf{delR:}
	If $\ell$ is an instance of $delR(r)$, then $x' = x \setminus (R(r) \cup \{UR(u,r) \mid UR(u,r) \in x\} \cup \{PA(r,p) \mid PA(r,p) \in x\})$.
	Let $T = \{(u,c,sig) \mid \langle \rolekey, u, (r,v_r), c, sig\rangle \in FS\}$ and $F =  \{fn \mid \exists(op,v_{fn},c_{fn},sig).(\langle \filekey$, $(r,v_r)$, $\langle fn,op\rangle$, $v_{fn}$, $c_{fn}$, $SU$, $sig\rangle \in FS)\}$.
	For each $fn \in F$, let $T_{fn} = \{(op',v,c_v,sig) \mid \langle \filekey$, $(r,v_r)$, $\langle fn,op'\rangle$, $v$, $c_v$, $SU$, $sig\rangle \in FS\}$.
	Let $\{fn_1, fn_2, \ldots, fn_n\}$ be the elements of $F$ in arbitrary order.
	Then
	\begin{align*}
		\sigma'&(x') = \sigma'\big(x \setminus (R(r) \cup \{UR(u,r) \mid UR(u,r) \in x\} \\
		&\quad \cup \{PA(r,p) \mid PA(r,p) \in x\}\big) \\
		&= \sigma'(x) \setminus ROLES(r,v_r) \setminus \big\{FS\big(\langle \rolekey, u, (r,v_r), \\
		&\qquad c, sig\rangle\big) \mid (u,c,sig) \in T\big\} \setminus \big\{FS\big(\langle \filekey, (r,v_r), \\
		&\qquad \langle fn,op'\rangle, v, c_v, SU, sig\rangle\big) \mid \big(fn \in F \\
		&\qquad \land (op',v,c_v,sig) \in T_{fn}\big)\big\} \\
		&\cong terminal\Big(\sigma'(x) \setminus ROLES(r,v_r) \\
		&\quad \setminus \big\{FS\big(\langle \rolekey, u, (r,v_r), c, sig\rangle\big) \mid (u,c,sig) \in T\big\}, \\
		&\qquad revokeP\big(r, \langle fn_1, \rwp\rangle\big) \circ revokeP\big(r, \langle fn_2, \rwp\rangle\big) \\
		&\qquad \circ \cdots \circ revokeP\big(r, \langle fn_n, \rwp\rangle\big)\Big) \\
		&= next\big(\sigma'(x), delR(r)\big) \\
		&= terminal\big(\sigma'(x),\alpha(\sigma'(x),\ell)\big).
	\end{align*}

	\item \textbf{addP:}
	If $\ell$ is an instance of $addP(p)$ with $p = \langle fn,op\rangle$ and $fn$ the name of file $f$, then $x' = x \cup P(p)$.
	Thus for $k \leftarrow \SymGen(m)$,
	\begin{align*}
		\sigma'&(x') = \sigma'\big(x \cup P(p)\big) \\
		&= \sigma'(x) \cup \text{FILES}(fn,1) \\
		&\quad \cup FS\Big(\Big\langle \file, fn, 1, \SymEnc_k(f)\Big\rangle\Big) \cup FS\Big(\Big\langle \filekey, SU, \\
		&\qquad \langle fn,\rwp\rangle, 1, \IBEEnc_{SU}(k), SU, \IBSSign_{SU}\Big\rangle\Big) \\
		&= next\big(\sigma'(x), addP(fn,f)\big) \\
		&= terminal\big(\sigma'(x),\alpha(\sigma'(x),\ell)\big).
	\end{align*}

	\item \textbf{delP:}
	If $\ell$ is an instance of $delP(p)$ with $p = \langle fn,op\rangle$, then $x' = x \setminus (P(p) \cup \{PA(r,p) \mid PA(r,p) \in x\})$.
	Let $T = \{(v,c) \mid \langle \file$, $fn$, $v$, $c\rangle \in FS\}$ and $T' = \{(r,op',v,c',id,sig) \mid \langle \filekey$, $r$, $\langle fn,op'\rangle$, $v$, $c'$, $id$, $sig\rangle \in FS\}.$
	Then
	\begin{align*}
		\sigma'(x') &= \sigma'\big(x \setminus (P(p) \cup \{PA(r,p) \mid PA(r,p) \in x\})\big) \\
		&= \sigma'(x) \setminus \text{FILES}(fn,v_{fn}) \\
		&\quad \setminus \big\{FS\big(\langle \file, fn, v, c\rangle\big) \mid (v,c) \in T\big\} \\
		&\quad \setminus \big\{FS\big(\langle \filekey, r, \langle fn, op'\rangle, v, c', id, sig\rangle\big) \mid (r,op', \\
		&\qquad v,c',id,sig) \in T\big\} \\
		&= next\big(\sigma'(x), delP(fn)\big) \\
		&= terminal\big(\sigma'(x),\alpha(\sigma'(x),\ell)\big).
	\end{align*}

	\item \textbf{assignU:}
	If $\ell$ is an instance of $assignU(u,r)$, then $x' = x \cup UR(u,r)$.
	Thus for $\langle \rolekey, SU, (r,1), c, sig\rangle \in FS$ in $\sigma'(x)$,
	\begin{align*}
		\sigma'(x') &= \sigma'\big(x \cup UR(u,r)\big) \\
		&= \sigma'(x) \cup FS\Big(\Big\langle \rolekey, SU, (r,1), \\
		&\quad\: \IBEEnc_u\left(\IBEDec_{k_{SU}}(c)\right), \IBSSign_{SU}\Big\rangle\Big) \\
		&= next\big(\sigma'(x), assignU(u,r)\big) \\
		&= terminal\big(\sigma'(x),\alpha(\sigma'(x),\ell)\big).
	\end{align*}

	\item \textbf{revokeU:}
	If $\ell$ is an instance of $revokeUser(u,r)$, then $x' = x \setminus UR(u,r)$.
	Let $k_{(r,v_r+1)} \leftarrow \IBEKeyGen((r$,$v_r+1))$ and $s_{(r,v_r+1)} \leftarrow \IBSKeyGen((r$,$v_r+1))$.
	Let $T = \{(u',c_{u'},sig) \mid \langle \rolekey$, $u'$, $(r,v_r)$, $c_{u'}$, $sig\rangle \in FS\}$ and $F =  \{fn \mid \exists(op,v_{fn},c_{fn},sig).(\langle \filekey$, $(r,v_r)$, $\langle fn,op\rangle$, $v_{fn}$, $c_{fn}$, $SU$, $sig\rangle \in FS)$.
	For each $fn \in F$, let $k_{fn} \leftarrow \SymGen$, $T_{fn} = \{(op',v,c_v,sig) \mid \langle \filekey$, $(r,v_r)$, $\langle fn,op'\rangle$, $v$, $c_v$, $SU$, $sig\rangle \in FS\}$ and $T'_{fn} = \{id,op',c_{id},sig) \mid \langle \filekey$, $id$, $\langle fn,op'\rangle$, $v_{fn}$, $c_{id}$, $SU$, $sig\rangle \in FS\}$.
	Then
	\begin{align*}
		\sigma'&(x') = \sigma'\big(x \setminus UR(u,r)\big) \\
		&= \sigma'(x) \setminus \big\{FS\big(\langle\rolekey, u, (r,v_r), c_u, \\
		&\qquad sig\rangle\big) \mid (u,c_u,sig) \in T\big\} \\
		&\cong \sigma'(x) \setminus \big\{FS\big(\langle\rolekey, u', (r,v_r), c_{u'}, \\
		&\qquad sig\rangle\big) \mid (u',c_{u'},sig) \in T\big\} \\
		&\quad \cup \Big\{FS\Big(\Big\langle\rolekey, u', (r,v_r), \IBEEnc_{u'}\left(k_{(r,v_r+1)}, s_{(r,v_r+1)}\right), \\
		&\qquad \IBSSign_{SU}\Big\rangle\Big) \mid (u',c_{u'},sig) \in T \land u' \neq u\Big\} \\
		&\quad \setminus \big\{FS\big(\langle \filekey, (r,v_r), \langle fn,op'\rangle, v, c_v, \\
		&\qquad SU, sig\rangle\big) \mid fn \in F \land (op',v,c_v,sig) \in T_{fn}\big\} \\
		&\quad \cup \Big\{FS\Big(\Big\langle \filekey, (r,v_r+1), \langle fn,op'\rangle, v, \\
		&\qquad \IBEEnc_{(r,v_r+1)}\left(\IBEDec_{k_{(r,v_r)}}(c_v)\right), SU, \\
		&\qquad \IBSSign_{SU}\Big\rangle\Big) \mid fn \in F \land (op',v,c_v,sig) \in T_{fn}\Big\} \\
		&\quad \cup \Big\{FS\Big(\Big\langle \filekey, id, \langle fn,op'\rangle, v_{fn}+1, \IBEEnc_{id}(k'_p), \\
		&\qquad SU, \IBSSign_{SU}\Big\rangle\Big) \mid fn \in F \land (id,c_{id},sig) \in T'_{fn}\Big\} \\
		&\quad \cup \big\{\text{FILES}(fn,v_{fn}+1) \mid fn \in F\big\} \\
		&\quad \setminus \big\{\text{FILES}(fn,v_{fn}) \mid fn \in F\big\} \\
		&\quad \cup \text{ROLES}(r,v_r+1) \setminus \text{ROLES}(r,v_r) \\
		&= next\big(\sigma'(x), revokeU(u,r)\big) \\
		&= terminal\big(\sigma'(x),\alpha(\sigma'(x),\ell)\big).
	\end{align*}

	\item \textbf{assignP:}
	If $\ell$ is an instance of $assignP(r,p)$ with $p = \langle fn,op\rangle$, then $x' = x \cup PA(r,p)$.
	We have two cases where $assignP(r,p)$ has an effect on $x$:
	\begin{itemize}
		\item If $op = \rwp$ and there exists $\langle \filekey$, $(r,v_r)$, $\langle fn, \readp\rangle$, $v_{fn}$, $c$, $SU$, $sig\rangle$, then let $T = \{(v,c_v,sig) \mid \langle \filekey, (r,v_r),\langle fn, \readp\rangle, v, c_v, SU, sig\rangle \in FS\}$. Then
		\begin{align*}
			\sigma'(x') &= \sigma'\big(x \cup PA(r,p)\big) \\
			&= \sigma'(x) \setminus \big\{FS\big(\langle \filekey, (r,v_r), \langle fn, \readp\rangle, v, c_v, \\
			&\qquad SU, sig\rangle\big) \mid (v,c_v,sig) \in T\big\} \\
			&\quad \cup \big\{FS\Big(\Big\langle \filekey, (r,v_r), \langle fn, \rwp\rangle, v, c_v, SU, \\
			&\qquad \IBSSign_{SU}\Big\rangle\Big) \mid (v,c_v,sig) \in T\big\} \\
			&= next\big(\sigma'(x), assignP(r,p)\big) \\
			&= terminal\big(\sigma'(x),\alpha(\sigma'(x),\ell)\big).
		\end{align*}
		\item If there does not exist $\langle \filekey$, $(r,v_r)$, $\langle fn, op'\rangle$, $v_{fn}$, $c$, $SU$, $sig\rangle$, then let $T = \{(v,c_v) \mid \exists(id,sig).(\langle \filekey, SU,\langle fn, \rwp\rangle, v, c_v, id, sig\rangle \in FS)\}$. Then
		\begin{align*}
			\sigma'(x') &= \sigma'\big(x \cup PA(r,p)\big) \\
			&= \sigma'(x) \cup \Big\{FS\Big(\Big\langle \filekey, (r,v_r), \langle fn, op\rangle, v, \\
			&\quad \IBEEnc_{(r,v_r)}\left(\IBEDec_{k_{SU}}(c_v)\right), SU, \\
			&\quad \IBSSign_{SU}\Big\rangle\Big) \mid (v,c_v) \in T\Big\} \\
			&= next\big(\sigma'(x), assignP(r,p)\big) \\
			&= terminal\big(\sigma'(x),\alpha(\sigma'(x),\ell)\big).
		\end{align*}
	\end{itemize}

	\item \textbf{revokeP:}
	If $\ell$ is an instance of $revokeP(r,p)$ with $p = \langle fn,op\rangle$, then $x' = x \setminus PA(r,p)$.
	\begin{itemize}
		\item If $op = \writep$, then let $T = \{(v,c_v,sig) \mid \langle \filekey, (r,v_r), \langle fn, \rwp\rangle, v, c_v, SU, sig\rangle \in FS\}$. Then
		\begin{align*}
			\sigma'(x') &= \sigma'\big(x \setminus PA(r,p)\big) \\
			&= \sigma'(x) \setminus \big\{FS\big(\langle \filekey, (r,v_r), \langle fn, \rwp\rangle, v, c_v, \\
			&\qquad SU, sig\rangle\big) \mid (v,c_v,sig) \in T\big\} \\
			& \quad \cup \Big\{FS\Big(\Big\langle \filekey, (r,v_r), \langle fn, \readp\rangle, v, c_v, SU, \\
			&\qquad \IBSSign_{SU}\Big\rangle\Big) \mid (v,c_v,sig) \in T\Big\} \\
			&= next\big(\sigma'(x), assignP(r,p)\big) \\
			&= terminal\big(\sigma'(x),\alpha(\sigma'(x),\ell)\big).
		\end{align*}
		\item If $op = \readp$, then let $k' \leftarrow \SymGen$, $T = \{(op',v,c_v,sig) \mid \langle \filekey$, $(r,v_r)$, $\langle fn, op'\rangle$, $v$, $c_v$, $SU$, $sig\rangle \in FS\}$, and $T' = \{(id,op') \mid id \neq r \land \exists (c_{id},sig).(\langle \filekey$, $id$, $\langle fn, op'\rangle$, $v_{fn}$, $c_{id}$, $SU$, $sig\rangle \in FS)\}$. Then
		\begin{align*}
			\sigma'&(x') = \sigma'\big(x \setminus PA(r,p)\big) \\
			&= \sigma'(x) \setminus \big\{FS\big(\langle \filekey, (r,v_r), \langle fn, op'\rangle, v, c_v, \\
			&\quad SU, sig\rangle) \mid (op',v,c_v,sig) \in T\big\} \\
			&\cong \sigma'(x) \setminus \big\{FS\big(\langle \filekey, (r,v_r), \langle fn, op'\rangle, v, c_v, \\
			&\qquad SU, sig\rangle\big) \mid (op',v,c_v,sig) \in T\big\} \\
			&\quad \cup \Big\{FS\Big(\Big\langle \filekey, id, \langle fn, op'\rangle, v_{fn}+1, \\
			&\qquad \IBEEnc_{id}(k'), SU, \IBSSign_{SU}\Big\rangle\Big) \mid (id,op') \in T\Big\} \\
			&\quad \cup \text{FILES}(fn,v_{fn}+1) \setminus \text{FILES}(fn,v_{fn}) \\
			&= next\big(\sigma'(x), assignP(r,p)\big) \\
			&= terminal\big(\sigma'(x),\alpha(\sigma'(x),\ell)\big).
		\end{align*}
	\end{itemize}

\end{itemize}

\subsubsection{Safety}

The label mapping $\alpha$ is safe by inspection---for any $\rbac_0$ state $x$ and label $\ell$, the IBE/IBS label $\alpha(\sigma(x),\ell)$ never revokes or grants authorizations except the images of those that are revoked or granted by $\ell$.\end{IEEEproof}

%


\section{PKI Proof}
\label[appendix]{sec:pkiproof}
\allowdisplaybreaks

We first provide a formal definition of an access control system that uses PKI and symmetric-key cryptography, and then show it implements $\rbac_0$, proving \cref{thm:pki}.

\subsection{Our PKI System}

\subsubsection{Preliminaries}
\begin{itemize}
\item We use $m$ as the symmetric-key size.
\item For signatures, we assume that hash-and-sign is used, where the message is hashed with a collision-resistant hash function and then digitally signed.
\end{itemize}

\subsubsection{States}
\begin{itemize}
	\item USERS: a list of $(u,\enckey_u,\verkey_u)$ tuples containing user names and their corresponding public keys
	\item ROLES: a list of $(r,v_r,\enckey_{(r,v_r)},\verkey_{(r,v_r)})$ tuples containing role names, version numbers, and their public keys
	\item FILES: a list of $(fn,v_{fn})$ pairs containing file names and version numbers
	\item $FS$: the set of tuples (\rolekey, \filekey, or \file) stored on the filestore
\end{itemize}

\subsubsection{Request}
\begin{itemize}
	\item $u,p$ for whether user $u$ has permission $p$
\end{itemize}

\subsubsection{Queries}
\begin{itemize}
	\item $RK$ returns whether a user is in a role.
	Note that we do not verify the validity of the encrypted keys because the encryption is performed by the trusted admin, and the signature ensures integrity.
	\begin{multline*}
	RK(u,r) \triangleq \exists (c, sig).(\langle \rolekey, u, (r,v_r), c, sig\rangle \in FS \\
	\land sig = \SigSign_{\sigkey_{SU}}(\langle \rolekey, u, (r,v_r), c\rangle))
	\end{multline*}
	Checking $RK$ requires one instance of \SigVerify.
	\item $FK$ returns whether a role has a permission for the latest version of a file.
	As is the case $RK$, we do not need to verify the validity of the encrypted key.
	\begin{multline*}
	FK(r,\langle fn,op\rangle) \triangleq \; \exists (c, sig).( \\
	\langle \filekey, r, \langle fn,op\rangle, v_{fn}, c, SU, sig\rangle \in F \\
	\land sig = \SigSign_{\sigkey_{SU}}(\langle \filekey, r, \langle fn,op\rangle, v_r, c, SU\rangle))
	\end{multline*}
	Checking $FK$ requires one instance of \SigVerify.
	\item $Role(r) \triangleq \exists (v,k_1,k_2).((r,v,k_1,k_2) \in ROLES)$
	\item $auth$ returns whether a user has a permission.
	\[auth(u,p) \triangleq \exists r.(RK(u,r) \land FK(r,p))\]
	Checking $auth$ requires two instances of \SigVerify.
\end{itemize}

\subsubsection{Labels}
The labels used in this system are simply the operations in \cref{fig:rbacpki}.

\subsection{Implementing \texorpdfstring{$\rbac_0$}{RBAC0} using PKI}

We use the definitions of congruence-preservation and congruence-correctness found in \cref{sec:versioning}.

\begin{theorem}
There exists an implementation $\langle \alpha,\sigma,\pi \rangle$ of $\rbac_0$ using PKI where:
\begin{itemize}
	\item $\alpha$ congruence-preserves $\sigma$ and preserves safety
	\item $\sigma$ preserves $\pi$
	\item $\pi$ is AC-preserving
\end{itemize}
Thus there exists a congruence-correct, AC-preserving, safe implementation of $\rbac_0$ using PKI.
\end{theorem}

\begin{IEEEproof}

The notation and conventions used here are listed in \cref{sec:implementation-pki}.

\subsubsection{State mapping \texorpdfstring{$\sigma$}{sigma}}\mbox{}

{
\setlength\parindent{0pt}
For each $u \in U \cup \{SU\}$:
\begin{itemize}
	\item Generate $\left(\enckey_u, \deckey_u\right) \leftarrow \PKGen$ and $\left(\verkey_u, \sigkey_u\right) \leftarrow \SigGen$.
	\item Add $\left(u,\enckey_u,\verkey_u\right)$ to USERS.
\end{itemize}

Let $FS = \{\}$. \\
Let ROLES and FILES be blank.

For each $R(r) \in M$:
\begin{itemize}
	\item Generate encryption key pair $\left(\enckey_{(r,1)}, \deckey_{(r,1)}\right) \leftarrow \PKGen$ and signature key pair $\left(\verkey_{(r,1)}, \sigkey_{(r,1)}\right) \leftarrow \SigGen$.
	\item Add $\left(r,1,\enckey_{(r,1)},\verkey_{(r,1)}\right)$ to ROLES.
	\item Let $FS = FS \cup \{\langle \rolekey$, $SU$, $(r,1)$, $\PKEnc_{\enckey_{SU}}\left(\deckey_{(r,1)}, \sigkey_{(r,1)}\right)$, $\SigSign_{SU}\rangle\}$.
\end{itemize}

For each $P(fn) \in M$ where $fn$ is the name of file $f$:
\begin{itemize}
	\item Add $(fn,1)$ to FILES.
	\item Produce a symmetric key $k = \SymGen(m)$.
	\item Let $FS = FS \cup \{\langle \file, fn, 1, \SymEnc_k(f), SU, \SigSign_{SU}\rangle\}$.
	\item Let $FS = FS \cup \{\langle \filekey$, $SU$, $\langle fn,\rwp\rangle$, $1$, $\PKEnc_{\enckey_{SU}}(k)$, $SU$, $\SigSign_{SU}\rangle\}$.
\end{itemize}

For each $UR(u,r) \in M$:
\begin{itemize}
	\item Find $\langle \rolekey, SU, (r,1), c, sig\rangle \in FS$.
	\item Let $FS = FS \cup \{\langle \rolekey$, $SU$, $(r,1)$, $\PKEnc_{\enckey_u}\left(\PKDec_{\deckey_{SU}}(c)\right)$, $\SigSign_{SU}\rangle\}$.
\end{itemize}

For each $PA(r,\langle fn,op\rangle)$:
\begin{itemize}
	\item Find $\langle \filekey, SU, \langle fn,\rwp\rangle, 1, c, SU, sig\rangle$.
	\item Let $FS = FS \cup \{\langle \filekey$, $(r,1)$, $\langle fn,op\rangle$, $1$, $\PKEnc_{\enckey_{(r,1)}}\left(\PKDec_{\deckey_{SU}}(c)\right)$, $SU$, $\SigSign_{SU}\rangle\}$.
\end{itemize}

$output(FS, \text{ROLES}, \text{FILES})$
} 

\subsubsection{Query mapping \texorpdfstring{$\pi$}{pi}}

\begin{align*}
	\pi_{UR(u,r)}(T) &= RK(u,r) \in T \\
	\pi_{PA(r,p)}(T) &= FK(r,p) \in T \\
	\pi_{R(r)}(T) &= Role(r) \in T \\
	\pi_{auth(u,p)}(T) &= auth(u,p) \in T
\end{align*}

The query mapping $\pi$ is AC-preserving because it maps $auth(u,p)$ to \true{} for theory $T$ if and only if $T$ contains $auth(u,p)$.

\subsubsection{\texorpdfstring{$\sigma$}{sigma} preserves \texorpdfstring{$\pi$}{pi}}

This means that for every $RBAC_0$ state $x$, $Th(x) = \pi(Th(\sigma(x)))$.
To prove this, we show that for each $RBAC_0$ state $x$ and query $q$, $x \vdash q$ if and only if $\pi_q(Th(\sigma(x))) = \true$.

We consider each type of query separately.

\begin{itemize}

	\item \textbf{UR:}
	If $x \vdash UR(u,r)$ then $UR(u,r) \in Th(x)$, meaning that in $x$, $\langle u,r\rangle \in UR$.
	Thus in $\sigma(x)$, $v_r = 1$ and $\exists (c,sig).(\langle \rolekey, u, (r,1), c, sig\rangle \in FS \land sig = \SigSign_{\sigkey_{SU}}(\langle \rolekey, u, (r,v_r), c\rangle))$.
	Hence $RK(u,r) \in Th(\sigma(x))$, so $\pi_{UR(u,r)}(Th(\sigma(x))) = \true.$

	If $x \nvdash UR(u,r)$ then $UR(u,r) \notin Th(x)$, meaning that in $x$, $\langle u,r\rangle \notin UR$.
	Thus in $\sigma(x)$, $v_r = 1$  and $\nexists (c,sig).(\langle \rolekey, u, (r,1), c, sig\rangle \in FS)$.
	Hence $RK(u,r) \notin Th(\sigma(x))$, so $\pi_{UR(u,r)}(Th(\sigma(x))) = \false.$

	\item \textbf{PA:}
	If $x \vdash PA(r,p)$ with $p = \langle fn,op\rangle$, then $PA(r,p) \in Th(x)$, meaning that in $x$, $\langle r,p\rangle \in PA$.
	Thus in $\sigma(x)$, $v_{fn} = 1$ and $\exists (c,sig).(\langle \filekey, r, \langle fn,op\rangle, v_{fn}, c, SU, sig\rangle \in FS \land sig = \SigSign_{\sigkey_{SU}}(\langle \filekey, r, \langle fn,op\rangle, v_{fn}, c, SU\rangle))$.
	Hence $FK(r,p) \in Th(\sigma(x))$, so $\pi_{PA(r,p)}(Th(\sigma(x))) = \true.$

	If $x \nvdash PA(r,p)$ with $p = \langle fn,op\rangle$, then $PA(r,p) \notin Th(x)$, meaning that in $x$, $\langle r,p\rangle \notin PA$.
	Thus in $\sigma(x)$, $v_{fn} = 1$ and $\nexists (c,sig).(\langle \filekey, r, \langle fn,op\rangle, v_{fn}, c, SU, sig\rangle \in FS)$.
	Hence $FK(r,p) \notin Th(\sigma(x))$, so $\pi_{PA(r,p)}(Th(\sigma(x))) = \false.$

	\item \textbf{R:}
	If $x \vdash R(r)$ then $R(r) \in Th(x)$, meaning that in $x$, $r \in R$.
	Thus in $\sigma(x)$, $\exists (k_1,k_2).(r,1,k_1,k_2) \in \text{ROLES}$.
	Hence $Role(r) \in Th(\sigma(x))$, so $\pi_{R(r)}(Th(\sigma(x))) = \true.$

	If $x \nvdash R(r)$, then $R(r) \notin Th(x)$, meaning that in $x$, $r \notin R$.
	Thus in $\sigma(x)$, $\nexists (v,k_1,k_2).((r,v,k_1,k_2) \in \text{ROLES})$.
	Hence $Role(r) \notin Th(\sigma(x))$, so $\pi_{R(r)}(Th(\sigma(x))) = \false.$

	\item \textbf{auth:}
	If $x \vdash auth(u,p)$ then $auth(u,p) \in Th(x)$, so there exists $r$ such that $UR(u,r) \in Th(x) \land PA(r,p) \in Th(x)$.
	Since $\sigma$ preserves $\pi$ for $UR$ and $PA$ queries, $RK(u,r) \in Th(\sigma(x)) \land FK(r,p) \in Th(\sigma(x))$.
	Hence $auth(u,p) \in Th(\sigma(x))$, so $\pi_{auth(u,p)}(Th(\sigma(x))) = \true.$

	If $x \nvdash auth(u,p)$ then $auth(u,p) \notin Th(x)$, so $\nexists r.(UR(u,r) \in Th(x) \land PA(r,p) \in Th(x))$.
	Since $\sigma$ preserves $\pi$ for $UR$ and $PA$ queries, $\nexists r.(RK(u,r) \in Th(\sigma(x)) \land FK(r,p) \in Th(\sigma(x)))$.
	Hence $auth(u,p) \notin Th(\sigma(x))$, so $\pi_{auth(u,p)}(Th(\sigma(x))) = \false.$

\end{itemize}

\subsubsection{Label mapping \texorpdfstring{$\alpha$}{alpha}} The label mapping $\alpha$ simply maps any $\rbac_0$ label, regardless of the state, to the PKI label of the same name found in \cref{fig:rbacpki}. The only difference is that in PKI, $addP$ takes as input a filename and file instead of a permission and $delP$ takes as input a filename instead of a permission.

\subsubsection{\texorpdfstring{$\alpha$}{alpha} congruence-preserves \texorpdfstring{$\sigma$}{sigma}}

We consider each type of $RBAC_0$ label separately.
We let $\sigma'$ be a state mapping congruent to $\sigma$ and let $x' = next(x,\ell)$ be the result of executing label $\ell$ in state $x$.
While key generation and encryption algorithms are normally randomized, for determining equality of states we assume that they are deterministic.

\begin{itemize}

	\item \textbf{addU:}
	If $\ell$ is an instance of $addU(u)$, then $x' = x \cup U(u)$.
	Thus there exists $(\enckey_u, \deckey_u) \leftarrow \PKGen$ and $(\verkey_u, \sigkey_u) \leftarrow \SigGen$ such that
	\begin{align*}
	\sigma'(x') &= \sigma'\big(x \cup U(u)\big) = \sigma'(x) \cup \text{USERS}(u,\enckey_u,\verkey_u) \\
	&= next\big(\sigma'(x), addU(u)\big) \\
	&= terminal\big(\sigma'(x),\alpha(\sigma'(x),\ell)\big).
	\end{align*}

	\item \textbf{delU:}
	If $\ell$ is an instance of $delU(u)$, then $x' = x \setminus (U(u) \cup \{UR(u,r) \mid UR(u,r) \in x\})$.
	Let $T = \{(r,c,sig) \mid \langle \rolekey, u, (r,v_r), c, sig\rangle \in FS\}$ and $T' = \{r \mid \exists(c,sig).((r,c,sig) \in T)\}$.
	Let $\{r_1, r_2, \ldots, r_n\}$ be the elements of $T'$ in arbitrary order.
	Then
	\begin{align*}
		\sigma'(x') &= \sigma'\big(x \setminus (U(u) \cup \{UR(u,r) \mid UR(u,r) \in x\})\big) \\
		&= \sigma'(x) \setminus \text{USERS}\big(u,\enckey_u,\verkey_u\big) \\
		&\quad \setminus \big\{FS\big(\langle \rolekey, u, (r,v_r), c, sig\rangle\big) \mid (r,c,sig) \in T\big\} \\
		&\cong terminal\big(\sigma'(x) \setminus \text{USERS}\big(u,\enckey_u,\verkey_u\big), \\
		&\quad revokeU(u,r_1) \circ revokeU(u,r_2) \\
		&\quad \circ \cdots \circ revokeU(u,r_n)\big) \\
		&= next\big(\sigma'(x), delU(u)\big) \\
		&= terminal\big(\sigma'(x),\alpha(\sigma'(x),\ell)\big).
	\end{align*}

	\item \textbf{addR:}
	If $\ell$ is an instance of $addR(r)$, then $x' = x \cup R(r)$.
	Thus there exists $(\enckey_{(r,1)}, \deckey_{(r,1)}) \leftarrow \PKGen$ and $(\verkey_{(r,1)}, \sigkey_{(r,1)}) \leftarrow \SigGen$ such that
	\begin{align*}
		\sigma'(x') &= \sigma'\big(x \cup R(r)\big) \\
		&= \sigma'(x) \cup \text{ROLES}\Big(r,1,\enckey_{(r,1)},\verkey_{(r,1)}\Big) \\
		&\quad \cup FS\Big(\Big\langle \rolekey, SU, (r,1), \\
		&\quad \PKEnc_{\sigkey_{SU}}\Big(\deckey_{(r,1)},\sigkey_{(r,1)}\Big),\SigSign_{SU}\Big\rangle\Big) \\
		&= next\big(\sigma'(x), addR(r)\big) \\
		&= terminal\big(\sigma'(x),\alpha(\sigma'(x),\ell)\big).
	\end{align*}

	\item \textbf{delR:}
	If $\ell$ is an instance of $delR(r)$, then $x' = x \setminus (R(r) \cup \{UR(u,r) \mid UR(u,r) \in x\} \cup \{PA(r,p) \mid PA(r,p) \in x\})$.
	Let $T = \{(u,c,sig) \mid \langle \rolekey, u, (r,v_r), c, sig\rangle \in FS\}$ and $F =  \{fn \mid \exists(op,v_{fn},c_{fn},sig).(\langle \filekey$, $(r,v_r)$, $\langle fn,op\rangle$, $v_{fn}$, $c_{fn}$, $SU$, $sig\rangle \in FS)\}$.
	For each $fn \in F$, let $T_{fn} = \{(op',v,c_v,sig) \mid \langle \filekey$, $(r,v_r)$, $\langle fn,op'\rangle$, $v$, $c_v$, $SU$, $sig\rangle \in FS\}$.
	Let $\{fn_1, fn_2, \ldots, fn_n\}$ be the elements of $F$ in arbitrary order.
	Then
	\begin{align*}
		\sigma'&(x') = \sigma'\big(x \setminus (R(r) \cup \{UR(u,r) \mid UR(u,r) \in x\} \\
		&\quad \cup \{PA(r,p) \mid PA(r,p) \in x\}\big) \\
		&= \sigma'(x) \setminus ROLES\Big(r,v_r,\enckey_{(r,v_r)},\verkey_{(r,v_r)}\Big) \\
		&\qquad \setminus \big\{FS\big(\langle \rolekey, u, (r,v_r), c, sig\rangle\big) \mid (u,c,sig) \in T\big\} \\
		&\qquad \setminus \big\{FS\big(\langle \filekey, (r,v_r), \langle fn,op'\rangle, v, c_v, \\
		&\qquad SU, sig\rangle\big) \mid \big(fn \in F \land (op',v,c_v,sig) \in T_{fn}\big)\big\} \\
		&\cong terminal\Big(\sigma'(x) \setminus ROLES\Big(r,v_r,\enckey_{(r,v_r)},\verkey_{(r,v_r)}\Big) \\
		&\quad \setminus \big\{FS\big(\langle \rolekey, u, (r,v_r), c, sig\rangle\big) \mid (u,c,sig) \in T\big\}, \\
		&\qquad revokeP\big(r, \langle fn_1, \rwp\rangle\big) \circ revokeP\big(r, \langle fn_2, \rwp\rangle\big) \\
		&\qquad \circ \cdots \circ revokeP\big(r, \langle fn_n, \rwp\rangle\big)\Big) \\
		&= next\big(\sigma'(x), delR(r)\big) \\
		&= terminal\big(\sigma'(x),\alpha(\sigma'(x),\ell)\big).
	\end{align*}

	\item \textbf{addP:}
	If $\ell$ is an instance of $addP(p)$ with $p = \langle fn,op\rangle$ and $fn$ the name of file $f$, then $x' = x \cup P(p)$.
	Thus for $k \leftarrow \SymGen(m)$,
	\begin{align*}
		\sigma'&(x') = \sigma'\big(x \cup P(p)\big) \\
		&= \sigma'(x) \cup \text{FILES}(fn,1) \\
		&\quad \cup FS\Big(\Big\langle \file, fn, 1, \SymEnc_k(f)\Big\rangle\Big) \cup FS\Big(\Big\langle \filekey, SU, \\
		&\qquad \langle fn,\rwp\rangle, 1, \PKEnc_{\enckey_{SU}}(k), SU, \SigSign_{SU}\Big\rangle\Big) \\
		&= next\big(\sigma'(x), addP(fn,f)\big) \\
		&= terminal\big(\sigma'(x),\alpha(\sigma'(x),\ell)\big).
	\end{align*}

	\item \textbf{delP:}
	If $\ell$ is an instance of $delP(p)$ with $p = \langle fn,op\rangle$, then $x' = x \setminus (P(p) \cup \{PA(r,p) \mid PA(r,p) \in x\})$.
	Let $T = \{(v,c) \mid \langle \file$, $fn$, $v$, $c\rangle \in FS\}$ and $T' = \{(r,op',v,c',id,sig) \mid \langle \filekey$, $r$, $\langle fn,op'\rangle$, $v$, $c'$, $id$, $sig\rangle \in FS\}.$
	Then
	\begin{align*}
		\sigma'(x') &= \sigma'\big(x \setminus (P(p) \cup \{PA(r,p) \mid PA(r,p) \in x\})\big) \\
		&= \sigma'(x) \setminus \text{FILES}(fn,v_{fn}) \\
		&\quad \setminus \big\{FS\big(\langle \file, fn, v, c\rangle\big) \mid (v,c) \in T\big\} \\
		&\quad \setminus \big\{FS\big(\langle \filekey, r, \langle fn, op'\rangle, v, c', id, sig\rangle\big) \mid (r,op', \\
		&\qquad v,c',id,sig) \in T\big\} \\
		&= next\big(\sigma'(x), delP(fn)\big) \\
		&= terminal\big(\sigma'(x),\alpha(\sigma'(x),\ell)\big).
	\end{align*}

	\item \textbf{assignU:}
	If $\ell$ is an instance of $assignU(u,r)$, then $x' = x \cup UR(u,r)$.
	Thus for $\langle \rolekey, SU, (r,1), c, sig\rangle \in FS$ in $\sigma'(x)$,
	\begin{align*}
		\sigma'(x') &= \sigma'\big(x \cup UR(u,r)\big) \\
		&= \sigma'(x) \cup FS\Big(\Big\langle \rolekey, SU, (r,1), \\
		&\quad\: \PKEnc_{\enckey_u}\left(\PKDec_{\deckey_{SU}}(c)\right), \SigSign_{SU}\Big\rangle\Big) \\
		&= next\big(\sigma'(x), assignU(u,r)\big) \\
		&= terminal\big(\sigma'(x),\alpha(\sigma'(x),\ell)\big).
	\end{align*}

	\item \textbf{revokeU:}
	If $\ell$ is an instance of $revokeUser(u,r)$, then $x' = x \setminus UR(u,r)$.
	Let $(\enckey_{(r,v_r+1)}, \deckey_{(r,v_r+1)}) \leftarrow \PKGen$ and $(\verkey_{(r,v_r+1)}, \sigkey_{(r,v_r+1)}) \leftarrow \SigGen$.
	Let $T = \{(u',c_{u'},sig) \mid \langle \rolekey$, $u'$, $(r,v_r)$, $c_{u'}$, $sig\rangle \in FS\}$ and $F =  \{fn \mid \exists(op,v_{fn},c_{fn},sig).(\langle \filekey$, $(r,v_r)$, $\langle fn,op\rangle$, $v_{fn}$, $c_{fn}$, $SU$, $sig\rangle \in FS)$.
	For each $fn \in F$, let $k_{fn} \leftarrow \SymGen$, $T_{fn} = \{(op',v,c_v,sig) \mid \langle \filekey$, $(r,v_r)$, $\langle fn,op'\rangle$, $v$, $c_v$, $SU$, $sig\rangle \in FS\}$ and $T'_{fn} = \{id,op',c_{id},sig) \mid \langle \filekey$, $id$, $\langle fn,op'\rangle$, $v_{fn}$, $c_{id}$, $SU$, $sig\rangle \in FS\}$.
	Then
	\begin{align*}
		\sigma'&(x') = \sigma'\big(x \setminus UR(u,r)\big) \\
		&= \sigma'(x) \setminus \big\{FS\big(\langle\rolekey, u, (r,v_r), c_u, \\
		&\qquad sig\rangle\big) \mid (u,c_u,sig) \in T\big\} \\
		&\cong \sigma'(x) \setminus \big\{FS\big(\langle\rolekey, u', (r,v_r), c_{u'}, \\
		&\qquad sig\rangle\big) \mid (u',c_{u'},sig) \in T\big\} \cup \Big\{FS\Big(\Big\langle\rolekey, u', \\
		&\qquad (r,v_r), \PKEnc_{\enckey_{u'}}\left(\deckey_{(r,v_r+1)}, \sigkey_{(r,v_r+1)}\right), \\
		&\qquad \SigSign_{SU}\Big\rangle\Big) \mid (u',c_{u'},sig) \in T \land u' \neq u\Big\} \\
		&\quad \setminus \big\{FS\big(\langle \filekey, (r,v_r), \langle fn,op'\rangle, v, c_v, \\
		&\qquad SU, sig\rangle\big) \mid fn \in F \land (op',v,c_v,sig) \in T_{fn}\big\} \\
		&\quad \cup \Big\{FS\Big(\Big\langle \filekey, (r,v_r+1), \langle fn,op'\rangle, v, \\
		&\qquad \PKEnc_{\enckey_{(r,v_r+1)}}\left(\PKDec_{\deckey_{(r,v_r)}}(c_v)\right), SU, \\
		&\qquad \SigSign_{SU}\Big\rangle\Big) \mid fn \in F \land (op',v,c_v,sig) \in T_{fn}\Big\} \\
		&\quad \cup \Big\{FS\Big(\Big\langle \filekey, id, \langle fn,op'\rangle, v_{fn}+1, \PKEnc_{\enckey_{id}}(k'_p), \\
		&\qquad SU, \SigSign_{SU}\Big\rangle\Big) \mid fn \in F \land (id,c_{id},sig) \in T'_{fn}\Big\} \\
		&\quad \cup \big\{\text{FILES}(fn,v_{fn}+1) \mid fn \in F\big\} \\
		&\quad \setminus \big\{\text{FILES}(fn,v_{fn}) \mid fn \in F\big\} \\
		&\quad \cup \text{ROLES}\left(r,v_r+1,\enckey_{(r,v_r+1)},\verkey_{(r,v_r+1)}\right) \\
		&\quad \setminus \text{ROLES}\left(r,v_r,\enckey_{(r,v_r)},\verkey_{(r,v_r)}\right) \\
		&= next\big(\sigma'(x), revokeU(u,r)\big) \\
		&= terminal\big(\sigma'(x),\alpha(\sigma'(x),\ell)\big).
	\end{align*}

	\item \textbf{assignP:}
	If $\ell$ is an instance of $assignP(r,p)$ with $p = \langle fn,op\rangle$, then $x' = x \cup PA(r,p)$.
	We have two cases where $assignP(r,p)$ has an effect on $x$:
	\begin{itemize}
		\item If $op = \rwp$ and there exists $\langle \filekey$, $(r,v_r)$, $\langle fn, \readp\rangle$, $v_{fn}$, $c$, $SU$, $sig\rangle$, then let $T = \{(v,c_v,sig) \mid \langle \filekey, (r,v_r),\langle fn, \readp\rangle, v, c_v, SU, sig\rangle \in FS\}$. Then
		\begin{align*}
			\sigma'(x') &= \sigma'\big(x \cup PA(r,p)\big) \\
			&= \sigma'(x) \setminus \big\{FS\big(\langle \filekey, (r,v_r), \langle fn, \readp\rangle, v, c_v, \\
			&\qquad SU, sig\rangle\big) \mid (v,c_v,sig) \in T\big\} \\
			&\quad \cup \big\{FS\Big(\Big\langle \filekey, (r,v_r), \langle fn, \rwp\rangle, v, c_v, SU, \\
			&\qquad \SigSign_{SU}\Big\rangle\Big) \mid (v,c_v,sig) \in T\big\} \\
			&= next\big(\sigma'(x), assignP(r,p)\big) \\
			&= terminal\big(\sigma'(x),\alpha(\sigma'(x),\ell)\big).
		\end{align*}
		\item If there does not exist $\langle \filekey$, $(r,v_r)$, $\langle fn, op'\rangle$, $v_{fn}$, $c$, $SU$, $sig\rangle$, then let $T = \{(v,c_v) \mid \exists(id,sig).(\langle \filekey, SU,\langle fn, \rwp\rangle, v, c_v, id, sig\rangle \in FS)\}$. Then
		\begin{align*}
			\sigma'(x') &= \sigma'\big(x \cup PA(r,p)\big) \\
			&= \sigma'(x) \cup \Big\{FS\Big(\Big\langle \filekey, (r,v_r), \langle fn, op\rangle, v, \\
			&\quad \PKEnc_{\enckey_{(r,v_r)}}\left(\PKDec_{\deckey_{SU}}(c_v)\right), SU, \\
			&\quad \SigSign_{SU}\Big\rangle\Big) \mid (v,c_v) \in T\Big\} \\
			&= next\big(\sigma'(x), assignP(r,p)\big) \\
			&= terminal\big(\sigma'(x),\alpha(\sigma'(x),\ell)\big).
		\end{align*}
	\end{itemize}

	\item \textbf{revokeP:}
	If $\ell$ is an instance of $revokeP(r,p)$ with $p = \langle fn,op\rangle$, then $x' = x \setminus PA(r,p)$.
	\begin{itemize}
		\item If $op = \writep$, then let $T = \{(v,c_v,sig) \mid \langle \filekey, (r,v_r), \langle fn, \rwp\rangle, v, c_v, SU, sig\rangle \in FS\}$. Then
		\begin{align*}
			\sigma'(x') &= \sigma'\big(x \setminus PA(r,p)\big) \\
			&= \sigma'(x) \setminus \big\{FS\big(\langle \filekey, (r,v_r), \langle fn, \rwp\rangle, v, c_v, \\
			&\qquad SU, sig\rangle\big) \mid (v,c_v,sig) \in T\big\} \\
			& \quad \cup \Big\{FS\Big(\Big\langle \filekey, (r,v_r), \langle fn, \readp\rangle, v, c_v, SU, \\
			&\qquad \SigSign_{SU}\Big\rangle\Big) \mid (v,c_v,sig) \in T\Big\} \\
			&= next\big(\sigma'(x), assignP(r,p)\big) \\
			&= terminal\big(\sigma'(x),\alpha(\sigma'(x),\ell)\big).
		\end{align*}
		\item If $op = \readp$, then let $k' \leftarrow \SymGen$, $T = \{(op',v,c_v,sig) \mid \langle \filekey$, $(r,v_r)$, $\langle fn, op'\rangle$, $v$, $c_v$, $SU$, $sig\rangle \in FS\}$, and $T' = \{(id,op') \mid id \neq r \land \exists (c_{id},sig).(\langle \filekey$, $id$, $\langle fn, op'\rangle$, $v_{fn}$, $c_{id}$, $SU$, $sig\rangle \in FS)\}$. Then
		\begin{align*}
			\sigma'&(x') = \sigma'\big(x \setminus PA(r,p)\big) \\
			&= \sigma'(x) \setminus \big\{FS\big(\langle \filekey, (r,v_r), \langle fn, op'\rangle, v, c_v, \\
			&\quad SU, sig\rangle) \mid (op',v,c_v,sig) \in T\big\} \\
			&\cong \sigma'(x) \setminus \big\{FS\big(\langle \filekey, (r,v_r), \langle fn, op'\rangle, v, c_v, \\
			&\qquad SU, sig\rangle\big) \mid (op',v,c_v,sig) \in T\big\} \\
			&\quad \cup \Big\{FS\Big(\Big\langle \filekey, id, \langle fn, op'\rangle, v_{fn}+1, \\
			&\qquad \PKEnc_{\enckey_{id}}(k'), SU, \SigSign_{SU}\Big\rangle\Big) \mid (id,op') \in T\Big\} \\
			&\quad \cup \text{FILES}(fn,v_{fn}+1) \setminus \text{FILES}(fn,v_{fn}) \\
			&= next\big(\sigma'(x), assignP(r,p)\big) \\
			&= terminal\big(\sigma'(x),\alpha(\sigma'(x),\ell)\big).
		\end{align*}
	\end{itemize}

\end{itemize}

\subsubsection{Safety}

The label mapping $\alpha$ is safe by inspection---for any $\rbac_0$ state $x$ and label $\ell$, the PKI label $\alpha(\sigma(x),\ell)$ never revokes or grants authorizations except the images of those that are revoked or granted by $\ell$.\end{IEEEproof}

\end{document}